\documentclass[11pt,a4paper]{article}
\usepackage{jheppub}
\usepackage{amssymb,amsmath}
\usepackage{float}
\expandafter\ifx\csname ProvidesPackage\endcsname\relax\toks0=\expandafter{%
\fi\ProvidesPackage{diagrams}[2008/10/15 v3.93 Paul Taylor's commutative
diagrams]
\toks0=\bgroup}
\ifx\diagram\isundefined\else\expandafter\endinput
\fi

\edef\cdrestoreat{
\noexpand\catcode`\noexpand\@=\the\catcode`\@
\noexpand\catcode`\noexpand\#=\the\catcode`\#
\noexpand\catcode`\noexpand\$=\the\catcode`\$
\noexpand\catcode`\noexpand\<=\the\catcode`\<
\noexpand\catcode`\noexpand\>=\the\catcode`\>
\noexpand\catcode`\noexpand\:=\the\catcode`\:
\noexpand\catcode`\noexpand\;=\the\catcode`\;
\noexpand\catcode`\noexpand\!=\the\catcode`\!
\noexpand\catcode`\noexpand\?=\the\catcode`\?
\noexpand\catcode`\noexpand\+=\the\catcode'53
}\catcode`\@=11 \catcode`\#=6 \catcode`\<=12 \catcode`\>=12 \catcode'53=12
\catcode`\:=12 \catcode`\;=12 \catcode`\!=12 \catcode`\?=12

\ifx\diagram@help@messages\CD@qK\let\diagram@help@messages y\fi

\def\cdps@Rokicki#1{\special{ps:#1}}\let\cdps@dvips\cdps@Rokicki\let
\cdps@RadicalEye\cdps@Rokicki\let\CD@HB\cdps@Rokicki\let\CD@IK\cdps@Rokicki
\let\CD@HB\cdps@Rokicki
\def\cdps@Bechtolsheim#1{\special{dvitps: Literal "#1"}}%
\let\cdps@dvitps\cdps@Bechtolsheim\let\cdps@IntegratedComputerSystems
\cdps@Bechtolsheim
\def\cdps@Clark#1{\special{dvitops: inline #1}}
\let\cdps@dvitops\cdps@Clark
\let\cdps@OzTeX\empty\let\cdps@oztex\empty\let\cdps@Trevorrow\empty
\def\cdps@Coombes#1{\special{ps-string #1}}

\count@=\year\multiply\count@12 \advance\count@\month
\ifnum\count@>24144 
\ifnum\count@>24147 
\errhelp{You may press RETURN and carry on for the time being.}\errmessage{! remain
compatible with your files. Please get it NOW.}\fi\fi

\def\CD@DE{\global\let}\def\CD@RH{\outer\def}

{\escapechar\m@ne\xdef\CD@o{\string\{}\xdef\CD@yC{\string\}}
\catcode`\&=4 \CD@DE\CD@Q=&\xdef\CD@S{\string\&}
\catcode`\$=3 \CD@DE\CD@k=$\CD@DE\CD@ND=$
\xdef\CD@nC{\string\$}\gdef\CD@LG{$$}
\catcode`\_=8 \CD@DE\CD@lJ=_
\obeylines\catcode`\^=7 \CD@DE\@super=^
\ifnum\newlinechar=10 \gdef\CD@uG{^^J}
\else\ifnum\newlinechar=13 \gdef\CD@uG{^^M}
\else\ifnum\newlinechar=-1 \gdef\CD@uG{^^J}
\else\CD@DE\CD@uG\space\expandafter\message{! input error: \noexpand
\newlinechar\space is ASCII \the\newlinechar, not LF=10 or CR=13.}
\fi\fi\fi}

\mathchardef\lessthan='30474 \mathchardef\greaterthan='30476

\ifx\tenln\CD@qK
\font\tenln=line10\relax
\fi\ifx\tenlnw\CD@qK\ifx\tenln\nullfont\let\tenlnw\nullfont\else
\font\tenlnw=linew10\relax
\fi\fi

\ifx\inputlineno\CD@qK\csname newcount\endcsname\inputlineno\inputlineno\m@ne
\fi

\def\cd@shouldnt#1{\CD@KB{* THIS (#1) SHOULD NEVER HAPPEN! *}}

\def\get@round@pair#1(#2,#3){#1{#2}{#3}}
\def\get@square@arg#1[#2]{#1{#2}}
\def\CD@AE#1{\CD@PK\let\CD@DH\CD@@E\CD@@E#1,],}
\def\CD@m{[}\def\CD@RD{]}\def\commdiag#1{{\let\enddiagram\relax\diagram[]#1%
\enddiagram}}

\def\CD@BF{{\ifx\CD@EH[\aftergroup\get@square@arg\aftergroup\CD@YH\else
\aftergroup\CD@JH\fi}}
\def\CD@CF#1#2{\def\CD@YH{#1}\def\CD@JH{#2}\futurelet\CD@EH\CD@BF}

\def\CD@KK{|}

\def\CD@PB{
\tokcase\CD@DD:\CD@y\break@args;\catcase\@super:\upper@label;\catcase\CD@lJ:%
\lower@label;\tokcase{~}:\middle@label;
\tokcase<:\CD@iF;
\tokcase>:\CD@iI;
\tokcase(:\CD@BC;
\tokcase[:\optional@;
\tokcase.:\CD@JJ;
\catcase\space:\eat@space;\catcase\bgroup:\positional@;\default:\CD@@A
\break@args;\endswitch}

\def\switch@arg{
\catcase\@super:\upper@label;\catcase\CD@lJ:\lower@label;\tokcase[:\optional@
;
\tokcase.:\CD@JJ;
\catcase\space:\eat@space;\catcase\bgroup:\positional@;\tokcase{~}:%
\middle@label;
\default:\CD@y\break@args;\endswitch}

\let\CD@tJ\relax\ifx\protect\CD@qK\let\protect\relax\fi\ifx\AtEndDocument
\CD@qK\def\CD@PG{\CD@gB}\def\CD@GF#1#2{}\else\def\CD@PG#1{\edef\CD@CH{#1}%
\expandafter\CD@oC\CD@CH\CD@OD}\def\CD@oC#1\CD@OD{\AtEndDocument{\typeout{%
\CD@tA: #1}}}\def\CD@GF#1#2{\gdef#1{#2}\AtEndDocument{#1}}\fi\def\CD@ZA#1#2{%
\def#1{\CD@PG{#2\CD@mD\CD@W}\CD@DE#1\relax}}\def\CD@uF#1\repeat{\def\CD@p{#1}%
\CD@OF}\def\CD@OF{\CD@p\relax\expandafter\CD@OF\fi}\def\CD@sF#1\repeat{\def
\CD@q{#1}\CD@PF}\def\CD@PF{\CD@q\relax\expandafter\CD@PF\fi}\def\CD@tF#1%
\repeat{\def\CD@r{#1}\CD@QF}\def\CD@QF{\CD@r\relax\expandafter\CD@QF\fi}\def
\CD@tG#1#2#3{\def#2{\let#1\iftrue}\def#3{\let#1\iffalse}#3}\if y%
\diagram@help@messages\def\CD@rG#1#2{\csname newtoks\endcsname#1#1=%
\expandafter{\csname#2\endcsname}}\else\csname newtoks\endcsname\no@cd@help
\no@cd@help{See the manual}\def\CD@rG#1#2{\let#1\no@cd@help}\fi\chardef\CD@lF
=1 \chardef\CD@lI=2 \chardef\CD@MH=5 \chardef\CD@tH=6 \chardef\CD@sH=7
\chardef\CD@PC=9 \dimendef\CD@hI=2 \dimendef\CD@hF=3 \dimendef\CD@mF=4
\dimendef\CD@mI=5 \dimendef\CD@wJ=6 \dimendef\CD@tI=8 \dimendef\CD@sI=9
\skipdef\CD@uB=1 \skipdef\CD@NF=2 \skipdef\CD@tB=3 \skipdef\CD@ZE=4 \skipdef
\countdef\CD@JC=9 \countdef\CD@gD=8 \countdef\CD@A=7 \def\sdef#1#2{\def#1{#2}%
}\def\CD@L#1{\expandafter\aftergroup\csname#1\endcsname}\def\CD@RC#1{%
\expandafter\def\csname#1\endcsname}\def\CD@sD#1{\expandafter\gdef\csname#1%
\endcsname}\def\CD@vC#1{\expandafter\edef\csname#1\endcsname}\def\CD@nF#1#2{%
\expandafter\let\csname#1\expandafter\endcsname\csname#2\endcsname}\def\CD@EE
#1#2{\expandafter\CD@DE\csname#1\expandafter\endcsname\csname#2\endcsname}%
\def\CD@AK#1{\csname#1\endcsname}\def\CD@XJ#1{\expandafter\show\csname#1%
\endcsname}\def\CD@ZJ#1{\expandafter\showthe\csname#1\endcsname}\def\CD@WJ#1{%
\expandafter\showbox\csname#1\endcsname}\def\CD@tA{Commutative Diagram}\edef
\CD@kH{\string\par}\edef\CD@dC{\string\diagram}\edef\CD@HD{\string\enddiagram
}\edef\CD@EC{\string\\}\def\CD@eF{LaTeX}\ifx\@ignoretrue\CD@qK\expandafter
\CD@tG\csname if@ignore\endcsname\ignore@true\ignore@false\def\@ignoretrue{%
\global\ignore@true}\def\@ignorefalse{\global\ignore@false}\fi

\def\CD@g{{\ifnum0=`}\fi}\def\CD@wC{\ifnum0=`{\fi}}\def\catcase#1:{\ifcat
\noexpand\CD@EH#1\CD@tJ\expandafter\CD@kC\else\expandafter\CD@dJ\fi}\def
\tokcase#1:{\ifx\CD@EH#1\CD@tJ\expandafter\CD@kC\else\expandafter\CD@dJ\fi}%
\def\CD@kC#1;#2\endswitch{#1}\def\CD@dJ#1;{}\let\endswitch\relax\def\default:%
#1;#2\endswitch{#1}\ifx\at@\CD@qK\def\at@{@}\fi\edef\CD@P{\CD@o pt\CD@yC}%
\CD@RC{\CD@P>}#1>#2>{\CD@z\rTo\sp{#1}\sb{#2}\CD@z}\CD@RC{\CD@P<}#1<#2<{\CD@z
\lTo\sp{#1}\sb{#2}\CD@z}\CD@RC{\CD@P)}#1)#2){\CD@z\rTo\sp{#1}\sb{#2}\CD@z}%
\CD@RC{\CD@P(}#1(#2({\CD@z\lTo\sp{#1}\sb{#2}\CD@z}
\def\CD@O{\def\endCD{\enddiagram}\CD@RC{\CD@P A}##1A##2A{\uTo<{##1}>{##2}%
\CD@z\CD@z}\CD@RC{\CD@P V}##1V##2V{\dTo<{##1}>{##2}\CD@z\CD@z}\CD@RC{\CD@P=}{%
\CD@z\hEq\CD@z}\CD@RC{\CD@P\CD@KK}{\vEq\CD@z\CD@z}\CD@RC{\CD@P\string\vert}{%
\vEq\CD@z\CD@z}\CD@RC{\CD@P.}{\CD@z\CD@z}\let\CD@z\CD@Q}\def\CD@IE{\let\tmp
\CD@JE\ifcat A\noexpand\CD@CH\else\ifcat=\noexpand\CD@CH\else\ifcat\relax
\noexpand\CD@CH\else\let\tmp\at@\fi\fi\fi\tmp}\def\CD@JE#1{\CD@nF{tmp}{\CD@P
\string#1}\ifx\tmp\relax\def\tmp{\at@#1}\fi\tmp}\def\CD@z{}\begingroup
\aftergroup\def\aftergroup\CD@T\aftergroup{\aftergroup\def\catcode`\@\active
\aftergroup @\endgroup{\futurelet\CD@CH\CD@IE}}\newcount\CD@uA\newcount\CD@vA
\newcount\CD@wA\newcount\CD@xA\newdimen\CD@OA\newdimen\CD@PA\CD@tG\CD@gE
\CD@@A\CD@y\CD@tG\CD@hE\CD@EA\CD@BA\newdimen\CD@RA\newdimen\CD@SA\newcount
\CD@yA\newcount\CD@zA\newdimen\CD@QA\newbox\CD@DA\CD@tG\CD@lE\CD@dA\CD@bA
\newcount\CD@LH\newcount\CD@TC\def\CD@V#1#2{\ifdim#1<#2\relax#1=#2\relax\fi}%
\def\CD@X#1#2{\ifdim#1>#2\relax#1=#2\relax\fi}\newdimen\CD@XH\CD@XH=1sp
\newdimen\CD@zC\CD@zC\z@\def\CD@cJ{\ifdim\CD@zC=1em\else\CD@nJ\fi}\def\CD@nJ{%
\CD@zC1em\def\CD@NC{\fontdimen8\textfont3 }\CD@@J\CD@NJ\setbox0=\vbox{\CD@t
\noindent\CD@k\null\penalty-9993\null\CD@ND\null\endgraf\setbox0=\lastbox
\unskip\unpenalty\setbox1=\lastbox\global\setbox\CD@IG=\hbox{\unhbox0\unskip
\unskip\unpenalty\setbox0=\lastbox}\global\setbox\CD@KG=\hbox{\unhbox1\unskip
\unpenalty\setbox1=\lastbox}}}\newdimen\CD@@I\CD@@I=1true in \divide\CD@@I300
\def\CD@zH#1{\multiply#1\tw@\advance#1\ifnum#1<\z@-\else+\fi\CD@@I\divide#1%
\tw@\divide#1\CD@@I\multiply#1\CD@@I}\def\MapBreadth{\afterassignment\CD@gI
\CD@LF}\newdimen\CD@LF\newdimen\CD@oI\def\CD@gI{\CD@oI\CD@LF\CD@V\CD@@I{4%
\CD@XH}\CD@X\CD@@I\p@\CD@zH\CD@oI\ifdim\CD@LF>\z@\CD@V\CD@oI\CD@@I\fi\CD@cJ}%
\def\CD@RJ#1{\CD@zD\count@\CD@@I#1\ifnum\count@>\z@\divide\CD@@I\count@\fi
\CD@gI\CD@NJ}\def\CD@NJ{\dimen@\CD@QC\count@\dimen@\divide\count@5\divide
\count@\CD@@I\edef\CD@OC{\the\count@}}\def\CD@AJ{\CD@QJ\z@}\def\CD@QJ#1{%
\CD@tI\axisheight\advance\CD@tI#1\relax\advance\CD@tI-.5\CD@oI\CD@zH\CD@tI
\CD@sI-\CD@tI\advance\CD@tI\CD@LF}\newdimen\CD@DC\CD@DC\z@\newdimen\CD@eJ
\CD@eJ\z@\def\CD@CJ#1{\CD@sI#1\relax\CD@tI\CD@sI\advance\CD@tI\CD@LF\relax}%
\def\horizhtdp{height\CD@tI depth\CD@sI}\def\axisheight{\fontdimen22\the
\textfont\tw@}\def\script@axisheight{\fontdimen22\the\scriptfont\tw@}\def
\ss@axisheight{\fontdimen22\the\scriptscriptfont\tw@}\def\CD@NC{0.4pt}\def
\CD@VK{\fontdimen3\textfont\z@}\def\CD@UK{\fontdimen3\textfont\z@}\newdimen
\PileSpacing\newdimen\CD@nA\CD@nA\z@\def\CD@RG{\ifincommdiag1.3em\else2em\fi}%
\newdimen\CD@YB\def\CellSize{\afterassignment\CD@kB\DiagramCellHeight}%
\newdimen\DiagramCellHeight\DiagramCellHeight-\maxdimen\newdimen
\DiagramCellWidth\DiagramCellWidth-\maxdimen\def\CD@kB{\DiagramCellWidth
\DiagramCellHeight}\def\CD@QC{3em}\newdimen\MapShortFall\def\MapsAbut{%
\MapShortFall\z@\objectheight\z@\objectwidth\z@}\newdimen\CD@iA\CD@iA\z@
\CD@tG\CD@vE\CD@aB\CD@ZB\expandafter\ifx\expandafter\iftrue\csname
ifUglyObsoleteDiagrams\endcsname\CD@ZB\else\CD@aB\fi\CD@nF{%
ifUglyObsoleteDiagrams}{relax}\newif\ifUglyObsoleteDiagrams\def\CD@nK{\CD@aB
\UglyObsoleteDiagramsfalse}\def\CD@oK{\CD@ZB\UglyObsoleteDiagramstrue}\CD@vE
\CD@nK\else\CD@oK\fi\CD@tG\CD@hK\CD@dK\CD@cK\CD@cK\def\CD@sK{\ifx\pdfoutput
\CD@qK\else\ifx\pdfoutput\relax\else\ifnum\pdfoutput>\z@\CD@pK\fi\fi\fi} \def
\CD@pK{\global\CD@dK\global\CD@aB\global\UglyObsoleteDiagramsfalse\global\let
\CD@n\empty\global\let\CD@oK\relax\global\let\CD@pK\relax\global\let\CD@sK
\relax}\def\CD@tK#1{}\ifx\pdfliteral\CD@qK\else\ifx
\pdfliteral\relax\else\let\CD@tK\pdfliteral\fi\fi\ifx\XeTeXrevision\CD@qK
\else\ifx\XeTeXrevision\relax\else\ifdim\XeTeXrevision pt<.996pt \expandafter
\message{! XeTeX version \XeTeXrevision\space does not support PDF literals,
so diagonals will not work!}\else\expandafter\message{RUNNING UNDER XETEX
\XeTeXrevision}\CD@pK\fi\fi\fi\CD@sK\def\newarrowhead{\CD@mG h\CD@BG\CD@GG>}%
\def\newarrowtail{\CD@mG t\CD@BG\CD@GG>}\def\newarrowmiddle{\CD@mG m\CD@BG
\hbox@maths\empty}\def\newarrowfiller{\CD@mG f\CD@bE\CD@MK-}\def\CD@mG#1#2#3#%
4#5#6#7#8#9{\CD@RC{r#1:#5}{#2{#6}}\CD@RC{l#1:#5}{#2{#7}}\CD@RC{d#1:#5}{#3{#8}%
}\CD@RC{u#1:#5}{#3{#9}}\CD@vC{-#1:#5}{\expandafter\noexpand\csname-#1:#4%
\endcsname\noexpand\CD@MC}\CD@vC{+#1:#5}{\expandafter\noexpand\csname+#1:#4%
\endcsname\noexpand\CD@MC}}\CD@ZA\CD@MC{\CD@eF\space diagonals are used unless
PostScript is set}\def\defaultarrowhead#1{\edef\CD@sJ{#1}\CD@@J}\def\CD@@J{%
\CD@IJ\CD@sJ<>ht\CD@IJ\CD@sJ<>th}\def\CD@IJ#1#2#3#4#5{\CD@HJ{r#4}{#3}{l#5}{#2%
}{r#4:#1}\CD@HJ{r#5}{#2}{l#4}{#3}{l#4:#1}\CD@HJ{d#4}{#3}{u#5}{#2}{d#4:#1}%
\CD@HJ{d#5}{#2}{u#4}{#3}{u#4:#1}}\def\CD@HJ#1#2#3#4#5{\begingroup\aftergroup
\CD@GJ\CD@L{#1+:#2}\CD@L{#1:#2}\CD@L{#3:#4}\CD@L{#5}\endgroup}\def\CD@GJ#1#2#%
3#4{\csname newbox\endcsname#1\def#2{\copy#1}\def#3{\copy#1}\setbox#1=\box
\voidb@x}\def\CD@sJ{}\CD@@J\def\CD@GJ#1#2#3#4{\setbox#1=#4}\ifx\tenln
\nullfont\def\CD@sJ{vee}\else\let\CD@sJ\CD@eF\fi\def\CD@xF#1#2#3{\begingroup
\aftergroup\CD@wF\CD@L{#1#2:#3#3}\CD@L{#1#2:#3}\aftergroup\CD@yF\CD@L{#1#2:#3%
-#3}\CD@L{#1#2:#3}\endgroup}\def\CD@wF#1#2{\def#1{\hbox{\rlap{#2}\kern.4%
\CD@zC#2}}}\def\CD@yF#1#2{\def#1{\hbox{\rlap{#2}\kern.4\CD@zC#2\kern-.4\CD@zC
}}}\CD@xF lh>\CD@xF rt>\CD@xF rh<\CD@xF rt<\def\CD@yF#1#2{\def#1{\hbox{\kern-%
.4\CD@zC\rlap{#2}\kern.4\CD@zC#2}}}\CD@xF rh>\CD@xF lh<\CD@xF lt>\CD@xF lt<%
\def\CD@wF#1#2{\def#1{\vbox{\vbox to\z@{#2\vss}\nointerlineskip\kern.4\CD@zC#%
2}}}\def\CD@yF#1#2{\def#1{\vbox{\vbox to\z@{#2\vss}\nointerlineskip\kern.4%
\CD@zC#2\kern-.4\CD@zC}}}\CD@xF uh>\CD@xF dt>\CD@xF dh<\CD@xF dt<\def\CD@yF#1%
#2{\def#1{\vbox{\kern-.4\CD@zC\vbox to\z@{#2\vss}\nointerlineskip\kern.4%
\CD@zC#2}}}\CD@xF dh>\CD@xF ut>\CD@xF uh<\CD@xF ut<\def\CD@BG#1{\hbox{%
\mathsurround\z@\offinterlineskip\CD@k\mkern-1.5mu{#1}\mkern-1.5mu\CD@ND}}%
\def\hbox@maths#1{\hbox{\CD@k#1\CD@ND}}\def\CD@GG#1{\hbox to\CD@LF{\setbox0=%
\hbox{\offinterlineskip\mathsurround\z@\CD@k{#1}\CD@ND}\dimen0.5\wd0\advance
\dimen0-.5\CD@oI\CD@zH{\dimen0}\kern-\dimen0\unhbox0\hss}}\def\CD@sB#1{\hbox
to2\CD@LF{\hss\offinterlineskip\mathsurround\z@\CD@k{#1}\CD@ND\hss}}\def
\CD@vF#1{\hbox{\mathsurround\z@\CD@k{#1}\CD@ND}}\def\CD@bE#1{\hbox{\kern-.15%
\CD@zC\CD@k{#1}\CD@ND\kern-.15\CD@zC}}\def\CD@MK#1{\vbox{\offinterlineskip
\kern-.2ex\CD@GG{#1}\kern-.2ex}}\def\@fillh{\xleaders\vrule\horizhtdp}\def
\@fillv{\xleaders\hrule width\CD@LF}\CD@nF{rf:-}{@fillh}\CD@nF{lf:-}{@fillh}%
\CD@nF{df:-}{@fillv}\CD@nF{uf:-}{@fillv}\CD@nF{rh:}{null}\CD@nF{rm:}{null}%
\CD@nF{rt:}{null}\CD@nF{lh:}{null}\CD@nF{lm:}{null}\CD@nF{lt:}{null}\CD@nF{dh%
:}{null}\CD@nF{dm:}{null}\CD@nF{dt:}{null}\CD@nF{uh:}{null}\CD@nF{um:}{null}%
\CD@nF{ut:}{null}\CD@nF{+h:}{null}\CD@nF{+m:}{null}\CD@nF{+t:}{null}\CD@nF{-h%
:}{null}\CD@nF{-m:}{null}\CD@nF{-t:}{null}\def\CD@@D{\hbox{\vrule height 1pt
depth-1pt width 1pt}}\CD@RC{rf:}{\CD@@D}\CD@nF{lf:}{rf:}\CD@nF{+f:}{rf:}%
\CD@RC{df:}{\CD@@D}\CD@nF{uf:}{df:}\CD@nF{-f:}{df:}\def\CD@BD{\CD@U\null
\CD@@D\null\CD@@D\null}\edef\CD@lG{\string\newarrow}\def\newarrow#1#2#3#4#5#6%
{\begingroup\edef\@name{#1}\edef\CD@oJ{#2}\edef\CD@iD{#3}\edef\CD@QG{#4}\edef
\CD@jD{#5}\edef\CD@LE{#6}\let\CD@HE\CD@sG\let\CD@FK\CD@BH\let\@x\CD@AH\ifx
\CD@oJ\CD@iD\let\CD@oJ\empty\fi\ifx\CD@LE\CD@jD\let\CD@LE\empty\fi\def\CD@LI{%
r}\def\CD@SF{l}\def\CD@IC{d}\def\CD@yJ{u}\def\CD@gH{+}\def\@m{-}\ifx\CD@iD
\CD@jD\ifx\CD@QG\CD@iD\let\CD@QG\empty\fi\ifx\CD@LE\empty\ifx\CD@iD\CD@aE\let
\@x\CD@yG\else\let\@x\CD@zG\fi\fi\else\edef\CD@a{\CD@iD\CD@oJ}\ifx\CD@a\empty
\ifx\CD@QG\CD@jD\let\CD@QG\empty\fi\fi\fi\ifmmode\aftergroup\CD@kG\else\CD@@A
\CD@oB rh{head\space\space}\CD@LE\CD@oB rf{filler}\CD@iD\CD@oB rm{middle}%
\CD@QG\ifx\CD@jD\CD@iD\else\CD@oB rf{filler}\CD@jD\fi\CD@oB rt{tail\space
\space}\CD@oJ\CD@gE\CD@HE\CD@FK\@x\CD@nG l-2+2{lu}{nw}\NorthWest\CD@nG r+2+2{%
ru}{ne}\NorthEast\CD@nG l-2-2{ld}{sw}\SouthWest\CD@nG r+2-2{rd}{se}\SouthEast
\else\aftergroup\CD@b\CD@L{r\@name}\fi\fi\endgroup}\def\CD@sG{\CD@vG\CD@LI
\CD@SF rl\Horizontal@Map}\def\CD@BH{\CD@vG\CD@IC\CD@yJ du\Vertical@Map}\def
\CD@AH{\CD@vG\CD@gH\@m+-\Vector@Map}\def\CD@yG{\CD@vG\CD@gH\@m+-\Slant@Map}%
\def\CD@zG{\CD@vG\CD@gH\@m+-\Slope@Map}\catcode`\/=\active\def\CD@vG#1#2#3#4#%
5{\CD@jG#1#3#5t:\CD@oJ/f:\CD@iD/m:\CD@QG/f:\CD@jD/h:\CD@LE//\CD@jG#2#4#5h:%
\CD@LE/f:\CD@jD/m:\CD@QG/f:\CD@iD/t:\CD@oJ//}\def\CD@jG#1#2#3#4//{\edef\CD@fG
{#2}\aftergroup\sdef\CD@L{#1\@name}\aftergroup{\aftergroup#3\CD@M#4//%
\aftergroup}}\def\CD@M#1/{\edef\CD@EH{#1}\ifx\CD@EH\empty\else\CD@L{\CD@fG#1}%
\expandafter\CD@M\fi}\catcode`\/=12 \def\CD@nG#1#2#3#4#5#6#7#8{\aftergroup
\sdef\CD@L{#6\@name}\aftergroup{\CD@L{#2\@name}\if#2#4\aftergroup\CD@CI\else
\aftergroup\CD@BI\fi\CD@L{#1\@name}%
\aftergroup(\aftergroup#3\aftergroup,\aftergroup#5\aftergroup)\aftergroup}}%
\def\CD@oB#1#2#3#4{\expandafter\ifx\csname#1#2:#4\endcsname\relax\CD@y\CD@gB{%
arrow#3 "#4" undefined}\fi}\CD@rG\CD@VE{All five components must be defined
before an arrow.}\CD@rG\CD@SE{\CD@lG, unlike \string\HorizontalMap, is a
declaration.}\def\CD@b#1{\CD@YA{Arrows \string#1 etc could not be defined}%
\CD@VE}\def\CD@kG{\CD@YA{misplaced \CD@lG}\CD@SE}\def\newdiagramgrid#1#2#3{%
\CD@RC{cdgh@#1}{#2,],}
\CD@RC{cdgv@#1}{#3,],}}
\CD@tG\ifincommdiag\incommdiagtrue\incommdiagfalse\CD@tG\CD@@F\CD@IF\CD@HF
\newcount\CD@VA\CD@VA=0 \def\CD@yH{\CD@VA6 }\def\CD@OB{\CD@VA1 \global\CD@yA1
\CD@DE\CD@YF\empty}\def\CD@YF{}\def\CD@nB#1{\relax\CD@MD\edef\CD@vJ{#1}%
\begingroup\CD@rE\else\ifcase\CD@VA\ifmmode\else\CD@YG\CD@E0\fi\or\CD@cE5\or
\CD@YG\CD@F5\or\CD@YG\CD@B5\or\CD@YG\CD@B5\or\CD@YG\CD@C5\or\CD@cE7\or\CD@YG
\CD@D7\fi\fi\endgroup\xdef\CD@YF{#1}}\def\CD@pB#1#2#3#4#5{\relax\CD@MD\xdef
\CD@vJ{#4}\begingroup\ifnum\CD@VA<#1 \expandafter\CD@cE\ifcase\CD@VA0\or#2\or
#3\else#2\fi\else\ifnum\CD@VA<6 \CD@tJ\CD@YG\CD@B#2\else\CD@YG\CD@G#2\fi\fi
\endgroup\CD@DE\CD@YF\CD@vJ\ifincommdiag\let\CD@ZD#5\else\let\CD@ZD\CD@LK\fi}%
\def\CD@yI{\global\CD@yA=\ifnum\CD@VA<5 1\else2\fi\relax}\def\CD@OI{\CD@VA
\CD@yA}\def\CD@cE#1{\aftergroup\CD@VA\aftergroup#1\aftergroup\relax}\def
\CD@HH{\def\CD@nB##1{\relax}\let\CD@pB\CD@FH\let\CD@yH\relax\let\CD@OB\relax
\let\CD@yI\relax\let\CD@OI\relax}\def\CD@FH#1#2#3#4#5{\ifincommdiag\let\CD@ZD
#5\else\xdef\CD@vJ{#4}\let\CD@ZD\CD@LK\fi}\def\CD@YG#1{\aftergroup#1%
\aftergroup\relax\CD@cE}\def\CD@B{\CD@YE\CD@S\CD@ME\CD@Q}\def\CD@G{\CD@YE{%
\CD@yC\CD@S}\CD@XE\CD@QD\CD@Q}\def\CD@F{\CD@YE{*\CD@S}\CD@RE\clubsuit\CD@Q}%
\def\CD@C{\CD@YE{\CD@S*\CD@S}\CD@RE\CD@Q\clubsuit\CD@Q}\def\CD@D{\CD@YE\CD@EC
\CD@TE\\}\def\CD@E{\CD@YE\CD@nC\CD@QE\CD@k}\def\CD@LK{\CD@YA{\CD@vJ\space
ignored \CD@dH}\CD@WE}\def\CD@FE{}\def\CD@d{\CD@YA{maps must never be enclosed
in braces}\CD@OE}\def\CD@dH{outside diagram}\def\CD@FC{\string\HonV, \string
\VonH\space and \string\HmeetV}\CD@rG\CD@ME{The way that horizontal and
vertical arrows are terminated implicitly means\CD@uG that they cannot be
mixed with each other or with \CD@FC.}\CD@rG\CD@XE{\string\pile\space is for
parallel horizontal arrows; verticals can just be put together in\CD@uG a cell%
. \CD@FC\space are not meaningful in a \string\pile.}\CD@rG\CD@RE{The
horizontal maps must point to an object, not each other (I've put in\CD@uG one
which you're unlikely to want). Use \string\pile\space if you want them
parallel.}\CD@rG\CD@TE{Parallel horizontal arrows must be in separate layers
of a \string\pile.}\CD@rG\CD@QE{Horizontal arrows may be used \CD@dH s, but
must still be in maths.}\CD@rG\CD@WE{Vertical arrows, \CD@FC\space\CD@dH s don%
't know where\CD@uG where to terminate.}\CD@rG\CD@OE{This prevents them from
stretching correctly.}\def\CD@YE#1{\CD@YA{"#1" inserted \ifx\CD@YF\empty
before \CD@vJ\else between \CD@YF\ifx\CD@YF\CD@vJ s\else\space and \CD@vJ\fi
\fi}}\count@=\year\multiply\count@12 \advance\count@\month\ifnum\count@>24151
\expandafter\endinput\fi\def
\Horizontal@Map{\CD@nB{horizontal map}\CD@LC\CD@TJ\CD@qD}\def\CD@TJ{\CD@GB-%
9999 \let\CD@ZD\CD@XD\ifincommdiag\else\CD@cJ\ifinpile\else\skip2\z@ plus 1.5%
\CD@VK minus .5\CD@UK\skip4\skip2 \fi\fi\let\CD@kD\@fillh\CD@nF{fill@dot}{rf:%
.}}\def\Vector@Map{\CD@HK4}\def\Slant@Map{\CD@HK{\CD@EF255\else6\fi}}\def
\Slope@Map{\CD@HK\CD@OC}\def\CD@HK#1#2#3#4#5#6{\CD@LC\def\CD@WK{2}\def\CD@aK{%
2}\def\CD@ZK{1}\def\CD@bK{1}\let\Horizontal@Map\CD@nI\def\CD@OG{#1}\def\CD@NI
{\CD@U#2#3#4#5#6}}\def\CD@nI{\CD@TJ\CD@JB\let\CD@ZD\CD@TD\CD@qD}\CD@tG\CD@pE
\CD@rA\CD@qA\CD@rA\def\cds@missives{\CD@rA}\def\CD@TD{\CD@vE\let\CD@OG\CD@OC
\CD@x\CD@zE\CD@WF\fi\setbox0\hbox{\incommdiagfalse\CD@HI}\CD@pE\CD@aD\else
\global\CD@YC\CD@bD\fi\ifvoid6 \ifvoid7 \CD@eE\fi\fi\CD@zE\else\CD@BD\global
\CD@YC\let\CD@CG\CD@IH\CD@YD\fi\else\CD@NI\CD@MI\global\CD@YC\CD@YD\fi}\def
\CD@LC{\begingroup\dimen1=\MapShortFall\dimen2=\CD@RG\dimen5=\MapShortFall
\setbox3=\box\voidb@x\setbox6=\box\voidb@x\setbox7=\box\voidb@x\CD@pD
\mathsurround\z@\skip2\z@ plus1fill minus 1000pt\skip4\skip2 \CD@TB}\CD@tG
\CD@tE\CD@UB\CD@TB\def\CD@U#1#2#3#4#5{\let\CD@oJ#1\let\CD@iD#2\let\CD@QG#3%
\let\CD@jD#4\let\CD@LE#5\CD@TB\ifx\CD@iD\CD@jD\CD@UB\fi}\def\CD@qD#1#2#3#4#5{%
\CD@U#1#2#3#4#5\CD@tD}\def\Vertical@Map{\CD@pB433{vertical map}\CD@cD\CD@LC
\CD@GB-9995 \let\CD@kD\@fillv\CD@nF{fill@dot}{df:.}\CD@qD}\def\break@args{%
\def\CD@tD{\CD@ZD}\CD@ZD\endgroup\aftergroup\CD@FE}\def\CD@MJ{\setbox1=\CD@oJ
\setbox5=\CD@LE\ifvoid3 \ifx\CD@QG\null\else\setbox3=\CD@QG\fi\fi\CD@@G2%
\CD@iD\CD@@G4\CD@jD}\def\CD@pF#1{\ifvoid1\else\CD@oF1#1\fi\ifvoid2\else\CD@oF
2#1\fi\ifvoid3\else\CD@oF3#1\fi\ifvoid4\else\CD@oF4#1\fi\ifvoid5\else\CD@oF5#%
1\fi} \def\CD@oF#1#2{\setbox#1\vbox{\offinterlineskip\box#1\dimen@\prevdepth
\advance\dimen@-#2\relax\setbox0\null\dp0\dimen@\ht0-\dimen@\box0}}\def\CD@@G
#1#2{\ifx#2\CD@kD\setbox#1=\box\voidb@x\else\setbox#1=#2\def#2{\xleaders\box#%
1}\fi}\CD@ZA\CD@BK{\string\HorizontalMap, \string\VerticalMap\space and
\string\DiagonalMap\CD@uG are obsolete - use \CD@lG\space to pre-define maps}%
\def\HorizontalMap#1#2#3#4#5{\CD@BK\CD@nB{old horizontal map}\CD@LC\CD@TJ\def
\CD@oJ{\CD@UH{#1}}\CD@SH\CD@iD{#2}\def\CD@QG{\CD@UH{#3}}\CD@SH\CD@jD{#4}\def
\CD@LE{\CD@UH{#5}}\CD@tD}\def\VerticalMap#1#2#3#4#5{\CD@BK\CD@pB433{vertical
map}\CD@cD\CD@LC\CD@GB-9995 \let\CD@kD\@fillv\def\CD@oJ{\CD@GG{#1}}\CD@VH
\CD@iD{#2}\def\CD@QG{\CD@GG{#3}}\CD@VH\CD@jD{#4}\def\CD@LE{\CD@GG{#5}}\CD@tD}%
\def\DiagonalMap#1#2#3#4#5{\CD@BK\CD@LC\def\CD@OG{4}\let\CD@kD\CD@qK\let
\CD@ZD\CD@YD\def\CD@WK{2}\def\CD@aK{2}\def\CD@ZK{1}\def\CD@bK{1}\def\CD@QG{%
\CD@vF{#3}}\ifPositiveGradient\let\mv\raise\def\CD@oJ{\CD@vF{#5}}\def\CD@iD{%
\CD@vF{#4}}\def\CD@jD{\CD@vF{#2}}\def\CD@LE{\CD@vF{#1}}\else\let\mv\lower\def
\CD@oJ{\CD@vF{#1}}\def\CD@iD{\CD@vF{#2}}\def\CD@jD{\CD@vF{#4}}\def\CD@LE{%
\CD@vF{#5}}\fi\CD@tD}\def\CD@aE{-}\def\CD@AD{\empty}\def\CD@SH{\CD@EG\CD@bE
\CD@aE\@fillh}\def\CD@VH{\CD@EG\CD@MK\CD@KK\@fillv}\def\CD@EG#1#2#3#4#5{\def
\CD@CH{#5}\ifx\CD@CH#2\let#4#3\else\let#4\null\ifx\CD@CH\empty\else\ifx\CD@CH
\CD@AD\else\let#4\CD@CH\fi\fi\fi}\def\CD@UH#1{\hbox{\mathsurround\z@
\offinterlineskip\def\CD@CH{#1}\ifx\CD@CH\empty\else\ifx\CD@CH\CD@AD\else
\CD@k\mkern-1.5mu{\CD@CH}\mkern-1.5mu\CD@ND\fi\fi}}\def\CD@yD#1#2{\setbox#1=%
\hbox\bgroup\setbox0=\hbox{\CD@k\labelstyle()\CD@ND}
\setbox1=\null\ht1\ht0\dp1\dp0\box1 \kern.1\CD@zC\CD@k\bgroup\labelstyle
\aftergroup\CD@LD\CD@xD}\def\CD@LD{\CD@ND\kern.1\CD@zC\egroup\CD@tD}\def
\CD@xD{\futurelet\CD@EH\CD@mJ}\def\CD@mJ{
\catcase\bgroup:\CD@v;\catcase\egroup:\missing@label;\catcase\space:\CD@TF;%
\tokcase[:\CD@XF;
\default:\CD@zJ;\endswitch}\def\CD@v{\let\CD@MD\CD@c\let\CD@CH}\def\CD@zJ#1{%
\let\CD@UF\egroup{\let\actually@braces@missing@around@macro@in@label\CD@ZH
\let\CD@MD\CD@xC\let\CD@UF\CD@VF#1%
\actually@braces@missing@around@macro@in@label}\CD@UF}\def
\actually@braces@missing@around@macro@in@label{\let\CD@CH=}\def\missing@label
{\egroup\CD@YA{missing label}\CD@PE}\def\CD@xC{\egroup\missing@label}\outer
\def\CD@ZH{}\def\CD@UF{}\def\CD@VF{\CD@wC\CD@UF}\def\CD@MD{}\def\CD@XF{\let
\CD@N\CD@xD\get@square@arg\CD@AE}\CD@rG\CD@PE{The text which has just been
read is not allowed within map labels.}\def\CD@c{\egroup\CD@YA{missing \CD@yC
\space inserted after label}\CD@PE}\def\upper@label{\CD@oD\CD@yD6}\def
\lower@label{\def\positional@{\CD@@A\break@args}\CD@yD7}\def\middle@label{%
\CD@yD3}\CD@tG\CD@yE\CD@pD\CD@oD\def\CD@iF{\ifPositiveGradient\CD@tJ
\expandafter\upper@label\else\expandafter\lower@label\fi}\def\CD@iI{%
\ifPositiveGradient\CD@tJ\expandafter\lower@label\else\expandafter
\upper@label\fi}\def\positional@{\CD@gB{labels as positional arguments are
obsolete}\CD@yE\CD@tJ\expandafter\upper@label\else\expandafter\lower@label\fi
-}\def\CD@tD{\futurelet\CD@EH\switch@arg}\def\eat@space{\afterassignment
\CD@tD\let\CD@EH= }\def\CD@TF{\afterassignment\CD@xD\let\CD@EH= }\def\CD@BC{%
\get@round@pair\CD@uD}\def\CD@uD#1#2{\def\CD@WK{#1}\def\CD@aK{#2}\CD@tD}\def
\optional@{\let\CD@N\CD@tD\get@square@arg\CD@AE}\def\CD@JJ.{\CD@sC\CD@tD}\def
\CD@sC{\let\CD@iD\fill@dot\let\CD@jD\fill@dot\def\CD@MI{\let\CD@iD\dfdot\let
\CD@jD\dfdot}}\def\CD@MI{}\def\CD@@E#1,{\CD@nH#1,\begingroup\ifx\@name\CD@RD
\CD@FF\aftergroup\CD@e\fi\aftergroup\CD@jC\else\expandafter\def\expandafter
\CD@RF\expandafter{\csname\@name\endcsname}\expandafter\CD@vD\CD@RF\CD@KD\ifx
\CD@RF\empty\aftergroup\CD@pC\expandafter\aftergroup\csname\CD@FB\@name
\endcsname\expandafter\aftergroup\csname\CD@FB @\@name\endcsname\else\gdef
\CD@GE{#1}\CD@gB{\string\relax\space inserted before `[\CD@GE'}\message{(I was
trying to read this as a \CD@tA\ option.)}\aftergroup\CD@H\fi\fi\endgroup}%
\def\CD@vD#1#2\CD@KD{\def\CD@RF{#2}}\def\CD@jC{\let\CD@CH\CD@N\let\CD@N\relax
\CD@CH}\def\CD@H#1],{
\CD@jC\relax\def\CD@RF{#1}\ifx\CD@RF\empty\def\CD@RF{[\CD@GE]}%
\else\def\CD@RF{[\CD@GE,#1]}
\fi\CD@RF}\def\CD@pC#1#2{\ifx#2\CD@qK\ifx#1\CD@qK\CD@gB{option `\@name'
undefined}\else#1\fi\else\CD@FF\expandafter#2\CD@GK\CD@PK\else\CD@QK\fi\fi
\CD@DH}\CD@tG\CD@FF\CD@QK\CD@PK\def\CD@nH#1,{\CD@FF\ifx\CD@GK\CD@qK\CD@e\else
\expandafter\CD@oH\CD@GK,#1,(,),(,)[]%
\fi\fi\CD@FF\else\CD@mH#1==,\fi}\def\CD@e{\CD@gB{option `\@name' needs (x,y)
value}\CD@PK\let\@name\empty}\def\CD@mH#1=#2=#3,{\def\@name{#1}\def\CD@GK{#2}%
\def\CD@RF{#3}\ifx\CD@RF\empty\let\CD@GK\CD@qK\fi}%
\def\CD@oH#1(#2,#3)#4,(#5,#6)#7[]{\def\CD@GK{{#2}{#3}}\def\CD@RF{#1#4#5#6}%
\ifx\CD@RF\empty\def\CD@RF{#7}\ifx\CD@RF\empty\CD@e\fi\else\CD@e\fi}\def
\CD@FB{cds@}\let\CD@N\relax\def\CD@zD#1{\ifx\CD@GK\CD@qK\CD@gB{option `\@name
' needs a value}\else#1\CD@GK\relax\fi}\def\CD@BE#1#2{\ifx\CD@GK\CD@qK#1#2%
\relax\else#1\CD@GK\relax\fi}\def\cds@@showpair#1#2{\message{x=#1,y=#2}}\def
\cds@@diagonalbase#1#2{\edef\CD@ZK{#1}\edef\CD@bK{#2}}\def\CD@DI#1{\def\CD@CH
{#1}\CD@nF{@x}{cdps@#1}\ifx\CD@CH\empty\CD@f\CD@CH{cannot be used}\else\ifx
\CD@CH\relax\CD@f\CD@CH{unknown}\else\let\CD@IK\@x\fi\fi}\def\CD@f#1#2{\CD@gB
{PostScript translator `#1' #2}}\def\CD@PH{}\def\CD@PJ{\CD@fA\edef\CD@PH{%
\noexpand\CD@KB{\@name\space ignored within maths}}}\def\diagramstyle{\CD@cJ
\let\CD@N\relax\CD@CF\CD@AE\CD@AE}\CD@tG\CD@sE
\CD@SB\CD@RB\CD@tG\CD@qE\CD@EB\CD@DB\CD@tG\CD@oE\CD@pA\CD@oA\CD@tG\CD@iE
\CD@HA\CD@GA\CD@HA\CD@tG\CD@jE\CD@JA\CD@IA\CD@tG\CD@kE\CD@LA\CD@KA\CD@tG
\CD@EF\CD@DK\CD@CK\CD@tG\CD@rE\CD@JB\CD@IB\CD@tG\CD@mE\CD@gA\CD@fA\CD@tG
\CD@nE\CD@kA\CD@jA\CD@tG\CD@AF\CD@iG\CD@hG\CD@RC{cds@ }{}\CD@RC{cds@}{}\CD@RC
{cds@1em}{\CellSize1\CD@zC}\CD@RC{cds@1.5em}{\CellSize1.5\CD@zC}\CD@RC{cds@2%
em}{\CellSize2\CD@zC}\CD@RC{cds@2.5em}{\CellSize2.5\CD@zC}\CD@RC{cds@3em}{%
\CellSize3\CD@zC}\CD@RC{cds@3.5em}{\CellSize3.5\CD@zC}\CD@RC{cds@4em}{%
\CellSize4\CD@zC}\CD@RC{cds@4.5em}{\CellSize4.5\CD@zC}\CD@RC{cds@5em}{%
\CellSize5\CD@zC}\CD@RC{cds@6em}{\CellSize6\CD@zC}\CD@RC{cds@7em}{\CellSize7%
\CD@zC}\CD@RC{cds@8em}{\CellSize8\CD@zC}\def\cds@abut{\MapsAbut\dimen1\z@
\dimen5\z@}\def\cds@alignlabels{\CD@IA\CD@KA}\def\cds@amstex{\ifincommdiag
\CD@O\else\def\CD{\diagram[amstex]}
\fi\CD@T\catcode`\@\active}\def\cds@b{\let\CD@dB\CD@bB}\def\cds@balance{\let
\CD@hA\CD@AA}\let\cds@bottom\cds@b\def\cds@center{\cds@vcentre\cds@nobalance}%
\let\cds@centre\cds@center\def\cds@centerdisplay{\CD@HA\CD@PJ\cds@balance}%
\let\cds@centredisplay\cds@centerdisplay\def\cds@crab{\CD@BE\CD@DC{.5%
\PileSpacing}}\CD@RC{cds@crab-}{\CD@DC-.5\PileSpacing}\CD@RC{cds@crab+}{%
\CD@DC.5\PileSpacing}\CD@RC{cds@crab++}{\CD@DC1.5\PileSpacing}\CD@RC{cds@crab%
--}{\CD@DC-1.5\PileSpacing}\def\cds@defaultsize{\CD@BE{\let\CD@QC}{3em}\CD@NJ
}\def\cds@displayoneliner{\CD@DB}\let\cds@dotted\CD@sC\def\cds@dpi{\CD@RJ{1%
truein}}\def\cds@dpm{\CD@RJ{100truecm}}\let\CD@XA\CD@qK\def\cds@eqno{\let
\CD@XA\CD@GK\let\CD@EJ\empty}\def\cds@fixed{\CD@qA}\CD@tG\CD@fE\CD@J\CD@I\def
\cds@flushleft{\CD@I\CD@GA\CD@PJ\cds@nobalance\CD@BE\CD@nA\CD@nA}\def\cds@gap
{\CD@AJ\setbox3=\null\ht3=\CD@tI\dp3=\CD@sI\CD@BE{\wd3=}\MapShortFall} \def
\cds@grid{\ifx\CD@GK\CD@qK\let\h@grid\relax\let\v@grid\relax\else\CD@nF{%
h@grid}{cdgh@\CD@GK}\CD@nF{v@grid}{cdgv@\CD@GK}\ifx\h@grid\relax\CD@gB{%
unknown grid `\CD@GK'}\else\CD@WB\fi\fi}\let\h@grid\relax\let\v@grid\relax
\def\cds@gridx{\ifx\CD@GK\CD@qK\else\cds@grid\fi\let\CD@CH\h@grid\let\h@grid
\v@grid\let\v@grid\CD@CH}\def\cds@h{\CD@zD\DiagramCellHeight}\def\cds@hcenter
{\let\CD@hA\CD@aA}\let\cds@hcentre\cds@hcenter\def\cds@heads{\CD@BE{\let
\CD@sJ}\CD@sJ\CD@@J\CD@vE\else\ifx\CD@sJ\CD@eF\else\CD@MC\fi\fi}\let
\cds@height\cds@h\let\cds@hmiddle\cds@balance\def\cds@htriangleheight{\CD@BE
\DiagramCellHeight\DiagramCellHeight\DiagramCellWidth1.73205%
\DiagramCellHeight}\def\cds@htrianglewidth{\CD@BE\DiagramCellWidth
\DiagramCellWidth\DiagramCellHeight.57735\DiagramCellWidth}\CD@tG\CD@zE\CD@eE
\CD@dE\CD@eE\def\cds@hug{\CD@eE} \def\cds@inline{\CD@gA\let\CD@PH\empty}\def
\cds@inlineoneliner{\CD@EB}\CD@RC{cds@l>}{\CD@zD{\let\CD@RG}\dimen2=\CD@RG}%
\def\cds@labelstyle{\CD@zD{\let\labelstyle}}\def\cds@landscape{\CD@kA}\def
\cds@large{\CellSize5\CD@zC}\let\CD@EJ\empty\def\CD@FJ{\refstepcounter{%
equation}\def\CD@XA{\hbox{\@eqnnum}}}\def\cds@LaTeXeqno{\let\CD@EJ\CD@FJ}\def
\cds@lefteqno{\CD@pA}\def\cds@leftflush{\cds@flushleft\CD@J}\def
\cds@leftshortfall{\CD@zD{\dimen1 }}\def\cds@lowershortfall{%
\ifPositiveGradient\cds@leftshortfall\else\cds@rightshortfall\fi}\def
\cds@loose{\CD@VB}\def\cds@midhshaft{\CD@JA}\def\cds@midshaft{\CD@JA}\def
\cds@midvshaft{\CD@LA}\def\cds@moreoptions{\CD@@A}\let\cds@nobalance
\cds@hcenter\def\cds@nohcheck{\CD@HH}\def\cds@nohug{\CD@dE} \def
\cds@nooptions{\def\CD@aC{\CD@WD}}\let\cds@noorigin\cds@nobalance\def
\cds@nopixel{\CD@@I4\CD@XH\CD@cJ}\def\cds@UO{\CD@oK\global\let\CD@n\empty}%
\def\cds@UglyObsolete{\cds@UO\let\cds@PS\empty}\def\CD@rK#1{\CD@gB{option `#1%
' renamed as `UglyObsolete'}}\def\cds@noPostScript{\CD@rK{noPostScript}}\def
\cds@noPS{\CD@rK{noPostScript}}\def\cds@notextflow{\CD@RB}\def\cds@noTPIC{%
\CD@CK}\def\cds@objectstyle{\CD@zD{\let\objectstyle}}\def\cds@origin{\let
\CD@hA\CD@iB}\def\cds@p{\CD@zD\PileSpacing}\let\cds@pilespacing\cds@p\def
\cds@pixelsize{\CD@zD\CD@@I\CD@gI}\def\cds@portrait{\CD@jA}\def
\cds@PostScript{\CD@nK\global\let\CD@n\empty\CD@BE\CD@DI\empty}\def\cds@PS{%
\CD@nK\global\let\CD@n\empty}\CD@GF\CD@n{\typeout{\CD@tA: try the PostScript
option for better results}}\def\cds@repositionpullbacks{\let\make@pbk\CD@fH
\let\CD@qH\CD@pH}\def\cds@righteqno{\CD@oA}\def\cds@rightshortfall{\CD@zD{%
\dimen5 }}\def\cds@ruleaxis{\CD@zD{\let\axisheight}}\def\cds@cmex{\let\CD@GG
\CD@sB\let\CD@QJ\CD@CJ}\def\cds@s{\cds@height\DiagramCellWidth
\DiagramCellHeight}\def\cds@scriptlabels{\let\labelstyle\scriptstyle}\def
\cds@shortfall{\CD@zD\MapShortFall\dimen1\MapShortFall\dimen5\MapShortFall}%
\def\cds@showfirstpass{\CD@BE{\let\CD@nD}\z@}\def\cds@silent{\def\CD@KB##1{}%
\def\CD@gB##1{}}\let\cds@size\cds@s\def\cds@small{\CellSize2\CD@zC}\def
\cds@snake{\CD@BE\CD@eJ\z@}\def\cds@t{\let\CD@dB\CD@fB}\def\cds@textflow{%
\CD@SB\CD@PJ}\def\cds@thick{\let\CD@rF\tenlnw\CD@LF\CD@NC\CD@BE\MapBreadth{2%
\CD@LF}\CD@@J}\def\cds@thin{\let\CD@rF\tenln\CD@BE\MapBreadth{\CD@NC}\CD@@J}%
\def\cds@tight{\CD@WB}\let\cds@top\cds@t\def\cds@TPIC{\CD@DK}\def
\cds@uppershortfall{\ifPositiveGradient\cds@rightshortfall\else
\cds@leftshortfall\fi}\def\cds@vcenter{\let\CD@dB\CD@cB}\let\cds@vcentre
\cds@vcenter\def\cds@vtriangleheight{\CD@BE\DiagramCellHeight
\DiagramCellHeight\DiagramCellWidth.577035\DiagramCellHeight}\def
\cds@vtrianglewidth{\CD@BE\DiagramCellWidth\DiagramCellWidth
\DiagramCellHeight1.73205\DiagramCellWidth}\def\cds@vmiddle{\let\CD@dB\CD@eB}%
\def\cds@w{\CD@zD\DiagramCellWidth}\let\cds@width\cds@w\def\diagram{\relax
\protect\CD@bC}\def\enddiagram{\protect\CD@SG}\def\CD@bC{\CD@g\CD@uI
\incommdiagtrue\edef\CD@wI{\the\CD@NB}\global\CD@NB\z@\boxmaxdepth\maxdimen
\everycr{}\CD@sK\everymath{}\everyhbox{}\ifx\pdfsyncstop\CD@qK\else
\pdfsyncstop\fi\CD@aC}\def\CD@aC{\CD@y\let\CD@N\CD@ZC\CD@CF\CD@AE\CD@WD}\def
\CD@ZC{\CD@gE\expandafter\CD@aC\else\expandafter\CD@WD\fi}\def\CD@WD{\let
\CD@EH\relax\CD@nE\CD@vE\else\CD@hK\else\CD@KB{landscape ignored without
PostScript}\CD@jA\fi\fi\fi\CD@EJ\setbox2=\vbox\bgroup\CD@JF\CD@VD}\def\CD@cH{%
\CD@nE\CD@fB\else\CD@dB\fi\CD@hA\nointerlineskip\setbox0=\null\ht0-\CD@pI\dp0%
\CD@pI\wd0\CD@kI\box0 \global\CD@QA\CD@kF\global\CD@yA\CD@XB\ifx\CD@NK\CD@qK
\global\CD@RA\CD@kF\else\global\CD@RA\CD@NK\fi\egroup\CD@zF\CD@nE\setbox2=%
\hbox to\dp2{\vrule height\wd2 depth\CD@QA width\z@\global\CD@QA\ht2\ht2\z@
\dp2\z@\wd2\z@\CD@hK\CD@tK{q 0 1 -1 0 0 0 cm}\else\global\CD@iG\CD@IK{0 1
bturn}\fi\box2\CD@gK\hss}\CD@DB\fi\ifnum\CD@yA=1 \else\CD@DB\fi\global
\@ignorefalse\CD@mE\leavevmode\fi\ifvmode\CD@TA\else\ifmmode\CD@PH\CD@GI\else
\CD@qE\CD@gA\fi\ifinner\CD@gA\fi\CD@mE\CD@GI\else\CD@sE\CD@QB\else\CD@TA\fi
\fi\fi\fi\CD@dD}\def\CD@dD{\global\CD@NB\CD@wI\relax\CD@xE\global\CD@ID\else
\aftergroup\CD@mC\fi\if@ignore\aftergroup\ignorespaces\fi\CD@wC\ignorespaces}%
\def\CD@fB{\advance\CD@pI\dimen1\relax}\def\CD@eB{\advance\CD@pI.5\dimen1%
\relax}\def\CD@bB{}\def\CD@cB{\CD@fB\advance\CD@pI\CD@YB\divide\CD@pI2
\advance\CD@pI-\axisheight\relax}\def\CD@aA{}\def\CD@iB{\CD@kF\z@}\def\CD@AA{%
\ifdim\dimen2>\CD@kF\CD@kF\dimen2 \else\dimen2\CD@kF\CD@kI\dimen0 \advance
\CD@kI\dimen2 \fi}\def\CD@QB{\skip0\z@\relax\loop\skip1\lastskip\ifdim\skip1>%
\z@\unskip\advance\skip0\skip1 \repeat\vadjust{\prevdepth\dp\strutbox\penalty
\predisplaypenalty\vskip\abovedisplayskip\CD@UA\penalty\postdisplaypenalty
\vskip\belowdisplayskip}\ifdim\skip0=\z@\else\hskip\skip0 \global\@ignoretrue
\fi}\def\CD@TA{\CD@LG\kern-\displayindent\CD@UA\CD@LG\global\@ignoretrue}\def
\CD@UA{\hbox to\hsize{\CD@fE\ifdim\CD@RA=\z@\else\advance\CD@QA-\CD@RA\setbox
2=\hbox{\kern\CD@RA\box2}\fi\fi\setbox1=\hbox{\ifx\CD@XA\CD@qK\else\CD@k
\CD@XA\CD@ND\fi}\CD@oE\CD@iE\else\advance\CD@QA\wd1 \fi\wd1\z@\box1 \fi\dimen
0\wd2 \advance\dimen0\wd1 \advance\dimen0-\hsize\ifdim\dimen0>-\CD@nA\CD@HA
\fi\advance\dimen0\CD@QA\ifdim\dimen0>\z@\CD@KB{wider than the page by \the
\dimen0 }\CD@HA\fi\CD@iE\hss\else\CD@V\CD@QA\CD@nA\fi\CD@GI\hss\kern-\wd1\box
1 }}\def\CD@GI{\CD@AF\CD@@F\else\CD@SC\global\CD@hG\fi\fi\kern\CD@QA\box2 }%
\CD@tG\CD@wE\CD@YC\CD@XC\def\CD@JF{\CD@cJ\ifdim\DiagramCellHeight=-\maxdimen
\DiagramCellHeight\CD@QC\fi\ifdim\DiagramCellWidth=-\maxdimen
\DiagramCellWidth\CD@QC\fi\global\CD@XC\CD@IF\let\CD@FE\empty\let\CD@z\CD@Q
\let\overprint\CD@eH\let\CD@s\CD@rJ\let\enddiagram\CD@ED\let\\\CD@cC\let\par
\CD@jH\let\CD@MD\empty\let\switch@arg\CD@PB\let\shift\CD@iA\baselineskip
\DiagramCellHeight\lineskip\z@\lineskiplimit\z@\mathsurround\z@\tabskip\z@
\CD@OB}\def\CD@VD{\penalty-123 \begingroup\CD@jA\aftergroup\CD@K\halign
\bgroup\global\advance\CD@NB1 \vadjust{\penalty1}\global\CD@FA\z@\CD@OB\CD@j#%
#\CD@DD\CD@Q\CD@Q\CD@OI\CD@j##\CD@DD\cr}\def\CD@ED{\CD@MD\CD@GD\crcr\egroup
\global\CD@JD\endgroup}\def\CD@j{\global\advance\CD@FA1 \futurelet\CD@EH\CD@i
}\def\CD@i{\ifx\CD@EH\CD@DD\CD@tJ\hskip1sp plus 1fil \relax\let\CD@DD\relax
\CD@vI\else\hfil\CD@k\objectstyle\let\CD@FE\CD@d\fi}\def\CD@DD{\CD@MD\relax
\CD@yI\CD@vI\global\CD@QA\CD@iA\penalty-9993 \CD@ND\hfil\null\kern-2\CD@QA
\null}\def\CD@cC{\cr}\def\across#1{\span\omit\mscount=#1 \global\advance
\CD@FA\mscount\global\advance\CD@FA\m@ne\CD@sF\ifnum\mscount>2 \CD@fJ\repeat
\ignorespaces}\def\CD@fJ{\relax\span\omit\advance\mscount\m@ne}\def\CD@qJ{%
\ifincommdiag\ifx\CD@iD\@fillh\ifx\CD@jD\@fillh\ifdim\dimen3>\z@\else\ifdim
\dimen2>93\CD@@I\ifdim\dimen2>18\p@\ifdim\CD@LF>\z@\count@\CD@bJ\advance
\count@\m@ne\ifnum\count@<\z@\count@20\let\CD@aJ\CD@uJ\fi\xdef\CD@bJ{\the
\count@}\fi\fi\fi\fi\fi\fi\fi}\def\CD@cG#1{\vrule\horizhtdp width#1\dimen@
\kern2\dimen@}\def\CD@uJ{\rlap{\dimen@\CD@@I\CD@V\dimen@{.182\p@}\CD@zH
\dimen@\advance\CD@tI\dimen@\CD@cG0\CD@cG0\CD@cG2\CD@cG6\CD@cG6\CD@cG2\CD@cG0%
\CD@cG0\CD@cG2\CD@cG6\CD@cG0\CD@cG0\CD@cG2\CD@cG2\CD@cG6\CD@cG0\CD@cG0\CD@cG2%
\CD@cG6\CD@cG2\CD@cG2\CD@cG0\CD@cG0}}\def\CD@bJ{10}\def\CD@aJ{}\def\CD@XD{%
\CD@gE\CD@TB\fi\CD@x\CD@WF\CD@HI}\def\CD@x{\CD@QJ\CD@DC\CD@MJ\ifdim\CD@DC=\z@
\else\CD@pF\CD@DC\fi\ifvoid3 \setbox3=\null\ht3\CD@tI\dp3\CD@sI\else\CD@V{\ht
3}\CD@tI\CD@V{\dp3}\CD@sI\fi\dimen3=.5\wd3 \ifdim\dimen3=\z@\CD@tE\else\dimen
3-\CD@XH\fi\else\CD@TB\fi\CD@V{\dimen2}{\wd7}\CD@V{\dimen2}{\wd6}\CD@qJ
\advance\dimen2-2\dimen3 \dimen4.5\dimen2 \dimen2\dimen4 \advance\dimen2%
\CD@eJ\advance\dimen4-\CD@eJ\advance\dimen2-\wd1 \advance\dimen4-\wd5 \ifvoid
2 \else\CD@V{\ht3}{\ht2}\CD@V{\dp3}{\dp2}\CD@V{\dimen2}{\wd2}\fi\ifvoid4 \else
\CD@V{\ht3}{\ht4}\CD@V{\dp3}{\dp4}\CD@V{\dimen4}{\wd4}\fi\advance\skip2\dimen
2 \advance\skip4\dimen4 \CD@tE\advance\skip2\skip4 \dimen0\dimen5 \advance
\dimen0\wd5 \skip3-\skip4 \advance\skip3-\dimen0 \let\CD@jD\empty\else\skip3%
\z@\relax\dimen0\z@\fi}\def\CD@WF{\offinterlineskip\lineskip.2\CD@zC\ifvoid6
\else\setbox3=\vbox{\hbox to2\dimen3{\hss\box6\hss}\box3}\fi\ifvoid7 \else
\setbox3=\vtop{\box3 \hbox to2\dimen3{\hss\box7\hss}}\fi}\def\CD@HI{\kern
\dimen1 \box1 \CD@aJ\CD@iD\hskip\skip2 \kern\dimen0 \ifincommdiag\CD@jE
\penalty1\fi\kern\dimen3 \penalty\CD@GB\hskip\skip3 \null\kern-\dimen3 \else
\hskip\skip3 \fi\box3 \CD@jD\hskip\skip4 \box5 \kern\dimen5}\def\CD@MF{\ifnum
\CD@LH>\CD@TC\CD@V{\dimen1}\objectheight\CD@V{\dimen5}\objectheight\else\CD@V
{\dimen1}\objectwidth\CD@V{\dimen5}\objectwidth\fi}\def\CD@Y{\begingroup
\ifdim\dimen7=\z@\kern\dimen8 \else\ifdim\dimen6=\z@\kern\dimen9 \else\dimen5%
\dimen6 \dimen6\dimen9 \CD@KJ\dimen4\dimen2 \CD@dG{\dimen4}\dimen6\dimen5
\dimen7\dimen8 \CD@KJ\CD@iC{\dimen2}\ifdim\dimen2<\dimen4 \kern\dimen2 \else
\kern\dimen4 \fi\fi\fi\endgroup}\def\CD@jJ{\CD@JI\setbox\z@\hbox{\lower
\axisheight\hbox to\dimen2{\CD@DF\ifPositiveGradient\dimen8\ht\CD@MH\dimen9%
\CD@mI\else\dimen8\dp3 \dimen9\dimen1 \fi\else\dimen8 \ifPositiveGradient
\objectheight\else\z@\fi\dimen9\objectwidth\fi\advance\dimen8
\ifPositiveGradient-\fi\axisheight\CD@Y\unhbox\z@\CD@DF\ifPositiveGradient
\dimen8\dp3 \dimen9\dimen0 \else\dimen8\ht\CD@MH\dimen9\CD@mF\fi\else\dimen8
\ifPositiveGradient\z@\else\objectheight\fi\dimen9\objectwidth\fi\advance
\dimen8 \ifPositiveGradient\else-\fi\axisheight\CD@Y}}}\def\CD@bD{\dimen6
\CD@aK\DiagramCellHeight\dimen7 \CD@WK\DiagramCellWidth\CD@jJ
\ifPositiveGradient\advance\dimen7-\CD@ZK\DiagramCellWidth\else\dimen7 \CD@ZK
\DiagramCellWidth\dimen6\z@\fi\advance\dimen6-\CD@bK\DiagramCellHeight\CD@mK
\setbox0=\rlap{\kern-\dimen7 \lower\dimen6\box\z@}\ht0\z@\dp0\z@\raise
\axisheight\box0 }\def\CD@mK{\setbox0\hbox{\ht\z@\z@\dp\z@\z@\wd\z@\z@\CD@hK
\expandafter\CD@tK{q \CD@eK\space\CD@lK\space\CD@kK\space\CD@eK\space0 0 cm}%
\else\global\CD@iG\CD@eD{\the\CD@TC\space\ifPositiveGradient\else-\fi\the
\CD@LH\space bturn}\fi\box\z@\CD@gK}}\def\CD@vB{\advance\CD@hF-\CD@mI\CD@wJ
\CD@hF\advance\CD@wJ\CD@hI\ifvoid\CD@sH\ifdim\CD@wJ<.1em\ifnum\CD@gD=\@m\else
\CD@aG h\CD@wJ<.1em:objects overprint:\CD@FA\CD@gD\fi\fi\else\ifhbox\CD@sH
\CD@SK\else\CD@TK\fi\advance\CD@wJ\CD@mI\CD@bH{-\CD@mI}{\box\CD@sH}{\CD@wJ}%
\z@\fi\CD@hF-\CD@mF\CD@gD\CD@FA\CD@hI\z@}\def\CD@SK{\setbox\CD@sH=\hbox{%
\unhbox\CD@sH\unskip\unpenalty}\setbox\CD@tH=\hbox{\unhbox\CD@tH\unskip
\unpenalty}\setbox\CD@sH=\hbox to\CD@wJ{\CD@OA\wd\CD@sH\unhbox\CD@sH\CD@PA
\lastkern\unkern\ifdim\CD@PA=\z@\CD@UB\advance\CD@OA-\wd\CD@tH\else\CD@TB\fi
\ifnum\lastpenalty=\z@\else\CD@JA\unpenalty\fi\kern\CD@PA\ifdim\CD@hF<\CD@OA
\CD@JA\fi\ifdim\CD@hI<\wd\CD@tH\CD@JA\fi\CD@jE\CD@hI\CD@wJ\advance\CD@hI-%
\CD@OA\advance\CD@hI\wd\CD@tH\ifdim\CD@hI<2\wd\CD@tH\CD@aG h\CD@hI<2\wd\CD@tH
:arrow too short:\CD@FA\CD@gD\fi\divide\CD@hI\tw@\CD@hF\CD@wJ\advance\CD@hF-%
\CD@hI\fi\CD@tE\kern-\CD@hI\fi\hbox to\CD@hI{\unhbox\CD@tH}\CD@HG}}\CD@tG
\ifinpile\inpiletrue\inpilefalse\inpilefalse\def\pile{\protect\CD@UJ\protect
\CD@uH}\def\CD@uH#1{\CD@l#1\CD@QD}\def\CD@UJ{\CD@nB{pile}\setbox0=\vtop
\bgroup\aftergroup\CD@lD\inpiletrue\let\CD@FE\empty\let\pile\CD@KF\let\CD@QD
\CD@PD\let\CD@GD\CD@FD\CD@yH\baselineskip.5\PileSpacing\lineskip.1\CD@zC
\relax\lineskiplimit\lineskip\mathsurround\z@\tabskip\z@\let\\\CD@wH}\def
\CD@l{\CD@DE\CD@YF\empty\halign\bgroup\hfil\CD@k\let\CD@FE\CD@d\let\\\CD@vH##%
\CD@MD\CD@ND\hfil\CD@Q\CD@R##\cr}\CD@rG\CD@NE{pile only allows one column.}%
\CD@rG\CD@UE{you left it out!}\def\CD@R{\CD@QD\CD@Q\relax\CD@YA{missing \CD@yC
\space inserted after \string\pile}\CD@NE}\def\CD@PD{\CD@MD\crcr\egroup
\egroup}\def\CD@GD{\CD@MD}\def\CD@FD{\CD@MD\relax\CD@QD\CD@YA{missing \CD@yC
\space inserted between \string\pile\space and \CD@HD}\CD@UE}\def\CD@QD{%
\CD@MD}\def\CD@lD{\vbox{\dimen1\dp0 \unvbox0 \setbox0=\lastbox\advance\dimen1%
\dp0 \nointerlineskip\box0 \nointerlineskip\setbox0=\null\dp0.5\dimen1\ht0-%
\dp0 \box0}\ifincommdiag\CD@tJ\penalty-9998 \fi\xdef\CD@YF{pile}}\def\CD@vH{%
\cr}\def\CD@wH{\noalign{\skip@\prevdepth\advance\skip@-\baselineskip
\prevdepth\skip@}}\def\CD@KF#1{#1}\def\CD@TK{\setbox\CD@sH=\vbox{\unvbox
\CD@sH\setbox1=\lastbox\setbox0=\box\voidb@x\CD@tF\setbox\CD@sH=\lastbox
\ifhbox\CD@sH\CD@rC\repeat\unvbox0 \global\CD@QA\CD@ZE}\CD@ZE\CD@QA}\def
\CD@rC{\CD@jE\setbox\CD@sH=\hbox{\unhbox\CD@sH\unskip\setbox\CD@sH=\lastbox
\unskip\unhbox\CD@sH}\ifdim\CD@wJ<\wd\CD@sH\CD@aG h\CD@wJ<\wd\CD@sH:arrow in
pile too short:\CD@FA\CD@gD\else\setbox\CD@sH=\hbox to\CD@wJ{\unhbox\CD@sH}%
\fi\else\CD@gJ\fi\setbox0=\vbox{\box\CD@sH\nointerlineskip\ifvoid0 \CD@tJ\box
1 \else\vskip\skip0 \unvbox0 \fi}\skip0=\lastskip\unskip}\def\CD@gJ{\penalty7
\noindent\unhbox\CD@sH\unskip\setbox\CD@sH=\lastbox\unskip\unhbox\CD@sH
\endgraf\setbox\CD@tH=\lastbox\unskip\setbox\CD@tH=\hbox{\CD@JG\unhbox\CD@tH
\unskip\unskip\unpenalty}\ifcase\prevgraf\cd@shouldnt P\or\ifdim\CD@wJ<\wd
\CD@tH\CD@aG h\CD@wJ<\wd\CD@sH:object in pile too wide:\CD@FA\CD@gD\setbox
\CD@sH=\hbox to\CD@wJ{\hss\unhbox\CD@tH\hss}\else\setbox\CD@sH=\hbox to\CD@wJ
{\hss\kern\CD@hF\unhbox\CD@tH\kern\CD@hI\hss}\fi\or\setbox\CD@sH=\lastbox
\unskip\CD@SK\else\cd@shouldnt Q\fi\unskip\unpenalty}\def\CD@cD{\CD@MJ\ifvoid
3 \setbox3=\null\ht3\axisheight\dp3-\ht3 \dimen3.5\CD@LF\else\dimen4\dp3
\dimen3.5\wd3 \setbox3=\CD@GG{\box3}\dp3\dimen4 \ifdim\ht3=-\dp3 \else\CD@TB
\fi\fi\dimen0\dimen3 \advance\dimen0-.5\CD@LF\setbox0\null\ht0\ht3\dp0\dp3\wd
0\wd3 \ifvoid6\else\setbox6\hbox{\unhbox6\kern\dimen0\kern2pt}\dimen0\wd6 \fi
\ifvoid7\else\setbox7\hbox{\kern2pt\kern\dimen3\unhbox7}\dimen3\wd7 \fi
\setbox3\hbox{\ifvoid6\else\kern-\dimen0\unhbox6\fi\unhbox3 \ifvoid7\else
\unhbox7\kern-\dimen3\fi}\ht3\ht0\dp3\dp0\wd3\wd0 \CD@tE\dimen4=\ht\CD@MH
\advance\dimen4\dp5 \advance\dimen4\dimen1 \let\CD@jD\empty\else\dimen4\ht3
\fi\setbox0\null\ht0\dimen4 \offinterlineskip\setbox8=\vbox spread2ex{\kern
\dimen5 \box1 \CD@iD\vfill\CD@tE\else\kern\CD@eJ\fi\box0}\ht8=\z@\setbox9=%
\vtop spread2ex{\kern-\ht3 \kern-\CD@eJ\box3 \CD@jD\vfill\box5 \kern\dimen1}%
\dp9=\z@\hskip\dimen0plus.0001fil \box9 \kern-\CD@LF\box8 \CD@kE\penalty2 \fi
\CD@tE\penalty1 \fi\kern\PileSpacing\kern-\PileSpacing\kern-.5\CD@LF\penalty
\CD@GB\null\kern\dimen3}\def\CD@cI{\ifhbox\CD@VA\CD@KB{clashing verticals}\ht
\CD@MH.5\dp\CD@VA\dp\CD@MH-\ht5 \CD@yB\ht\CD@MH\z@\dp\CD@MH\z@\fi\dimen1\dp
\CD@VA\CD@xA\prevgraf\unvbox\CD@VA\CD@wA\lastpenalty\unpenalty\setbox\CD@VA=%
\null\setbox\CD@lI=\hbox{\CD@JG\unhbox\CD@lI\unskip\unpenalty\dimen0\lastkern
\unkern\unkern\unkern\kern\dimen0 \CD@HG}\setbox\CD@lF=\hbox{\unhbox\CD@lF
\dimen0\lastkern\unkern\unkern\global\CD@QA\lastkern\unkern\kern\dimen0 }%
\CD@tF\ifnum\CD@xA>4 \CD@zI\repeat\unskip\unskip\advance\CD@mF.5\wd\CD@VA
\advance\CD@mF\wd\CD@lF\advance\CD@mI.5\wd\CD@VA\advance\CD@mI\wd\CD@lI\ifnum
\CD@FA=\CD@lA\CD@OA.5\wd\CD@VA\edef\CD@NK{\the\CD@OA}\fi\setbox\CD@VA=\hbox{%
\kern-\CD@mF\box\CD@lF\unhbox\CD@VA\box\CD@lI\kern-\CD@mI\penalty\CD@wA
\penalty\CD@NB}\ht\CD@VA\dimen1 \dp\CD@VA\z@\wd\CD@VA\CD@tB\CD@vB}\def\CD@zI{%
\ifdim\wd\CD@lF<\CD@QA\setbox\CD@lF=\hbox to\CD@QA{\CD@JG\unhbox\CD@lF}\fi
\advance\CD@xA\m@ne\setbox\CD@VA=\hbox{\box\CD@lF\unhbox\CD@VA}\unskip\setbox
\CD@lF=\lastbox\setbox\CD@lF=\hbox{\unhbox\CD@lF\unskip\unpenalty\dimen0%
\lastkern\unkern\unkern\global\CD@QA\lastkern\unkern\kern\dimen0 }}\def\CD@yB
{\dimen1\dp\CD@VA\ifhbox\CD@VA\CD@xB\else\CD@zB\fi\setbox\CD@VA=\vbox{%
\penalty\CD@NB}\dp\CD@VA-\dp\CD@MH\wd\CD@VA\CD@tB}\def\CD@zB{\unvbox\CD@VA
\CD@wA\lastpenalty\unpenalty\ifdim\dimen1<\ht\CD@MH\CD@aG v\dimen1<\ht\CD@MH:%
rows overprint:\CD@NB\CD@wA\fi}\def\CD@xB{\dimen0=\ht\CD@VA\setbox\CD@VA=%
\hbox\bgroup\advance\dimen1-\ht\CD@MH\unhbox\CD@VA\CD@xA\lastpenalty
\unpenalty\CD@wA\lastpenalty\unpenalty\global\CD@RA-\lastkern\unkern\setbox0=%
\lastbox\CD@tF\setbox\CD@VA=\hbox{\box0\unhbox\CD@VA}\setbox0=\lastbox\ifhbox
0 \CD@kJ\repeat\global\CD@SA-\lastkern\unkern\global\CD@QA\CD@JK\unhbox\CD@VA
\egroup\CD@JK\CD@QA\CD@bH{\CD@SA}{\box\CD@VA}{\CD@RA}{\dimen1}}\def\CD@kJ{%
\setbox0=\hbox to\wd0\bgroup\unhbox0 \unskip\unpenalty\dimen7\lastkern\unkern
\ifnum\lastpenalty=1 \unpenalty\CD@UB\else\CD@TB\fi\ifnum\lastpenalty=2
\unpenalty\dimen2.5\dimen0\advance\dimen2-.5\dimen1\advance\dimen2-%
\axisheight\else\dimen2\z@\fi\setbox0=\lastbox\dimen6\lastkern\unkern\setbox1%
=\lastbox\setbox0=\vbox{\unvbox0 \CD@tE\kern-\dimen1 \else\ifdim\dimen2=\z@
\else\kern\dimen2 \fi\fi}\ifdim\dimen0<\ht0 \CD@aG v\dimen0<\ht0:upper part of
vertical too short:{\CD@tE\CD@NB\else\CD@wA\fi}\CD@xA\else\setbox0=\vbox to%
\dimen0{\unvbox0}\fi\setbox1=\vtop{\unvbox1}\ifdim\dimen1<\dp1 \CD@aG v\dimen
1<\dp1:lower part of vertical too short:\CD@NB\CD@wA\else\setbox1=\vtop to%
\dimen1{\ifdim\dimen2=\z@\else\kern-\dimen2 \fi\unvbox1 }\fi\box1 \kern\dimen
6 \box0 \kern\dimen7 \CD@HG\global\CD@QA\CD@JK\egroup\CD@JK\CD@QA\relax}%
\countdef\CD@u=14 \newcount\CD@CA\newcount\CD@XB\newcount\CD@NB\let\CD@LB
\insc@unt\newcount\CD@FA\newcount\CD@lA\let\CD@mA\CD@XB\newcount\CD@MB\CD@tG
\CD@DF\CD@bI\CD@aI\CD@aI\def\CD@nD{-1}\def\CD@K{\ifnum\CD@nD<\z@\else
\begingroup\scrollmode\showboxdepth\CD@nD\showboxbreadth\maxdimen\showlists
\endgroup\fi\CD@bI\CD@zF\CD@CA=\CD@u\advance\CD@CA1 \CD@XB=\CD@CA\ifnum\CD@NB
=1 \CD@JA\fi\advance\CD@XB\CD@NB\dimen1\z@\skip0\z@\count@=\insc@unt\advance
\count@\CD@u\divide\count@2 \ifnum\CD@XB>\count@\CD@KB{The diagram has too
many rows! It can't be reformatted.}\else\CD@NG\CD@WI\fi\CD@cH}\def\CD@NG{%
\CD@NB\CD@CA\CD@uF\ifnum\CD@NB<\CD@XB\setbox\CD@NB\box\voidb@x\advance\CD@NB1%
\relax\repeat\CD@NB\CD@CA\skip\z@\z@\CD@uF\CD@GB\lastpenalty\unpenalty\ifnum
\CD@GB>\z@\CD@KE\repeat\ifnum\CD@GB=-123 \CD@tJ\unpenalty\else\cd@shouldnt D%
\fi\ifx\v@grid\relax\else\CD@NB\CD@XB\advance\CD@NB\m@ne\expandafter\CD@VJ
\v@grid\fi\CD@MB\CD@mA\CD@tB\z@\CD@XG\ifx\h@grid\relax\else\expandafter\CD@LJ
\h@grid\fi\count@\CD@XB\advance\count@\m@ne\CD@YB\ht\count@}\def\CD@KE{%
\ifcase\CD@GB\or\CD@MG\else\CD@uA-\lastpenalty\unpenalty\CD@vA\lastpenalty
\unpenalty\setbox0=\lastbox\CD@WG\fi\CD@wD}\def\CD@wD{\skip1\lastskip\unskip
\advance\skip0\skip1 \ifdim\skip1=\z@\else\expandafter\CD@wD\fi}\def\CD@MG{%
\setbox0=\lastbox\CD@pI\dp0 \advance\CD@pI\skip\z@\skip\z@\z@\advance\CD@NF
\CD@pI\CD@uE\ifnum\CD@NB>\CD@CA\CD@NF\DiagramCellHeight\CD@pI\CD@NF\advance
\CD@pI-\CD@qI\fi\fi\CD@qI\ht0 \CD@NF\CD@qI\setbox\CD@NB\hbox{\unhbox\CD@NB
\unhbox0}\dp\CD@NB\CD@pI\ht\CD@NB\CD@qI\advance\CD@NB1 }\def\CD@WG{\ifnum
\CD@uA<\z@\advance\CD@uA\CD@XB\ifnum\CD@uA<\CD@CA\CD@UG\else\CD@OA\dp\CD@uA
\CD@PA\ht\CD@uA\setbox\CD@uA\hbox{\box\z@\penalty\CD@vA\penalty\CD@GB\unhbox
\CD@uA}\dp\CD@uA\CD@OA\ht\CD@uA\CD@PA\fi\else\CD@UG\fi}\def\CD@UG{\CD@KB{%
diagonal goes outside diagram (lost)}}\def\CD@fI{\advance\CD@uA\CD@XB\ifnum
\CD@uA<\CD@CA\CD@UG\else\ifnum\CD@uA=\CD@NB\CD@VG\else\ifnum\CD@uA>\CD@NB
\cd@shouldnt M\else\CD@OA\dp\CD@uA\CD@PA\ht\CD@uA\setbox\CD@uA\hbox{\box\z@
\penalty\CD@vA\penalty\CD@GB\unhbox\CD@uA}\dp\CD@uA\CD@OA\ht\CD@uA\CD@PA\fi
\fi\fi}\def\CD@WI{\CD@t\CD@AJ\setbox\CD@PC=\hbox{\CD@k A\@super f\CD@lJ f%
\CD@ND}\CD@ZE\z@\CD@JK\z@\CD@kI\z@\CD@kF\z@\CD@NB=\CD@XB\CD@NF\z@\CD@uB\z@
\CD@uF\ifnum\CD@NB>\CD@CA\advance\CD@NB\m@ne\CD@qI\ht\CD@NB\CD@pI\dp\CD@NB
\advance\CD@NF\CD@qI\CD@rI\advance\CD@uB\CD@NF\CD@KC\CD@ZI\CD@w\ht\CD@NB
\CD@qI\dp\CD@NB\CD@pI\nointerlineskip\box\CD@NB\CD@NF\CD@pI\setbox\CD@NB\null
\ht\CD@NB\CD@uB\repeat\CD@wB\nointerlineskip\box\CD@NB\CD@gG\CD@ZE
\DiagramCellWidth{width}\CD@gG\CD@JK\DiagramCellHeight{height}\CD@VA\CD@LB
\advance\CD@VA-\CD@lA\advance\CD@VA\m@ne\advance\CD@VA\CD@mA\dimen0\wd\CD@VA
\CD@tI\axisheight\dimen1\CD@uB\advance\dimen1-\CD@YB\dimen2\CD@kI\advance
\dimen2-\dimen0 \advance\CD@XB-\CD@CA\advance\CD@LB-\CD@lA}\count@\year
\multiply\count@12 \advance\count@\month\ifnum\count@>24158 \loop\iftrue
\repeat\fi\def\CD@wB{\CD@qI-\CD@NF\CD@pI\CD@NF
\setbox\CD@MH=\null\dp\CD@MH\CD@NF\ht\CD@MH-\CD@NF\CD@mF\z@\CD@mI\z@\CD@lA
\CD@LB\advance\CD@lA-\CD@MB\advance\CD@lA\CD@mA\CD@FA\CD@LB\CD@VA\CD@MB\CD@sF
\ifnum\CD@FA>\CD@lA\advance\CD@FA\m@ne\advance\CD@VA\m@ne\CD@tB\wd\CD@VA
\setbox\CD@FA=\box\voidb@x\CD@yB\repeat\CD@w\ht\CD@NB\CD@qI\dp\CD@NB\CD@pI}%
\def\CD@gG#1#2#3{\ifdim#1>.01\CD@zC\CD@PA#2\relax\advance\CD@PA#1\relax
\advance\CD@PA.99\CD@zC\count@\CD@PA\divide\count@\CD@zC\CD@KB{increase cell #%
3 to \the\count@ em}\fi}\def\CD@rI{\CD@FA=\CD@LB\penalty4 \noindent\unhbox
\CD@NB\CD@sF\unskip\setbox0=\lastbox\ifhbox0 \advance\CD@FA\m@ne\setbox\CD@FA
\hbox to\wd0{\null\penalty-9990\null\unhbox0}\repeat\CD@lA\CD@FA\advance
\CD@FA\CD@MB\advance\CD@FA-\CD@mA\ifnum\CD@FA<\CD@LB\count@\CD@FA\advance
\count@\m@ne\dimen0=\wd\count@\count@\CD@MB\advance\count@\m@ne\CD@tB\wd
\count@\CD@sF\ifnum\CD@FA<\CD@LB\CD@DJ\CD@XG\dimen0\wd\CD@FA\advance\CD@FA1
\repeat\fi\CD@sF\CD@GB\lastpenalty\unpenalty\ifnum\CD@GB>\z@\CD@vA
\lastpenalty\unpenalty\CD@VG\repeat\endgraf\unskip\ifnum\lastpenalty=4
\unpenalty\else\cd@shouldnt S\fi}\def\CD@VG{\advance\CD@vA\CD@lA\advance
\CD@vA\m@ne\setbox0=\lastbox\ifnum\CD@vA<\CD@LB\setbox\CD@vA\hbox{\box0%
\penalty\CD@GB\unhbox\CD@vA}\else\CD@UG\fi}\def\CD@bG{}\CD@tG\CD@uE\CD@WB
\CD@VB\def\CD@DJ{\advance\dimen0\wd\CD@FA\divide\dimen0\tw@\CD@uE\dimen0%
\DiagramCellWidth\else\CD@V{\dimen0}\DiagramCellWidth\CD@pJ\fi\advance\CD@tB
\dimen0 }\def\CD@XG{\setbox\CD@MB=\vbox{}\dp\CD@MB=\CD@uB\wd\CD@MB\CD@tB
\advance\CD@MB1 }\def\CD@LJ#1,{\def\CD@GK{#1}\ifx\CD@GK\CD@RD\else\advance
\CD@tB\CD@GK\DiagramCellWidth\CD@XG\expandafter\CD@LJ\fi}\def\CD@VJ#1,{\def
\CD@GK{#1}\ifx\CD@GK\CD@RD\else\ifnum\CD@NB>\CD@CA\CD@NF\CD@GK
\DiagramCellHeight\advance\CD@NF-\dp\CD@NB\advance\CD@NB\m@ne\ht\CD@NB\CD@NF
\fi\expandafter\CD@VJ\fi}\def\CD@pJ{\CD@wE\CD@OA\dimen0 \advance\CD@OA-%
\DiagramCellWidth\ifdim\CD@OA>2\MapShortFall\CD@KB{badly drawn diagonals (see
manual)}\let\CD@pJ\empty\fi\else\let\CD@pJ\empty\fi}\def\CD@KC{\CD@VA\CD@mA
\CD@sF\ifnum\CD@VA<\CD@MB\dimen0\dp\CD@VA\advance\dimen0\CD@NF\dp\CD@VA\dimen
0 \advance\CD@VA1 \repeat}\def\CD@bH#1#2#3#4{\ifnum\CD@FA<\CD@LB\CD@OA=#1%
\relax\setbox\CD@FA=\hbox{\setbox0=#2\dimen7=#4\relax\dimen8=#3\relax\ifhbox
\CD@FA\unhbox\CD@FA\advance\CD@OA-\lastkern\unkern\fi\ifdim\CD@OA=\z@\else
\kern-\CD@OA\fi\raise\dimen7\box0 \kern-\dimen8 }\ifnum\CD@FA=\CD@lA\CD@V
\CD@kF\CD@OA\fi\else\cd@shouldnt O\fi}\def\CD@w{\setbox\CD@NB=\hbox{\CD@FA
\CD@lA\CD@VA\CD@mA\CD@PA\z@\relax\CD@sF\ifnum\CD@FA<\CD@LB\CD@tB\wd\CD@VA
\relax\CD@eI\advance\CD@FA1 \advance\CD@VA1 \repeat}\CD@V\CD@kI{\wd\CD@NB}\wd
\CD@NB\z@}\def\CD@eI{\ifhbox\CD@FA\CD@OA\CD@tB\relax\advance\CD@OA-\CD@PA
\relax\ifdim\CD@OA=\z@\else\kern\CD@OA\fi\CD@PA\CD@tB\advance\CD@PA\wd\CD@FA
\relax\unhbox\CD@FA\advance\CD@PA-\lastkern\unkern\fi}\def\CD@ZI{\setbox
\CD@sH=\box\voidb@x\CD@VA=\CD@MB\CD@FA\CD@LB\CD@VA\CD@mA\advance\CD@VA\CD@FA
\advance\CD@VA-\CD@lA\advance\CD@VA\m@ne\CD@tB\wd\CD@VA\count@\CD@LB\advance
\count@\m@ne\CD@hF.5\wd\count@\advance\CD@hF\CD@tB\CD@A\m@ne\CD@gD\@m\CD@sF
\ifnum\CD@FA>\CD@lA\advance\CD@FA\m@ne\advance\CD@hF-\CD@tB\CD@PI\wd\CD@VA
\CD@tB\advance\CD@hF\CD@tB\advance\CD@VA\m@ne\CD@tB\wd\CD@VA\repeat\CD@mF
\CD@kF\CD@mI-\CD@mF\CD@vB}\newcount\CD@GB\def\CD@s{}\def\CD@t{\mathsurround
\z@\hsize\z@\rightskip\z@ plus1fil minus\maxdimen\parfillskip\z@\linepenalty
9000 \looseness0 \hfuzz\maxdimen\hbadness10000 \clubpenalty0 \widowpenalty0
\displaywidowpenalty0 \interlinepenalty0 \predisplaypenalty0
\postdisplaypenalty0 \interdisplaylinepenalty0 \interfootnotelinepenalty0
\floatingpenalty0 \brokenpenalty0 \everypar{}\leftskip\z@\parskip\z@
\parindent\z@\pretolerance10000 \tolerance10000 \hyphenpenalty10000
\exhyphenpenalty10000 \binoppenalty10000 \relpenalty10000 \adjdemerits0
\doublehyphendemerits0 \finalhyphendemerits0 \CD@IA\prevdepth\z@}\newbox
\CD@KG\newbox\CD@IG\def\CD@JG{\unhcopy\CD@KG}\def\CD@HG{\unhcopy\CD@IG}\def
\CD@iJ{\hbox{}\penalty1\nointerlineskip}\def\CD@PI{\penalty5 \noindent\setbox
\CD@MH=\null\CD@mF\z@\CD@mI\z@\ifnum\CD@FA<\CD@LB\ht\CD@MH\ht\CD@FA\dp\CD@MH
\dp\CD@FA\unhbox\CD@FA\skip0=\lastskip\unskip\else\CD@OK\skip0=\z@\fi\endgraf
\ifcase\prevgraf\cd@shouldnt Y \or\cd@shouldnt Z \or\CD@RI\or\CD@XI\else
\CD@QI\fi\unskip\setbox0=\lastbox\unskip\unskip\unpenalty\noindent\unhbox0%
\setbox0\lastbox\unpenalty\unskip\unskip\unpenalty\setbox0\lastbox\CD@tF
\CD@GB\lastpenalty\unpenalty\ifnum\CD@GB>\z@\setbox\z@\lastbox\CD@lB\repeat
\endgraf\unskip\unskip\unpenalty}\def\CD@YJ{\CD@uA\CD@XB\advance\CD@uA-\CD@NB
\CD@vA\CD@FA\advance\CD@vA-\CD@lA\advance\CD@vA1 \expandafter\message{%
prevgraf=\the\prevgraf at (\the\CD@uA,\the\CD@vA)}}\def\CD@XI{\CD@CE\setbox
\CD@lI=\lastbox\setbox\CD@lI=\hbox{\unhbox\CD@lI\unskip\unpenalty}\unskip
\ifdim\ht\CD@lI>\ht\CD@PC\setbox\CD@MH=\copy\CD@lI\else\ifdim\dp\CD@lI>\dp
\CD@PC\setbox\CD@MH=\copy\CD@lI\else\CD@FG\CD@lI\fi\fi\advance\CD@mF.5\wd
\CD@lI\advance\CD@mI.5\wd\CD@lI\setbox\CD@lI=\hbox{\unhbox\CD@lI\CD@HG}\CD@bH
\CD@mF{\box\CD@lI}\CD@mI\z@\CD@yB\CD@vB}\def\CD@CE{\ifnum\CD@A>0 \advance
\dimen0-\CD@tB\CD@iA-.5\dimen0 \CD@A-\CD@A\else\CD@A0 \CD@iA\z@\fi\setbox
\CD@MH=\lastbox\setbox\CD@MH=\hbox{\unhbox\CD@MH\unskip\unskip\unpenalty
\setbox0=\lastbox\global\CD@QA\lastkern\unkern}\advance\CD@iA-.5\CD@QA\unskip
\setbox\CD@MH=\null\CD@mI\CD@iA\CD@mF-\CD@iA}\def\CD@Z{\ht\CD@MH\CD@tI\dp
\CD@MH\CD@sI}\def\CD@FG#1{\setbox\CD@MH=\hbox{\CD@V{\ht\CD@MH}{\ht#1}\CD@V{%
\dp\CD@MH}{\dp#1}\CD@V{\wd\CD@MH}{\wd#1}\vrule height\ht\CD@MH depth\dp\CD@MH
width\wd\CD@MH}}\def\CD@QI{\CD@CE\CD@Z\setbox\CD@lI=\lastbox\unskip\setbox
\CD@lF=\lastbox\unskip\setbox\CD@lF=\hbox{\unhbox\CD@lF\unskip\global\CD@yA
\lastpenalty\unpenalty}\advance\CD@yA9999 \ifcase\CD@yA\CD@VI\or\CD@YI\or
\CD@TI\or\CD@dI\or\CD@cI\or\CD@SI\else\cd@shouldnt9\fi}\def\CD@VI{\CD@FG
\CD@lI\CD@UI\setbox\CD@sH=\box\CD@lF\setbox\CD@tH=\box\CD@lI}\def\CD@YI{%
\CD@FG\CD@lF\setbox\CD@lI\hbox{\penalty8 \unhbox\CD@lI\unskip\unpenalty\ifnum
\lastpenalty=8 \else\CD@xH\fi}\CD@UI\setbox\CD@lF=\hbox{\unhbox\CD@lF\unskip
\unpenalty\global\setbox\CD@DA=\lastbox}\ifdim\wd\CD@lF=\z@\else\CD@xH\fi
\setbox\CD@sH=\box\CD@DA}\def\CD@xH{\CD@KB{extra material in \string\pile
\space cell (lost)}}\def\CD@UI{\CD@yB\ifvoid\CD@sH\else\CD@KB{Clashing
horizontal arrows}\CD@mI.5\CD@hF\CD@mF-\CD@mI\CD@vB\CD@mI\z@\CD@mF\z@\fi
\CD@hI\CD@hF\advance\CD@hI-\CD@mI\CD@hF-\CD@mF\CD@JC\CD@FA}\def\CD@RI{\setbox
0\lastbox\unskip\CD@iA\z@\CD@Z\ifdim\skip0>\z@\CD@tJ\CD@A0 \else\ifnum\CD@A<1
\CD@A0 \dimen0\CD@tB\fi\advance\CD@A1 \fi}\def\VonH{\CD@MA46\VonH{.5\CD@LF}}%
\def\HonV{\CD@MA57\HonV{.5\CD@LF}}\def\HmeetV{\CD@MA44\HmeetV{-\MapShortFall}%
}\def\CD@MA#1#2#3#4{\CD@pB34#1{\string#3}\CD@SD\CD@GB-999#2 \dimen0=#4\CD@tI
\dimen0\advance\CD@tI\axisheight\CD@sI\dimen0\advance\CD@sI-\axisheight\CD@CF
\CD@HC\CD@ZD}\def\CD@HC#1{\setbox0=\hbox{\CD@k#1\CD@ND}\dimen0.5\wd0 \CD@tI
\ht0 \CD@sI\dp0 \CD@ZD}\def\CD@SD{\setbox0=\null\ht0=\CD@tI\dp0=\CD@sI\wd0=%
\dimen0 \copy0\penalty\CD@GB\box0 }\def\CD@TI{\CD@GC\CD@yB}\def\CD@dI{\CD@GC
\CD@vB}\def\CD@SI{\CD@GC\CD@yB\CD@vB}\def\CD@GC{\setbox\CD@lI=\hbox{\unhbox
\CD@lI}\setbox\CD@lF=\hbox{\unhbox\CD@lF\global\setbox\CD@DA=\lastbox}\ht
\CD@MH\ht\CD@DA\dp\CD@MH\dp\CD@DA\advance\CD@mF\wd\CD@DA\advance\CD@mI\wd
\CD@lI}\CD@tG\ifPositiveGradient\CD@CI\CD@BI\CD@CI\CD@tG\ifClimbing\CD@rB
\CD@qB\CD@rB\newcount\DiagonalChoice\DiagonalChoice\m@ne\ifx\tenln\nullfont
\CD@tJ\def\CD@qF{\CD@KH\ifPositiveGradient/\else\CD@k\backslash\CD@ND\fi}%
\else\def\CD@qF{\CD@rF\char\count@}\fi\let\CD@rF\tenln\def\Use@line@char#1{%
\hbox{#1\CD@rF\ifPositiveGradient\else\advance\count@64 \fi\char\count@}}\def
\CD@cF{\Use@line@char{\count@\CD@TC\multiply\count@8\advance\count@-9\advance
\count@\CD@LH}}\def\CD@ZF{\Use@line@char{\ifcase\DiagonalChoice\CD@gF\or
\CD@fF\or\CD@fF\else\CD@gF\fi}}\def\CD@gF{\ifnum\CD@TC=\z@\count@'33 \else
\count@\CD@TC\multiply\count@\sixt@@n\advance\count@-9\advance\count@\CD@LH
\advance\count@\CD@LH\fi}\def\CD@fF{\count@'\ifcase\CD@LH55\or\ifcase\CD@TC66%
\or22\or52\or61\or72\fi\or\ifcase\CD@TC66\or25\or22\or63\or52\fi\or\ifcase
\CD@TC66\or16\or36\or22\or76\fi\or\ifcase\CD@TC66\or27\or25\or67\or22\fi\fi
\relax}\def\CD@uC#1{\hbox{#1\setbox0=\Use@line@char{#1}\ifPositiveGradient
\else\raise.3\ht0\fi\copy0 \kern-.7\wd0 \ifPositiveGradient\raise.3\ht0\fi
\box0}}\def\CD@jF#1{\hbox{\setbox0=#1\kern-.75\wd0 \vbox to.25\ht0{%
\ifPositiveGradient\else\vss\fi\box0 \ifPositiveGradient\vss\fi}}}\def\CD@jI#%
1{\hbox{\setbox0=#1\dimen0=\wd0 \vbox to.25\ht0{\ifPositiveGradient\vss\fi
\box0 \ifPositiveGradient\else\vss\fi}\kern-.75\dimen0 }}\CD@RC{+h:>}{%
\Use@line@char\CD@fF}\CD@RC{-h:>}{\Use@line@char\CD@gF}\CD@nF{+t:<}{-h:>}%
\CD@nF{-t:<}{+h:>}\CD@RC{+t:>}{\CD@jF{\Use@line@char\CD@fF}}\CD@RC{-t:>}{%
\CD@jI{\Use@line@char\CD@gF}}\CD@nF{+h:<}{-t:>}\CD@nF{-h:<}{+t:>}\CD@RC{+h:>>%
}{\CD@uC\CD@fF}\CD@RC{-h:>>}{\CD@uC\CD@gF}\CD@nF{+t:<<}{-h:>>}\CD@nF{-t:<<}{+%
h:>>}\CD@nF{+h:>->}{+h:>>}\CD@nF{-h:>->}{-h:>>}\CD@nF{+t:<-<}{-h:>>}\CD@nF{-t%
:<-<}{+h:>>}\CD@RC{+t:>>}{\CD@jF{\CD@uC\CD@fF}}\CD@RC{-t:>>}{\CD@jI{\CD@uC
\CD@gF}}\CD@nF{+h:<<}{-t:>>}\CD@nF{-h:<<}{+t:>>}\CD@nF{+t:>->}{+t:>>}\CD@nF{-%
t:>->}{-t:>>}\CD@nF{+h:<-<}{-t:>>}\CD@nF{-h:<-<}{+t:>>}\CD@RC{+f:-}{\CD@EF
\null\else\CD@cF\fi}\CD@nF{-f:-}{+f:-}\def\CD@tC#1#2{\vbox to#1{\vss\hbox to#%
2{\hss.\hss}\vss}}\def\hfdot{\CD@tC{2\axisheight}{.5em}}%
\def\vfdot{\CD@tC{1ex}\z@}
\def\CD@bF{\hbox{\dimen0=.3\CD@zC\dimen1\dimen0 \ifnum\CD@LH>\CD@TC\CD@iC{%
\dimen1}\else\CD@dG{\dimen0}\fi\CD@tC{\dimen0}{\dimen1}}}\newarrowfiller{.}%
\hfdot\hfdot\vfdot\vfdot\def\dfdot{\CD@bF\CD@CK}\CD@RC{+f:.}{\dfdot}\CD@RC{-f%
:.}{\dfdot}\def\CD@@K#1{\hbox\bgroup\def\CD@CH{#1\egroup}\afterassignment
\CD@CH
\count@='}\def\lnchar{\CD@@K\CD@qF}\def\CD@dF#1{\setbox#1=\hbox{\dimen5\dimen
#1 \setbox8=\box#1 \dimen1\wd8 \count@\dimen5 \divide\count@\dimen1 \ifnum
\count@=0 \box8 \ifdim\dimen5<.95\dimen1 \CD@gB{diagonal line too short}\fi
\else\dimen3=\dimen5 \advance\dimen3-\dimen1 \divide\dimen3\count@\dimen4%
\dimen3 \CD@dG{\dimen4}\ifPositiveGradient\multiply\dimen4\m@ne\fi\dimen6%
\dimen1 \advance\dimen6-\dimen3 \loop\raise\count@\dimen4\copy8 \ifnum\count@
>0 \kern-\dimen6 \advance\count@\m@ne\repeat\fi}}\def\CD@CG#1{\CD@EF\CD@xJ{#1%
}\else\CD@dF{#1}\fi}\def\CD@IH#1{}\newdimen\objectheight\objectheight1.8ex
\newdimen\objectwidth\objectwidth1em \def\CD@YD{\dimen6=\CD@aK
\DiagramCellHeight\dimen7=\CD@WK\DiagramCellWidth\CD@KJ\ifnum\CD@LH>0 \ifnum
\CD@TC>0 \CD@aF\else\aftergroup\CD@VC\fi\else\aftergroup\CD@UC\fi}\def\CD@VC{%
\CD@YA{diagonal map is nearly vertical}\CD@NA}\def\CD@UC{\CD@YA{diagonal map
is nearly horizontal}\CD@NA}\CD@rG\CD@NA{Use an orthogonal map instead}\def
\CD@aF{\CD@MJ\dimen3\dimen7\dimen7\dimen6\CD@iC{\dimen7}\advance\dimen3-%
\dimen7 \CD@MF\ifnum\CD@LH>\CD@TC\advance\dimen6-\dimen1\advance\dimen6-%
\dimen5 \CD@iC{\dimen1}\CD@iC{\dimen5}\else\dimen0\dimen1\advance\dimen0%
\dimen5\CD@dG{\dimen0}\advance\dimen6-\dimen0 \fi\dimen2.5\dimen7\advance
\dimen2-\dimen1 \dimen4.5\dimen7\advance\dimen4-\dimen5 \ifPositiveGradient
\dimen0\dimen5 \advance\dimen1-\CD@WK\DiagramCellWidth\advance\dimen1 \CD@ZK
\DiagramCellWidth\setbox6=\llap{\unhbox6\kern.1\ht2}\setbox7=\rlap{\kern.1\ht
2\unhbox7}\else\dimen0\dimen1 \advance\dimen1-\CD@ZK\DiagramCellWidth\setbox7%
=\llap{\unhbox7\kern.1\ht2}\setbox6=\rlap{\kern.1\ht2\unhbox6}\fi\setbox6=%
\vbox{\box6\kern.1\wd2}\setbox7=\vtop{\kern.1\wd2\box7}\CD@dG{\dimen0}%
\advance\dimen0-\axisheight\advance\dimen0-\CD@bK\DiagramCellHeight\dimen5-%
\dimen0 \advance\dimen0\dimen6 \advance\dimen1.5\dimen3 \ifdim\wd3>\z@\ifdim
\ht3>-\dp3\CD@TB\fi\fi\dimen3\dimen2 \dimen7\dimen2\advance\dimen7\dimen4
\ifvoid3 \else\CD@tE\else\dimen8\ht3\advance\dimen8-\axisheight\CD@iC{\dimen8%
}\CD@X{\dimen8}{.5\wd3}\dimen9\dp3\advance\dimen9\axisheight\CD@iC{\dimen9}%
\CD@X{\dimen9}{.5\wd3}\ifPositiveGradient\advance\dimen2-\dimen9\advance
\dimen4-\dimen8 \else\advance\dimen4-\dimen9\advance\dimen2-\dimen8 \fi\fi
\advance\dimen3-.5\wd3 \fi\dimen9=\CD@aK\DiagramCellHeight\advance\dimen9-2%
\DiagramCellHeight\CD@tE\advance\dimen2\dimen4 \CD@CG{2}\dimen2-\dimen0%
\advance\dimen2\dp2 \else\CD@CG{2}\CD@CG{4}\ifPositiveGradient\dimen2-\dimen0%
\advance\dimen2\dp2 \dimen4\dimen5\advance\dimen4-\ht4 \else\dimen4-\dimen0%
\advance\dimen4\dp4 \dimen2\dimen5\advance\dimen2-\ht2 \fi\fi\setbox0=\hbox to%
\z@{\kern\dimen1 \ifvoid1 \else\ifPositiveGradient\advance\dimen0-\dp1 \lower
\dimen0 \else\advance\dimen5-\ht1 \raise\dimen5 \fi\rlap{\unhbox1}\fi\raise
\dimen2\rlap{\unhbox2}\ifvoid3 \else\lower.5\dimen9\rlap{\kern\dimen3\unhbox3%
}\fi\kern.5\dimen7 \lower.5\dimen9\box6 \lower.5\dimen9\box7 \kern.5\dimen7
\CD@tE\else\raise\dimen4\llap{\unhbox4}\fi\ifvoid5 \else\ifPositiveGradient
\advance\dimen5-\ht5 \raise\dimen5 \else\advance\dimen0-\dp5 \lower\dimen0 \fi
\llap{\unhbox5}\fi\hss}\ht0=\axisheight\dp0=-\ht0\box0 }\def\NorthWest{\CD@BI
\CD@rB\DiagonalChoice0 }\def\NorthEast{\CD@CI\CD@rB\DiagonalChoice1 }\def
\SouthWest{\CD@CI\CD@qB\DiagonalChoice3 }\def\SouthEast{\CD@BI\CD@qB
\DiagonalChoice2 }\def\CD@aD{\vadjust{\CD@uA\CD@FA\advance\CD@uA
\ifPositiveGradient\else-\fi\CD@ZK\relax\CD@vA\CD@NB\advance\CD@vA-\CD@bK
\relax\hbox{\advance\CD@uA\ifPositiveGradient-\fi\CD@WK\advance\CD@vA\CD@aK
\hbox{\box6 \kern\CD@DC\kern\CD@eJ\penalty1 \box7 \box\z@}\penalty\CD@uA
\penalty\CD@vA}\penalty\CD@uA\penalty\CD@vA\penalty104}}\def\CD@eH#1{\relax
\vadjust{\hbox@maths{#1}\penalty\CD@FA\penalty\CD@NB\penalty\tw@}}\def\CD@lB{%
\ifcase\CD@GB\or\or\CD@bH{.5\wd0}{\box0}{.5\wd0}\z@\or\unhbox\z@\setbox\z@
\lastbox\CD@bH{.5\wd0}{\box0}{.5\wd0}\z@\unpenalty\unpenalty\setbox\z@
\lastbox\or\CD@TG\else\advance\CD@GB-100 \ifnum\CD@GB<\z@\cd@shouldnt B\fi
\setbox\z@\hbox{\kern\CD@mF\copy\CD@MH\kern\CD@mI\CD@uA\CD@XB\advance\CD@uA-%
\CD@NB\penalty\CD@uA\CD@uA\CD@FA\advance\CD@uA-\CD@lA\penalty\CD@uA\unhbox\z@
\global\CD@yA\lastpenalty\unpenalty\global\CD@zA\lastpenalty\unpenalty}\CD@uA
-\CD@yA\CD@vA\CD@zA\CD@fI\fi}\def\CD@TG{\unhbox\z@\setbox\z@\lastbox\CD@uA
\lastpenalty\unpenalty\advance\CD@uA\CD@mA\CD@vA\CD@XB\advance\CD@vA-%
\lastpenalty\unpenalty\dimen1\lastkern\unkern\setbox3\lastbox\dimen0\lastkern
\unkern\setbox0=\hbox to\z@{\unhbox0\setbox0\lastbox\setbox7\lastbox
\unpenalty\CD@eJ\lastkern\unkern\CD@DC\lastkern\unkern\setbox6\lastbox\dimen7%
\CD@tB\advance\dimen7-\wd\CD@uA\ifdim\dimen7<\z@\CD@CI\multiply\dimen7\m@ne
\let\mv\empty\else\CD@BI\def\mv{\raise\ht1}\kern-\dimen7 \fi\ifnum\CD@vA>%
\CD@NB\dimen6\CD@uB\advance\dimen6-\ht\CD@vA\else\dimen6\z@\fi\CD@jJ\CD@mK
\setbox1\null\ht1\dimen6\wd1\dimen7 \dimen7\dimen2 \dimen6\wd1 \CD@KJ\CD@uA
\CD@LH\CD@vA\CD@TC\dimen6\ht1 \CD@KJ\setbox2\null\divide\dimen2\tw@\advance
\dimen2\CD@eJ\CD@eG{\dimen2}\wd2\dimen2 \dimen0.5\dimen7 \advance\dimen0%
\ifPositiveGradient\else-\fi\CD@eJ\CD@dG{\dimen0}\advance\dimen0-\axisheight
\ht2\dimen0 \dimen0\CD@DC\CD@eG{\dimen0}\advance\dimen0\ht2\ht2\dimen0 \dimen
0\ifPositiveGradient-\fi\CD@DC\CD@dG{\dimen0}\advance\dimen0\wd2\wd2\dimen0
\setbox4\null\dimen0 .6\CD@zC\CD@eG{\dimen0}\ht4\dimen0 \dimen0 .2\CD@zC
\CD@dG{\dimen0}\wd4\dimen0 \dimen0\wd2 \ifvoid6\else\dimen1\ht4 \advance
\dimen1\ht2 \CD@CC6+-\raise\dimen1\rlap{\ifPositiveGradient\advance\dimen0-%
\wd6\advance\dimen0-\wd4 \else\advance\dimen0\wd4 \fi\kern\dimen0\box6}\fi
\dimen0\wd2 \ifvoid7\else\dimen1\ht4 \advance\dimen1-\ht2 \CD@CC7-+\lower
\dimen1\rlap{\ifPositiveGradient\advance\dimen0\wd4 \else\advance\dimen0-\wd7%
\advance\dimen0-\wd4 \fi\kern\dimen0\box7}\fi\mv\box0\hss}\ht0\z@\dp0\z@
\CD@bH{\z@}{\box\z@}{\z@}{\axisheight}}\def\CD@CC#1#2#3{\dimen4.5\wd#1 \ifdim
\dimen4>.25\dimen7\dimen4=.25\dimen7\fi\ifdim\dimen4>\CD@zC\dimen4.4\dimen4
\advance\dimen4.6\CD@zC\fi\CD@eG{\dimen4}\dimen5\axisheight\CD@dG{\dimen5}%
\advance\dimen4-\dimen5 \dimen5\dimen4\CD@eG{\dimen5}\advance\dimen0%
\ifPositiveGradient#2\else#3\fi\dimen5 \CD@dG{\dimen4}\advance\dimen1\dimen4 }
\def\CD@eD#1{\expandafter\CD@IK{#1}}\CD@ZA\CD@EK{output is PostScript
dependent}\def\CD@SC{\CD@IK{/bturn {gsave currentpoint currentpoint translate
4 2 roll neg exch atan rotate neg exch neg exch translate } def /eturn {%
currentpoint grestore moveto} def}}\def\CD@gK{\relax\CD@hK\CD@tK{Q}\else
\CD@IK{eturn}\fi} \def\CD@OJ#1{\count@#1\relax\multiply\count@7\advance
\count@16577\divide\count@33154 }\def\CD@fD#1{\expandafter\special{#1}} \def
\CD@xJ#1{\setbox#1=\hbox{\dimen0\dimen#1\CD@dG{\dimen0}\CD@OJ{\dimen0}\setbox
0=\null\ifPositiveGradient\count@-\count@\ht0\dimen0 \else\dp0\dimen0 \fi\box
0 \CD@uA\count@\CD@OJ\CD@LF\CD@fD{pn \the\count@}\CD@fD{pa 0 0}\CD@OJ{\dimen#%
1}\CD@fD{pa \the\count@\space\the\CD@uA}\CD@fD{fp}\kern\dimen#1}}\def\CD@JI{%
\CD@KJ\begingroup\ifdim\dimen7<\dimen6 \dimen2=\dimen6 \dimen6=\dimen7 \dimen
7=\dimen2 \count@\CD@LH\CD@LH\CD@TC\CD@TC\count@\else\dimen2=\dimen7 \fi
\ifdim\dimen6>.01\p@\CD@KI\global\CD@QA\dimen0 \else\global\CD@QA\dimen7 \fi
\endgroup\dimen2\CD@QA\CD@iK\CD@lK{\ifPositiveGradient\else-\fi\dimen6}\CD@iK
\CD@kK{\ifPositiveGradient-\fi\dimen6}\CD@iK\CD@eK{\dimen7}}\def\CD@KI{\CD@hJ
\ifdim\dimen7>1.73\dimen6 \divide\dimen2 4 \multiply\CD@TC2 \else\dimen2=0.%
353553\dimen2 \advance\CD@LH-\CD@TC\multiply\CD@TC4 \fi\dimen0=4\dimen2 \CD@ZG
4\CD@ZG{-2}\CD@ZG2\CD@ZG{-2.5}}\def\CD@AI{\begingroup\count@\dimen0 \dimen2 45%
pt \divide\count@\dimen2 \ifdim\dimen0<\z@\advance\count@\m@ne\fi\ifodd
\count@\advance\count@1\CD@@A\else\CD@y\fi\advance\dimen0-\count@\dimen2
\CD@gE\multiply\dimen0\m@ne\fi\ifnum\count@<0 \multiply\count@-7 \fi\dimen3%
\dimen1 \dimen6\dimen0 \dimen7 3754936sp \ifdim\dimen0<6\p@\def\CD@OG{4000}%
\fi\CD@KJ\dimen2\dimen3\CD@dG{\dimen2}\CD@hJ\multiply\CD@TC-6 \dimen0\dimen2
\CD@ZG1\CD@ZG{0.3}\dimen1\dimen0 \dimen2\dimen3 \dimen0\dimen3 \CD@ZG3\CD@ZG{%
1.5}\CD@ZG{0.3}\divide\count@2 \CD@gE\multiply\dimen1\m@ne\fi\ifodd\count@
\dimen2\dimen1\dimen1\dimen0\dimen0-\dimen2 \fi\divide\count@2 \ifodd\count@
\multiply\dimen0\m@ne\multiply\dimen1\m@ne\fi\global\CD@QA\dimen0\global
\CD@RA\dimen1\endgroup\dimen6\CD@QA\dimen7\CD@RA}\def\CD@OC{255}\let\CD@OG
\CD@OC\def\CD@KJ{\begingroup\ifdim\dimen7<\dimen6 \dimen9\dimen7\dimen7\dimen
6\dimen6\dimen9\CD@@A\else\CD@y\fi\dimen2\z@\dimen3\CD@XH\dimen4\CD@XH\dimen0%
\z@\dimen8=\CD@OG\CD@XH\CD@lC\global\CD@yA\dimen\CD@gE0\else3\fi\global\CD@zA
\dimen\CD@gE3\else0\fi\endgroup\CD@LH\CD@yA\CD@TC\CD@zA}\def\CD@lC{\count@
\dimen6 \divide\count@\dimen7 \advance\dimen6-\count@\dimen7 \dimen9\dimen4
\advance\dimen9\count@\dimen0 \ifdim\dimen9>\dimen8 \CD@@C\else\CD@AC\ifdim
\dimen6>\z@\dimen9\dimen6 \dimen6\dimen7 \dimen7\dimen9 \expandafter
\expandafter\expandafter\CD@lC\fi\fi}\def\CD@@C{\ifdim\dimen0=\z@\ifdim\dimen
9<2\dimen8 \dimen0\dimen8 \fi\else\advance\dimen8-\dimen4 \divide\dimen8%
\dimen0 \ifdim\count@\CD@XH<2\dimen8 \count@\dimen8 \dimen9\dimen4 \advance
\dimen9\count@\dimen0 \CD@AC\fi\fi}\def\CD@AC{\dimen4\dimen0 \dimen0\dimen9
\advance\dimen2\count@\dimen3 \dimen9\dimen2 \dimen2\dimen3 \dimen3\dimen9 }%
\def\CD@ZG#1{\CD@dG{\dimen2}\advance\dimen0 #1\dimen2 }\def\CD@dG#1{\divide#1%
\CD@TC\multiply#1\CD@LH}\def\CD@eG#1{\divide#1\CD@vA\multiply#1\CD@uA}\def
\CD@iC#1{\divide#1\CD@LH\multiply#1\CD@TC}\def\CD@hJ{\dimen6\CD@LH\CD@XH
\multiply\dimen6\CD@LH\dimen7\CD@TC\CD@XH\multiply\dimen7\CD@TC\CD@KJ}\def
\CD@iK#1#2{\begingroup\dimen@#2\relax\loop\ifdim\dimen2<.4\maxdimen\multiply
\dimen2\tw@\multiply\dimen@\tw@\repeat\divide\dimen2\@cclvi\divide\dimen@
\dimen2\relax\multiply\dimen@\@cclvi\expandafter\CD@jK\the\dimen@\endgroup
\let#1\CD@fK}{\catcode`p=12 \catcode`0=12 \catcode`.=12 \catcode`t=12 \gdef
\CD@jK#1pt{\gdef\CD@fK{#1}}}\ifx\errorcontextlines\CD@qK\CD@tJ\let\CD@GH
\relax\else\def\CD@GH{\errorcontextlines\m@ne}\fi\ifnum\inputlineno<0 \let
\CD@CD\empty\let\CD@W\empty\let\CD@mD\relax\let\CD@uI\relax\let\CD@vI\relax
\let\CD@zF\relax\else\def\CD@W{ at line \number\inputlineno}\def\CD@mD{ - first occurred%
}\def\CD@uI{\edef\CD@h{\the\inputlineno}\global\let\CD@jB\CD@h}\def\CD@h{9999%
}\def\CD@vI{\xdef\CD@jB{\the\inputlineno}}\def\CD@jB{\CD@h}\def\CD@zF{\ifnum
\CD@h<\inputlineno\edef\CD@CD{\space at lines \CD@h--\the\inputlineno}\else
\edef\CD@CD{\CD@W}\fi}\fi\let\CD@CD\empty\def\CD@YA#1#2{\CD@GH\errhelp=#2%
\expandafter\errmessage{\CD@tA: #1}}\def\CD@KB#1{\begingroup\expandafter
\message{! \CD@tA: #1\CD@CD}\ifnum\CD@XB>\CD@NB\ifnum\CD@CA>\CD@NB\else\ifnum
\CD@lA>\CD@FA\else\ifnum\CD@LB>\CD@FA\advance\CD@XB-\CD@NB\advance\CD@FA-%
\CD@lA\advance\CD@FA1\relax\expandafter\message{! (error detected at row \the
\CD@XB, column \the\CD@FA, but probably caused elsewhere)}\fi\fi\fi\fi
\endgroup}\def\CD@gB#1{{\expandafter\message{\CD@tA\space Warning: #1\CD@W}}}%
\def\CD@CB#1#2{\CD@gB{#1 \string#2 is obsolete\CD@mD}}\def\CD@AB#1{\CD@CB{%
Dimension}{#1}\CD@DE#1\CD@BB\CD@BB}\def\CD@BB{\CD@OA=}\def\CD@@B#1{\CD@CB{%
Count}{#1}\CD@DE#1\CD@OH\CD@OH}\def\CD@OH{\count@=}\def\HorizontalMapLength{%
\CD@AB\HorizontalMapLength}\def\VerticalMapHeight{\CD@AB\VerticalMapHeight}%
\def\VerticalMapDepth{\CD@AB\VerticalMapDepth}\def\VerticalMapExtraHeight{%
\CD@AB\VerticalMapExtraHeight}\def\VerticalMapExtraDepth{\CD@AB
\VerticalMapExtraDepth}\def\DiagonalLineSegments{\CD@@B\DiagonalLineSegments}%
\ifx\tenln\nullfont\CD@ZA\CD@KH{\CD@eF\space diagonal line and arrow font not
available}\else\let\CD@KH\relax\fi\def\CD@aG#1#2<#3:#4:#5#6{\begingroup\CD@PA
#3\relax\advance\CD@PA-#2\relax\ifdim.1em<\CD@PA\CD@uA#5\relax\CD@vA#6\relax
\ifnum\CD@uA<\CD@vA\count@\CD@vA\advance\count@-\CD@uA\CD@KB{#4 by \the\CD@PA
}\if#1v\let\CD@CH\CD@JK\edef\tmp{\the\CD@uA--\the\CD@vA,\the\CD@FA}\else
\advance\count@\count@\if#1l\advance\count@-\CD@A\else\if#1r\advance\count@
\CD@A\fi\fi\advance\CD@PA\CD@PA\let\CD@CH\CD@ZE\edef\tmp{\the\CD@NB,\the
\CD@uA--\the\CD@vA}\fi\divide\CD@PA\count@\ifdim\CD@CH<\CD@PA\global\CD@CH
\CD@PA\fi\fi\fi\endgroup}\CD@tG\CD@xE\CD@JD\CD@ID\CD@rG\CD@xI{See the message
above.}\CD@rG\CD@lH{Perhaps you've forgotten to end the diagram before
resuming the text, in\CD@uG which case some garbage may be added to the
diagram, but we should be ok now.\CD@uG Alternatively you've left a blank line
in the middle - TeX will now complain\CD@uG that the remaining \CD@S s are
misplaced - so please use comments for layout.}\CD@rG\CD@hD{You have already
closed too many brace pairs or environments; an \CD@HD\CD@uG command was (%
over)due.}\CD@rG\CD@hH{\CD@dC\space and \CD@HD\space commands must match.}%
\def\CD@jH{\ifnum\inputlineno=0 \else\expandafter\CD@iH\fi}\def\CD@iH{\CD@MD
\CD@GD\crcr\CD@YA{missing \CD@HD\space inserted before \CD@kH- type "h"}%
\CD@lH\enddiagram\CD@AG\CD@kH\par}\def\CD@AG#1{\edef\enddiagram{\noexpand
\CD@rD{#1\CD@W}}}\def\CD@rD#1{\CD@YA{\CD@HD\space(anticipated by #1) ignored}%
\CD@xI\let\enddiagram\CD@SG}\def\CD@SG{\CD@YA{misplaced \CD@HD\space ignored}%
\CD@hH}\def\CD@mC{\CD@YA{missing \CD@HD\space inserted.}\CD@hD\CD@AG{closing
group}}\ifx\ifx
\fi\fi

\catcode`\$=3 
\def\vboxtoz{\vbox to\z@}

\def\scriptaxis#1{\@scriptaxis{$\scriptstyle#1$}}
\def\ssaxis#1{\@ssaxis{$\scriptscriptstyle#1$}}
\def\@scriptaxis#1{\dimen0\axisheight\advance\dimen0-\ss@axisheight\raise
\dimen0\hbox{#1}}\def\@ssaxis#1{\dimen0\axisheight\advance\dimen0-%
\ss@axisheight\raise\dimen0\hbox{#1}}

\ifx\boldmath\CD@qK
\let\boldscriptaxis\scriptaxis
\def\boldscript#1{\hbox{$\scriptstyle#1$}}
\else\def\boldscriptaxis#1{\@scriptaxis{\boldmath$\scriptstyle#1$}}
\def\boldscript#1{\hbox{\boldmath$\scriptstyle#1$}}
\fi

\def\raisehook#1#2#3{\hbox{\setbox3=\hbox{#1$\scriptscriptstyle#3$}%
\dimen0\ss@axisheight
\dimen1\axisheight\advance\dimen1-\dimen0
\dimen2\ht3\advance\dimen2-\dimen0%
\advance\dimen2-0.021em\advance\dimen1 #2\dimen2%
\raise\dimen1\box3}}
\def\shifthook#1#2#3{\setbox1=\hbox{#1$\scriptscriptstyle#3$}\dimen0\wd1%
\divide\dimen0 12\CD@zH{\dimen0}
\dimen1\wd1\advance\dimen1-2\dimen0 \advance\dimen1-2\CD@oI\CD@zH{\dimen1}%
\kern#2\dimen1\box1}

\def\@cmex{\mathchar"03}

\def\make@pbk#1{\setbox\tw@\hbox to\z@{#1}\ht\tw@\z@\dp\tw@\z@\box\tw@}\def
\CD@fH#1{\overprint{\hbox to\z@{#1}}}\def\CD@qH{\kern0.11em}\def\CD@pH{\kern0%
.35em}

\def\dblvert{\def\CD@rH{\kern.5\PileSpacing}}\def\CD@rH{}

\def\SEpbk{\make@pbk{\CD@qH\CD@rH\vrule depth 2.87ex height -2.75ex width 0.%
95em \vrule height -0.66ex depth 2.87ex width 0.05em \hss}}

\def\SWpbk{\make@pbk{\hss\vrule height -0.66ex depth 2.87ex width 0.05em
\vrule depth 2.87ex height -2.75ex width 0.95em \CD@qH\CD@rH}}

\def\NEpbk{\make@pbk{\CD@qH\CD@rH\vrule depth -3.81ex height 4.00ex width 0.%
95em \vrule height 4.00ex depth -1.72ex width 0.05em \hss}}

\def\NWpbk{\make@pbk{\hss\vrule height 4.00ex depth -1.72ex width 0.05em
\vrule depth -3.81ex height 4.00ex width 0.95em \CD@qH\CD@rH}}

\def\puncture{{\setbox0\hbox{A}\vrule height.53\ht0 depth-.47\ht0 width.35\ht
0 \kern.12\ht0 \vrule height\ht0 depth-.65\ht0 width.06\ht0 \kern-.06\ht0
\vrule height.35\ht0 depth0pt width.06\ht0 \kern.12\ht0 \vrule height.53\ht0
depth-.47\ht0 width.35\ht0 }}

\def\NEclck{\overprint{\raise2.5ex\rlap{ \CD@rH$\scriptstyle\searrow$}}}
\def\NEanti{\overprint{\raise2.5ex\rlap{ \CD@rH$\scriptstyle\nwarrow$}}}
\def\NWclck{\overprint{\raise2.5ex\llap{$\scriptstyle\nearrow$ \CD@rH}}}
\def\NWanti{\overprint{\raise2.5ex\llap{$\scriptstyle\swarrow$ \CD@rH}}}
\def\SEclck{\overprint{\lower1ex\rlap{ \CD@rH$\scriptstyle\swarrow$}}}
\def\SEanti{\overprint{\lower1ex\rlap{ \CD@rH$\scriptstyle\nearrow$}}}
\def\SWclck{\overprint{\lower1ex\llap{$\scriptstyle\nwarrow$ \CD@rH}}}
\def\SWanti{\overprint{\lower1ex\llap{$\scriptstyle\searrow$ \CD@rH}}}

\def\rhvee{\mkern-10mu\greaterthan}
\def\lhvee{\lessthan\mkern-10mu}
\def\dhvee{\vboxtoz{\vss\hbox{$\vee$}\kern0pt}}
\def\uhvee{\vboxtoz{\hbox{$\wedge$}\vss}}
\newarrowhead{vee}\rhvee\lhvee\dhvee\uhvee

\def\dhlvee{\vboxtoz{\vss\hbox{$\scriptstyle\vee$}\kern0pt}}
\def\uhlvee{\vboxtoz{\hbox{$\scriptstyle\wedge$}\vss}}
\newarrowhead{littlevee}{\mkern1mu\scriptaxis\rhvee}{\scriptaxis\lhvee}%
\dhlvee\uhlvee\ifx\boldmath\CD@qK
\newarrowhead{boldlittlevee}{\mkern1mu\scriptaxis\rhvee}{\scriptaxis\lhvee}%
\dhlvee\uhlvee\else
\def\dhblvee{\vboxtoz{\vss\boldscript\vee\kern0pt}}
\def\uhblvee{\vboxtoz{\boldscript\wedge\vss}}
\newarrowhead{boldlittlevee}{\mkern1mu\boldscriptaxis\rhvee}{\boldscriptaxis
\lhvee}\dhblvee\uhblvee
\fi

\def\rhcvee{\mkern-10mu\succ}
\def\lhcvee{\prec\mkern-10mu}
\def\dhcvee{\vboxtoz{\vss\hbox{$\curlyvee$}\kern0pt}}
\def\uhcvee{\vboxtoz{\hbox{$\curlywedge$}\vss}}
\newarrowhead{curlyvee}\rhcvee\lhcvee\dhcvee\uhcvee

\def\rhvvee{\mkern-13mu\gg}
\def\lhvvee{\ll\mkern-13mu}
\def\dhvvee{\vboxtoz{\vss\hbox{$\vee$}\kern-.6ex\hbox{$\vee$}\kern0pt}}
\def\uhvvee{\vboxtoz{\hbox{$\wedge$}\kern-.6ex \hbox{$\wedge$}\vss}}
\newarrowhead{doublevee}\rhvvee\lhvvee\dhvvee\uhvvee

\def\rhtriangle{\triangleright\mkern1.2mu}
\def\lhtriangle{\triangleleft\mkern.8mu}
\def\uhtriangle{\vbox{\kern-.2ex \hbox{$\scriptscriptstyle\bigtriangleup$}%
\kern-.25ex}}
\def\dhtriangle{\vbox{\kern-.28ex \hbox{$\scriptscriptstyle\bigtriangledown$}%
\kern-.1ex}}
\def\dhblack{\vbox{\kern-.25ex\nointerlineskip\hbox{$\blacktriangledown$}}}%
\def\uhblack{\vbox{\kern-.25ex\nointerlineskip\hbox{$\blacktriangle$}}}%
\def\dhlblack{\vbox{\kern-.25ex\nointerlineskip\hbox{$\scriptstyle
\blacktriangledown$}}}
\def\uhlblack{\vbox{\kern-.25ex\nointerlineskip\hbox{$\scriptstyle
\blacktriangle$}}}
\newarrowhead{triangle}\rhtriangle\lhtriangle\dhtriangle\uhtriangle
\newarrowhead{blacktriangle}{\mkern-1mu\blacktriangleright\mkern.4mu}{%
\blacktriangleleft}\dhblack\uhblack\newarrowhead{littleblack}{\mkern-1mu%
\scriptaxis\blacktriangleright}{\scriptaxis\blacktriangleleft\mkern-2mu}%
\dhlblack\uhlblack

\def\rhla{\hbox{\setbox0=\lnchar55\dimen0=\wd0\kern-.6\dimen0\ht0\z@\raise
\axisheight\box0\kern.1\dimen0}}
\def\lhla{\hbox{\setbox0=\lnchar33\dimen0=\wd0\kern.05\dimen0\ht0\z@\raise
\axisheight\box0\kern-.5\dimen0}}
\def\dhla{\vboxtoz{\vss\rlap{\lnchar77}}}
\def\uhla{\vboxtoz{\setbox0=\lnchar66 \wd0\z@\kern-.15\ht0\box0\vss}}
\newarrowhead{LaTeX}\rhla\lhla\dhla\uhla

\def\lhlala{\lhla\kern.3em\lhla}
\def\rhlala{\rhla\kern.3em\rhla}
\def\uhlala{\hbox{\uhla\raise-.6ex\uhla}}
\def\dhlala{\hbox{\dhla\lower-.6ex\dhla}}
\newarrowhead{doubleLaTeX}\rhlala\lhlala\dhlala\uhlala

\def\hhO{\scriptaxis\bigcirc\mkern.4mu} \def\hho{{\circ}\mkern1.2mu}%
\newarrowhead{o}\hho\hho\circ\circ
\newarrowhead{O}\hhO\hhO{\scriptstyle\bigcirc}{\scriptstyle\bigcirc}

\def\rhtimes{\mkern-5mu{\times}\mkern-.8mu}\def\lhtimes{\mkern-.8mu{\times}%
\mkern-5mu}\def\uhtimes{\setbox0=\hbox{$\times$}\ht0\axisheight\dp0-\ht0%
\lower\ht0\box0 }\def\dhtimes{\setbox0=\hbox{$\times$}\ht0\axisheight\box0 }%
\newarrowhead{X}\rhtimes\lhtimes\dhtimes\uhtimes\newarrowhead+++++

\newarrowhead{Y}{\mkern-3mu\Yright}{\Yleft\mkern-3mu}\Ydown\Yup

\newarrowhead{->}\rightarrow\leftarrow\downarrow\uparrow

\newarrowhead{=>}\Rightarrow\Leftarrow{\@cmex7F}{\@cmex7E}

\newarrowhead{harpoon}\rightharpoonup\leftharpoonup\downharpoonleft
\upharpoonleft

\def\twoheaddownarrow{\rlap{$\downarrow$}\raise-.5ex\hbox{$\downarrow$}}
\def\twoheaduparrow{\rlap{$\uparrow$}\raise.5ex\hbox{$\uparrow$}}
\newarrowhead{->>}\twoheadrightarrow\twoheadleftarrow\twoheaddownarrow
\twoheaduparrow

\def\ltvee{\mkern-1mu{\lessthan}\mkern.4mu}
\newarrowtail{vee}\greaterthan\ltvee\vee\wedge

\newarrowtail{littlevee}{\scriptaxis\greaterthan}{\mkern-1mu\scriptaxis
\lessthan}{\scriptstyle\vee}{\scriptstyle\wedge}\ifx\boldmath\CD@qK
\newarrowtail{boldlittlevee}{\scriptaxis\greaterthan}{\mkern-1mu\scriptaxis
\lessthan}{\scriptstyle\vee}{\scriptstyle\wedge}\else\newarrowtail{%
boldlittlevee}{\boldscriptaxis\greaterthan}{\mkern-1mu\boldscriptaxis
\lessthan}{\boldscript\vee}{\boldscript\wedge}\fi

\newarrowtail{curlyvee}\succ{\mkern-1mu{\prec}\mkern.4mu}\curlyvee\curlywedge

\def\rttriangle{\mkern1.2mu\triangleright}
\newarrowtail{triangle}\rttriangle\lhtriangle\dhtriangle\uhtriangle
\newarrowtail{blacktriangle}\blacktriangleright{\mkern-1mu\blacktriangleleft
\mkern.4mu}\dhblack\uhblack\newarrowtail{littleblack}{\scriptaxis
\blacktriangleright\mkern-2mu}{\mkern-1mu\scriptaxis\blacktriangleleft}%
\dhlblack\uhlblack

\def\rtla{\hbox{\setbox0=\lnchar55\dimen0=\wd0\kern-.5\dimen0\ht0\z@\raise
\axisheight\box0\kern-.2\dimen0}}
\def\ltla{\hbox{\setbox0=\lnchar33\dimen0=\wd0\kern-.15\dimen0\ht0\z@\raise
\axisheight\box0\kern-.5\dimen0}}
\def\dtla{\vbox{\setbox0=\rlap{\lnchar77}\dimen0=\ht0\kern-.7\dimen0\box0%
\kern-.1\dimen0}}
\def\utla{\vbox{\setbox0=\rlap{\lnchar66}\dimen0=\ht0\kern-.1\dimen0\box0%
\kern-.6\dimen0}}
\newarrowtail{LaTeX}\rtla\ltla\dtla\utla

\def\rtvvee{\gg\mkern-3mu}
\def\ltvvee{\mkern-3mu\ll}
\def\dtvvee{\vbox{\hbox{$\vee$}\kern-.6ex \hbox{$\vee$}\vss}}
\def\utvvee{\vbox{\vss\hbox{$\wedge$}\kern-.6ex \hbox{$\wedge$}\kern\z@}}
\newarrowtail{doublevee}\rtvvee\ltvvee\dtvvee\utvvee

\def\ltlala{\ltla\kern.3em\ltla}
\def\rtlala{\rtla\kern.3em\rtla}
\def\utlala{\hbox{\utla\raise-.6ex\utla}}
\def\dtlala{\hbox{\dtla\lower-.6ex\dtla}}
\newarrowtail{doubleLaTeX}\rtlala\ltlala\dtlala\utlala

\def\utbar{\vrule height 0.093ex depth0pt width 0.4em}
\let\dtbar\utbar
\def\rtbar{\mkern1.5mu\vrule height 1.1ex depth.06ex width .04em\mkern1.5mu}%
\let\ltbar\rtbar
\newarrowtail{mapsto}\rtbar\ltbar\dtbar\utbar
\newarrowtail{|}\rtbar\ltbar\dtbar\utbar

\def\rthooka{\raisehook{}+\subset\mkern-1mu}
\def\lthooka{\mkern-1mu\raisehook{}+\supset}
\def\rthookb{\raisehook{}-\subset\mkern-2mu}
\def\lthookb{\mkern-1mu\raisehook{}-\supset}

\def\dthooka{\shifthook{}+\cap}
\def\dthookb{\shifthook{}-\cap}
\def\uthooka{\shifthook{}+\cup}
\def\uthookb{\shifthook{}-\cup}

\newarrowtail{hooka}\rthooka\lthooka\dthooka\uthooka\newarrowtail{hookb}%
\rthookb\lthookb\dthookb\uthookb

\ifx\boldmath\CD@qK\newarrowtail{boldhooka}\rthooka\lthooka\dthooka\uthooka
\newarrowtail{boldhookb}\rthookb\lthookb\dthookb\uthookb\newarrowtail{%
boldhook}\rthooka\lthooka\dthookb\uthooka\else\def\rtbhooka{\raisehook
\boldmath+\subset\mkern-1mu}
\def\ltbhooka{\mkern-1mu\raisehook\boldmath+\supset}
\def\rtbhookb{\raisehook\boldmath-\subset\mkern-2mu}
\def\ltbhookb{\mkern-1mu\raisehook\boldmath-\supset}
\def\dtbhooka{\shifthook\boldmath+\cap}
\def\dtbhookb{\shifthook\boldmath-\cap}
\def\utbhooka{\shifthook\boldmath+\cup}
\def\utbhookb{\shifthook\boldmath-\cup}
\newarrowtail{boldhooka}\rtbhooka\ltbhooka\dtbhooka\utbhooka\newarrowtail{%
boldhookb}\rtbhookb\ltbhookb\dtbhookb\utbhookb\newarrowtail{boldhook}%
\rtbhooka\ltbhooka\dtbhooka\utbhooka\fi

\def\dtsqhooka{\shifthook{}+\sqcap}
\def\ltsqhooka{\mkern-1mu\raisehook{}+\sqsupset}
\def\rtsqhooka{\raisehook{}+\sqsubset\mkern-1mu}
\def\utsqhooka{\shifthook{}+\sqcup}
\newarrowtail{sqhook}\rtsqhooka\ltsqhooka\dtsqhooka\utsqhooka

\newarrowtail{hook}\rthooka\lthookb\dthooka\uthooka\newarrowtail{C}\rthooka
\lthookb\dthooka\uthooka

\newarrowtail{o}\hho\hho\circ\circ
\newarrowtail{O}\hhO\hhO{\scriptstyle\bigcirc}{\scriptstyle\bigcirc}

\newarrowtail{X}\lhtimes\rhtimes\uhtimes\dhtimes\newarrowtail+++++

\newarrowtail{Y}\Yright\Yleft\Ydown\Yup

\newarrowtail{harpoon}\leftharpoondown\rightharpoondown\upharpoonright
\downharpoonright

\newarrowfiller{=}=={\@cmex77}{\@cmex77}
\def\vfthree{\mid\!\!\!\mid\!\!\!\mid}
\newarrowfiller{3}\equiv\equiv\vfthree\vfthree

\def\vfdashstrut{\vrule width0pt height1.3ex depth0.7ex}
\def\vfthedash{\vrule width\CD@LF height0.6ex depth 0pt}
\def\hfthedash{\CD@AJ\vrule\horizhtdp width 0.26em}
\def\hfdash{\mkern5.5mu\hfthedash\mkern5.5mu}
\def\vfdash{\vfdashstrut\vfthedash}
\newarrowfiller{dash}\hfdash\hfdash\vfdash\vfdash

\newarrowmiddle+++++

\iffalse
\newarrow{To}----{vee}
\newarrow{Arr}----{LaTeX}
\newarrow{Dotsto}....{vee}
\newarrow{Dotsarr}....{LaTeX}
\newarrow{Dashto}{}{dash}{}{dash}{vee}
\newarrow{Dasharr}{}{dash}{}{dash}{LaTeX}
\newarrow{Mapsto}{mapsto}---{vee}
\newarrow{Mapsarr}{mapsto}---{LaTeX}
\newarrow{IntoA}{hooka}---{vee}
\newarrow{IntoB}{hookb}---{vee}
\newarrow{Embed}{vee}---{vee}
\newarrow{Emarr}{LaTeX}---{LaTeX}
\newarrow{Onto}----{doublevee}
\newarrow{Dotsonarr}....{doubleLaTeX}
\newarrow{Dotsonto}....{doublevee}
\newarrow{Dotsonarr}....{doubleLaTeX}
\else
\newarrow{To}---->
\newarrow{Arr}---->
\newarrow{Dotsto}....>
\newarrow{Dotsarr}....>
\newarrow{Dashto}{}{dash}{}{dash}>
\newarrow{Dasharr}{}{dash}{}{dash}>
\newarrow{Mapsto}{mapsto}--->
\newarrow{Mapsarr}{mapsto}--->
\newarrow{IntoA}{hooka}--->
\newarrow{IntoB}{hookb}--->
\newarrow{Embed}>--->
\newarrow{Emarr}>--->
\newarrow{Onto}----{>>}
\newarrow{Dotsonarr}....{>>}
\newarrow{Dotsonto}....{>>}
\newarrow{Dotsonarr}....{>>}
\fi

\newarrow{Implies}===={=>}
\newarrow{Project}----{triangle}
\newarrow{Pto}----{harpoon}
\newarrow{Relto}{harpoon}---{harpoon}

\newarrow{Eq}=====
\newarrow{Line}-----
\newarrow{Dots}.....
\newarrow{Dashes}{}{dash}{}{dash}{}

\newarrow{SquareInto}{sqhook}--->

\newarrowhead{cmexbra}{\@cmex7B}{\@cmex7C}{\@cmex3B}{\@cmex38}
\newarrowtail{cmexbra}{\@cmex7A}{\@cmex7D}{\@cmex39}{\@cmex3A}
\newarrowmiddle{cmexbra}{\braceru\bracelu}{\bracerd\braceld}{\vcenter{%
\hbox@maths{\@cmex3D\mkern-2mu}}}
{\vcenter{\hbox@maths{\mkern2mu\@cmex3C}}}
\newarrow{@brace}{cmexbra}-{cmexbra}-{cmexbra}
\newarrow{@parenth}{cmexbra}---{cmexbra}
\def\rightBrace{\d@brace[thick,cmex]}
\def\leftBrace{\u@brace[thick,cmex]}
\def\upperBrace{\r@brace[thick,cmex]}
\def\lowerBrace{\l@brace[thick,cmex]}
\def\rightParenth{\d@parenth[thick,cmex]}
\def\leftParenth{\u@parenth[thick,cmex]}
\def\upperParenth{\r@parenth[thick,cmex]}
\def\lowerParenth{\l@parenth[thick,cmex]}

\let\dInto\dIntoB
\let\rdInto\rdIntoA
\let\hEq\rEq
\let\vEq\uEq
\let\hLine\rLine

\def\labelstyle{
\ifincommdiag
\textstyle
\else
\scriptstyle
\fi}
\let\objectstyle\displaystyle

\newdiagramgrid{pentagon}{0.618034,0.618034,1,1,1,1,0.618034,0.618034}{1.%
17557,1.17557,1.902113,1.902113}

\newdiagramgrid{perspective}{0.75,0.75,1.1,1.1,0.9,0.9,0.95,0.95,0.75,0.75}{0%
.75,0.75,1.1,1.1,0.9,0.9}

\diagramstyle[
dpi=300,
vmiddle,nobalance,
loose,
thin,
pilespacing=10pt,%
shortfall=4pt,
]

\ifx\CD@qK\else\CD@PK\relax\CD@FF\CD@e\fi\fi

\CD@vE\CD@hK\else\fi\else\fi\cdrestoreat
\newcommand{\matB}{{\mathcal B}}\newcommand{\ba}{\begin{eqnarray}}
\newcommand{\ea}{\end{eqnarray}}
\newcommand{\bege}{\begin{equation}}
\newcommand{\enge}{\end{equation}}
\newcommand{\benu}{\begin{enumerate}}
\newcommand{\enu}{\end{enumerate}}
\newcommand{\pa}{\partial}
\newcommand{\noi}{\noindent}
\newcommand{\bbbbox}{\mathop{\Box\kern -5pt\raisebox{.8pt}{$|$}}}

\newcommand{\mr}{\mathring}

\newcommand{\mt}{\mathcal}
\newcommand{\cl}{\mt{C}\ell}

\newcommand{\w}{\wedge}

\newcommand{\CC}{\mathbb{C}}
\newcommand{\OO}{\mathbb{O}}
\newcommand{\II}{\mathbb{I}}
\def\beq{\begin{eqnarray}}
\def\eeq{\end{eqnarray}}

\def\ua{\{-,+\}}
\def\da{\{+,-\}}

\def\ua{\{-,+\}}
\def\da{\{+,-\}}

\def\p{\mbox{\boldmath$\displaystyle\mathbf{p}$}}

\def\0{\mbox{\boldmath$\displaystyle\mathbf{0}$}}

\def\h00h{\mbox{\boldmath$\displaystyle\mathbf{(1/2,0)\oplus(0,1/2)}$}}

\def\ba{p \,e^{i\phi}  \sin(\theta)}

\title{Exotic Dark Spinor Fields}
\author[a]{Rold\~ao da Rocha}
\author[b]{Alex E. Bernardini}
\author[c]{J. M. Hoff da Silva}
\affiliation[a]{Centro de Matem\'atica, Computa\c c\~ao e Cogni\c c\~ao,
Universidade Federal do ABC\\
Rua Santa Ad\'elia, 166 09210-170, Santo Andr\'e, SP, Brazil}
\affiliation[b]{Departamento de F\'{\i}sica, Universidade Federal de S\~ao
Carlos\\
PO Box 676, 13565-905, S\~ao Carlos, SP, Brazil}
\affiliation[c]{UNESP - Campus de Guaratinguet\'a - DFQ\\
Av. Dr.
Ariberto Pereira da Cunha, 333 12516-410, Guaratinguet\'a-SP,
Brazil.}
\emailAdd{roldao.rocha@ufabc.edu.br}
\emailAdd{alexeb@ufscar.br}
\emailAdd{hoff@feg.unesp.br; hoff@ift.unesp.br}
\abstract{
Exotic dark spinor fields are introduced and investigated in the context of inequivalent spin structures on arbitrary curved spacetimes, which induces an additional term on the associated Dirac operator, related to a $\check{\rm C}$ech cohomology class. For the most kinds of spinor fields, any exotic term in the Dirac operator can be absorbed and encoded as a shift of the  electromagnetic vector potential representing an element of the cohomology group $H^1(M, \mathbb{Z}_2)$. The possibility of concealing such an exotic term does not exist in case of dark (ELKO) spinor fields, as they cannot carry electromagnetic charge, so that the full topological analysis must be evaluated.
Since exotic dark spinor fields also satisfy Klein-Gordon propagators, the dynamical constraints related to the exotic term in the Dirac equation can be explicitly calculated. It forthwith implies that the non-trivial topology associated to the spacetime can drastically engender --- from the dynamics of dark spinor fields --- constraints in the spacetime metric structure. Meanwhile, such constraints may be alleviated, at the cost of constraining the exotic spacetime topology. Besides being prime candidates to the dark matter problem, dark spinor fields are shown to be potential candidates to probe non-trivial topologies in spacetime, as well as probe the spacetime metric structure.}
\arxivnumber{arXiv:1103.4759v1 [hep-th]}
\begin{document}
\maketitle
\flushbottom

\section{Introduction}

ELKO --- \emph{Eigenspinoren des Ladungskonjugationsoperators} --- spinor
fields\footnote{ELKO is the German acronym for  eigenspinors of the charge conjugation
operator.}  describe a  non-standard Wigner class of fermions, for which charge conjugation and parity are commuting operators, rather than anticommuting ones \cite{allu,alu2,ahlu4}. 
They support two types of dispersion relations, accomplish dual-helicity eigenspinors of the spin-1/2 charge conjugation operator,
and carry mass dimension one, besides having non-local properties. At low-energy limits, ELKO behaves as a representation of the Lorentz group through the setup of a preferred frame related to its wave equation \cite{ahlu4,alex,alex1}.
Ahluwalia-Khalilova and Grumiller embedded ELKO \cite{allu} into the quantum field theory, from which large applications in cosmology and gravity can be outlined.
The corresponding ELKO Lagrangian neither predicts interactions with Standard Model (SM) fields nor shows coupling with gauge fields. Otherwise, exotic interactions with the Higgs boson can somehow be depicted in order to endow such spinor fields to be prime candidates to describe dark matter \cite{alu3}.
In particular, observational aspects on such a possibility has been proposed at LHC: dark (ELKO) spinor fields can be observed, at center of mass energy around 7 TeV and total luminosity from $1 fb^{-1}$ to $10 fb^{-1}$, indicating that the number of events is large enough to motivate a detailed analysis about ELKO particle at high energy experiments \cite{marcao}.



In addition, the embedding of dark spinor fields into the SM \cite{hof,osmano} was introduced. ELKO spinor fields
 dominant interaction via the gravitational field makes them naturally dark, and recently \cite{bo121} dark spinor fields were investigated in a cosmological setting, where interesting solutions and also models where the spinor is coupled conformally to gravity are provided. Some additional applications of ELKO spinor fields to cosmology can be seen, e.g., in  \cite{Gredat:2008qf,Shankaranarayanan:2009sz,Shankaranarayanan:2010st,Wei:2010ad,boe1,boe111,boe12,boe13,gau,schritt,clocks}. In particular, possible applications of ELKO spinor fields to more general $f(R)$ gravitational theories were accomplished in \cite{frabbri}, and supersymmetric models
concerning ELKO were introduced in \cite{Dick}.

The main aim of our manuscript is to investigate dark spinor fields in spacetimes with non-trivial topologies, in order to clarify how the dynamics of such dark spinor fields can induce constraints on the metric spacetime structure, as well as in the non-trivial topology itself.
Physical applications of non-trivial topologies on spacetime, including thermodynamics, superconductivity, and condensed matter have been extensively explored in the last years.
For instance, the quantum theory of fields propagating on a manifold $M$ not simply connected was investigated in \cite{isham0}. The existence of a nontrivial line bundle on a manifold $M$, whose sections may be regarded as a generalization of the concept
of a scalar field, is inherent in multiply connected manifolds, which in addition can imply in the existence of inequivalent spin structures.
It is well known that the set of real line bundles on $M$ and the set of inequivalent spin structures are both labeled by elements of the cohomology group $H^1(M,\mathbb{Z}_2)$ --- the group of homomorphisms of the fundamental group $\pi_1(M)$ into $\mathbb{Z}_2$.
Namely, there are many globally different spin structures which arise from inequivalent patchings of the local double
coverings, see, e.g., \cite{mil1}.
The use of generalized spin structures has been discussed in \cite{atiyah,forger,Hawking1,petry}, in particular the ones considered by \cite{avis0} for the case where the original fiber bundle has no spin structure at all.

Whereas the manifold $M$ may have no spinor bundle --- when it does not satisfy Geroch theorem hypotheses \cite{Geroch} --- it may have also many of them, which are split into equivalence classes --- called spin structure.
The formal aspects about the inequivalent spin structures
in spacetime in terms of the different possible spin connections \cite{isham0,isham1,petry} have been explored and reveal prominent physical applications.
In an arbitrary spacetime that admits spin structure we delve into the problem of how to select a
particular spin structure and its corresponding Dirac equation.
In \cite{isham0} it is discussed in what content the quantum field theory associated with a spinor field must involve some kind of ``average" of all the spin connections\footnote{As it must be summed over all topological sectors in instanton physics \cite{isham0}.}.

On simply connected spacetimes, the associated fundamental group satisfies $\pi_1(M) = 0$, therefore there is only one spin structure. Although compact  simply connected 4-dimensional manifolds admitting a spin structure can be classified in terms of the Euler and Pontryagin numbers \cite{Hawking2}, multiply connected spacetimes are devoid in general of such a classification.
As argued in \cite{isham0,isham1}, since Nature seems to use all mutually consistent degrees of
freedom in a physical system, the Feynman path integral formalism, for instance, should also include multiply
connected manifolds --- which is among other prominent motivations to investigate quantum field theory on such spacetimes, which in addition are used in quantum gravity at both cosmological and Planck length scales \cite{Hawking2,duff}.
Multiply connected manifolds are also elicited in the theory of superconductivity.
In fact, in \cite{petry} the inequivalent generalized spin structures are investigated in order to explain Cooper
pairing phenomena in superconductors.
In addition, such manifolds are used in the instanton compactification \cite{3} of 4-dimensional Euclidean spacetimes and also in t' Hooft's treatment of confinement.
In \cite{unwin}, the finite-temperature stress-energy-momentum for a conformally coupled massive scalar field in multiply connected spaces was calculated and cogently investigated from a thermodynamic viewpoint.
Also, in \cite{dewitt} and \cite{banach,Dowker} covariant Casimir calculations were performed
for the massless scalar field in several flat multiply connected spaces, expressing
the stress expectation values as the coincidence limit of a bilinear operator acting on the
Feynman propagator for the manifold, motivating the introduction of a robust mathematical approach \cite{banach1}.
Finally, in \cite{petry} exotic spinor fields provides pure geometrical explanation of the charge dependence on the quantized flux and also the Joseph current in superconductivity.

To summarize, one of the main outstanding exotic spinor fields features is that they must be taken into account and employed
in a variety of problems, wherein standard spinor fields cannot.
For instance, when the vacuum polarization tensor of spinor electrodynamics is calculated \cite{ford}, it was found that the two types of spinor fields --- standard and exotic --- generate different vacuum polarization effects, which are physically inequivalent.
When the effect of the vacuum polarization upon photon propagation is considered, it is shown that standard spinor fields give rise to non causal photon propagation, whereas exotic spinor fields do not. Even when the vacuum energy for a free spinor field is calculated, it is found that the exotic configuration gives rise to a vacuum state of lower energy than the standard one.
These prominent features make exotic spinor fields as a broad audience candidate for concrete physical problems.

We shall address to  the question about such inequivalent spin structures and their consequences to the coupled system of Dirac equations satisfied by the four types of dark spinor fields.
The physical assumption that --- under the exotic spin structure --- the exotic dark spinor fields satisfy the Klein-Gordon propagator brings up some constraint on the metric spacetime structure, as well as in the exotic topology, both arbitrary \emph{a priori}.
The characterization of dark (ELKO) spinor fields, and its inherent analysis is obtained through the natural introduction --- topologically impelled --- of an exotic term in the Dirac operator that, contrary to the case of the Dirac spinor field, cannot be absorbed in any external electromagnetic vector field.
For Dirac fields, such term can be concealed and encoded as a shift of the  electromagnetic potential\footnote{Representing an element of the cohomology group $H^1(M, \mathbb{Z}_2)$.}.
Therefore, besides addressing feasible aspect to the dark matter problem, dark spinor fields are also useful to probe non-trivial topological properties in spacetime.

The manuscript is organized as follows.
After briefly presenting some algebraic preliminaries in Section 2 regarding inequivalent spin structures, in Section 3, the ELKO properties are introduced, together with the bilinear covariants that completely characterize a spinor field through the Fierz identities.
In Section 4 the exotic structure is introduced and the corresponding implications on the behaviour of ELKO are depicted.
Dark spinor dynamics not only constrains the possibilities for the exotic topology but also induces constraints in the spacetime geometry through the exotic topology coming from the dynamics of dark spinor fields.
We prove that it brings up some subtle consequences on the spacetime geometry.
A brief summary of important useful results throughout the manuscript is
described in the Appendices A and B, and we draw our conclusions in
Section 5.

\section{Preliminaries: Exotic Spin Structures}
\label{w2}

In this Section we review some results concerning general inequivalent spin structures in spinor bundles. For more details see the Appendix.

One denotes by $(M, g,\nabla,\tau_g,\uparrow)$ the
spacetime structure \cite{moro,rod}: $M$ denotes a 4-dimensional manifold --- which we shall assume as a compact, paracompact, pseudo-Riemannian
manifold which is both space and time orientable and which admits spinor fields --- $g$ is the metric, $\nabla$ denotes the connection associated to $ g$, $\tau_g$ defines a
spacetime orientation and $\uparrow$ refers a time orientation. As usual $T^{\ast}M$ [$TM$] denotes the
cotangent [tangent] bundle over $M$, 
$F(M)$ denotes the principal bundle of frames, and  $P_{\mathrm{SO}_{1,3}^{e}}(M)$ denotes the orthonormal
coframe bundle. Such bundles do exist on spin manifolds.
Sections of $P_{\mathrm{SO}_{1,3}^{e}%
}(M\mathbf{)}$ are orthonormal coframes, and sections of $P_{\mathrm{Spin}%
_{1,3}^{e}}(M\mathbf{)}$ are also orthonormal coframes such that
although two coframes differing by a $2\pi$ rotation are distinct, two
coframes differing by a $4\pi$ rotation are identified.

A spin structure on
$M$ consists of a principal fiber bundle
$\mathbf{\pi}_{s}:P_{\mathrm{Spin}_{1,3}^{e}}(M)\rightarrow M$,
with group $\mathrm{Spin}_{1,3}^{e}$, and the fundamental map --- indeed a two-fold covering
\begin{equation}
s: P_{\mathrm{Spin}_{1,3}^{e}}(M)\rightarrow P_{\mathrm{SO}_{1,3}^{e}%
}(M),\nonumber
\end{equation}
satisfying the following conditions:\\
(i) $\mathbf{\pi}(s(p))=\mathbf{\pi}_{s}(p),\ \forall p\in P_{\mathrm{Spin}%
_{1,3}^{e}}(M);$ $\pi$ is the projection map of $P_{\mathrm{SO}
_{1,3}^{e}}(M)$ on $M$.\\
\noi
(ii) $s(p \phi)=s(p)\mathrm{Ad}_{\phi},\;\forall p\in
P_{\mathrm{Spin}_{1,3}^{e}}(M)$ and
$\mathrm{Ad}:\mathrm{Spin}_{1,3}^{e}\rightarrow\mathrm{Aut}(\cl_{1,3}),$
$\mathrm{Ad}_{\phi}:\Xi\mapsto
\phi\Xi\phi^{-1}\in\cl_{1,3}$ \cite{moro}.
Namely, the following diagram
\begin{diagram}
P_{\mathrm{Spin}_{1,3}^{e}}(M)&&\rTo^s&&P_{\mathrm{SO}_{1,3}^{e}}(M)\\
&\rdTo_{\pi_s} &&\ldTo_{\pi}& \\
&&M&&\\
\end{diagram}
commutes. The conditions for existence of a spin structure in a general manifold are discussed in \cite{42,43}.

It is well
known that a spin structure ($P_{\mathrm{Spin}_{1,3}^{e}}(M),s$) exists if and only if the second Stiefel-Whitney class associated to $M$ satisfies
certain properties.
If $H^1(M,\mathbb{Z}_2)$ is not trivial, the spin structure is not uniquely defined\footnote{Up to trivial
bundle isomorphisms.}, and all the other inequivalent spin structures can be provided from ($P_{\mathrm{Spin}_{1,3}^{e}}(M),s$).

Two spin structures  $P$:=($P_{\mathrm{Spin}_{1,3}^{e}}(M),s)$ and $\mathring{P}:=((\mathring{P}_{\mathrm{Spin}_{1,3}^{e}}(M),\mr{s})$ are said to be equivalent
if there exists a Spin$^e_{1,3}$-equivariant map $\zeta: P\rightarrow \mathring{P}$ compatible with $s$ and $\mathring{s}$:
\begin{diagram}
P&&\rTo^\zeta&&\mathring{P}\\
&\rdTo_{s} &&\ldTo_{\mathring{s}}& \\
&&P_{\mathrm{SO}_{1,3}^{e}}(M)&&\\
\end{diagram}

Now we briefly review some few definitions necessary to introduce exotic spinor fields.
Spinor fields are sections of vector bundles associated with  the
principal bundle of spinor coframes.
A complex spinor bundle for $M$ is a vector bundle
$S_{c}(M)=P_{\mathrm{Spin}_{1,3}^{e}}(M)\times_{\mu_{c}}\mathcal{M}_{c},%
$ 
where $\mathcal{M}_{c}$ is a complex left module for $\mathbb{C}\otimes
\cl_{1,3}\simeq {\rm M}(4,\mathbb{C})$, and where
$\mu_{c}$ is a representation of $\mathrm{Spin}_{1,3}^{e}$ in
$\mathrm{End}(\mathcal{M}_{c})$ given by left multiplication by elements of
$\mathrm{Spin}_{1,3}^{e}$. When $\mathcal{M}_{c}=\mathbb{C}^{4}$ and $\mu_{c}$ the $D^{(1/2,0)}%
\oplus D^{(0,1/2)}$ representation of $\mathrm{Spin}_{1,3}^{e}\simeq
{\rm SL}(2,\mathbb{C})$ in $\mathrm{End}(\mathbb{C}^{4})$, we immediately recognize
the usual definition of the covariant spinor bundle of $M$ as given, e.g., in
\cite{42} and \cite{43}.

 Classical spinor
fields\footnote{Quantum spinor fields are operator valued
distributions, as well known. It is not necessary to introduce
quantum fields in order to know the algebraic classification of
ELKO spinor fields.}
carrying a $D^{(1/2,0)}\oplus D^{(0,1/2)}$,
or $D^{(1/2,0)}$, or $D^{(0,1/2)}$ representation of
SL$(2,\mathbb{C})$ are sections of the vector bundle
$
{P}_{\mathrm{Spin}_{1,3}^{e}}(M)\times_{\rho}\mathbb{C}^{4},
$
where $\rho$ stands for the $D^{(1/2,0)}\oplus D^{(0,1/2)}$ (or
$D^{(1/2,0)}$,
or $D^{(0,1/2)}$) representation of SL$(2,\mathbb{C})$ in $\mathbb{C}^{4}$. {Other important spinor fields,
like Weyl spinor fields are obtained by imposing some constraints
on the sections of
${P}_{\mathrm{Spin}_{1,3}^{e}}(M)\times_{\rho}\mathbb{C}^{4}$, see, e.g., \cite{lou1,lou2} for details.}

Two spin structures ($P_{\mathrm{Spin}_{1,3}^{e}}(M),s$) and ($\mr{P}_{\mathrm{Spin}_{1,3}^{e}}(M),\mr{s}$) are respectively described by the maps ${h_{jk}}$ and ${\mr{h}_{jk}}$
from $U_i\cap U_j$ to Spin$_{1,3}^e$, both satisfying Eq.(\ref{jgij}), and also the property $\varsigma\circ {h_{jk}} = a_{jk} = \varsigma\circ{\mr{h}_{jk}}$.
The following diagram illustrates such relations, summarizing what should be emphasized heretofore:
\begin{diagram}
U_i\cap U_j \subset M& &\rTo^{h_{ij}}  & & {\rm Spin}_{1,3}^e & & \\
& \rdTo_{a_{ij}} & &\ldTo^\varsigma &  & \rdInto & \\
\dTo^{\mr{h}_{ij}} & & {\rm SO}^e_{1,3} &  & \dTo & & P_{\mathrm{Spin}_{1,3}^{e}}(M) \\
&\ruTo^\varsigma & \dInto & & & \ruTo_\zeta& \\
\mr{{\rm Spin}}_{1,3}^e &\hLine  & \VonH&\rTo  & \mr{P}_{\mathrm{Spin}_{1,3}^{e}}(M) & & \dTo_{{s}} \\
& & & & & \rdTo_{\mr{s}} & \\\
& & P_{\mathrm{SO}_{1,3}^{e}}(M) & & \rTo^{id} & & P_{\mathrm{SO}_{1,3}^{e}}(M) \\
\end{diagram}
\noi Here another identical copy  of ${\rm Spin}_{1,3}^e$ is denoted by   $\mr{{\rm Spin}}_{1,3}^e$, in order to become clearer the analysis about the inequivalent spin structures.

Now one defines a map $c_{jk}$ by the relation $h_{ij}(x) = \mr{h}_{ij}(x)c_{ij}$ such that $c_{ij}:U_i\cap U_j\rightarrow$ ker $\varsigma$ = $\mathbb{Z}_2\hookrightarrow$ Spin$^e_{1,3}$, satisfying $c_{ij}\circ c_{jk} = c_{ik}$. Such a map $c_{ij}$ defines an 1-dimensional real bundle denoted \cite{thomas}.
Given the irreducible representation $\rho:\cl_{1,3}\rightarrow$ M($4,\mathbb{C}$) in $
{P}_{\mathrm{Spin}_{1,3}^{e}}(M)\times_{\rho}\mathbb{C}^{4},
$ as the map $c_{ij}(x)$ is an element of $\mathbb{Z}_2$, it follows that $\rho(c_{ij}(x)) = \pm 1$, since $\rho$ is faithful. When $\rho$ is restricted to Spin$^e_{1,3}$, it is called Dirac representation.

We assume as in \cite{avis0,isham0,isham1,petry} that there is a set of functions $\xi_i: U_i\rightarrow\CC$
  such that $\|\xi_i(x)\| = 1$, namely $\xi_i(x)\in $ U(1), and
  \beq\label{xi1}
  \xi_i(x) (\xi_j(x))^{-1} = \rho(c_{ij}(x)) = \pm 1.\eeq\noi In the case where the second integral cohomology $H^2(M,\mathbb{Z}_2)$ has no 2-torsion, such functions always do exist \cite{isham0,isham1,petry,thomas}, and ${\xi}^2_i(x) = {\xi}^2_j(x)$, $x\in U_i\cap U_j$. Consequently the local functions $\xi_i$ define a unique unimodular function $\xi:M\rightarrow\CC$ such that for all $x\in U_i$ it follows that $\xi(x) = {\xi}_i^2(x)$.

Given now an arbitrary spinor field $\psi \in$ sec $
{P}_{\mathrm{Spin}_{1,3}^{e}}(M)\times_{\rho}\mathbb{C}^{4}$, to each element of $H^1(M,\mathbb{Z}_2)$, associate a covariant derivative  $\nabla$.
This construction provides indeed a one-to-one correspondence between elements of $H^1(M,\mathbb{Z}_2)$ and inequivalent spin structures.

A local component $\psi_i: U_i\rightarrow \CC^4$ of a spinor field in $
{P}_{\mathrm{Spin}_{1,3}^{e}}(M)\times_{\rho}\mathbb{C}^{4}$ is the unique function such that
$\rho(\ell_i,\psi_i(x)) = \psi(x)$, given local sections $\ell_i :U_i\rightarrow$ ($P_{\mathrm{Spin}_{1,3}^{e}}(M),s$),
we have the transition law
\[\psi_i(x) = \rho(h_{ij}(x))\psi_j(x),\qquad\text{where $x\in U_i\cap U_j$}.\]

A system of local sections $\mr\ell_i: U_i\rightarrow$ $\mr{P}_{\mathrm{Spin}_{1,3}^{e}}(M)$
can be constructed from the standard ones $\ell_i$ in such a way that $s\circ \ell_i$ = $\mr{s}\circ \mr{\ell}_i$,
as presented in the following diagram:
\begin{diagram}
&&U_i&&\\
&\ldTo^{\mr\ell_i}&\dTo^{m_i}&\rdTo^{\ell_i}&\\
\mr{P}_{\mathrm{Spin}_{1,3}^{e}}(M)&\rTo^{\mr{s}}&{P}_{\mathrm{SO}_{1,3}^{e}}(M)&\lTo^{s}&{P}_{\mathrm{Spin}_{1,3}^{e}}(M)\\
\end{diagram}
It enables the (exotic) local spinor field components to present the respective transition property
\beq\mr\psi_j(x) = \rho(\mr{h}_{ij}) = \rho(h_{ij}(x))\rho(c_{ij}(x))\mr\psi_i(x),\qquad\text{where $x\in U_i\cap U_j.$}\label{ccv}\eeq\noi

From Eq.(\ref{xi1}) it follows that $\rho(\xi_i) = \rho(c_{ij}(x))\rho(\xi_j)$ and if we compare it with  Eq.(\ref{ccv}), it is clear that $\rho(\xi_i)\mr\psi_i$  transforms as the local component $\psi_i$ of
${P}_{\mathrm{Spin}_{1,3}^{e}}(M)\times_{\rho}\mathbb{C}^{4}$, which subsequently induces a bundle map
\beq
f:\mr{P}_{\mathrm{Spin}_{1,3}^{e}}(M)\times_{\rho}\mathbb{C}^{4}&\rightarrow&{{P}}_{\mathrm{Spin}_{1,3}^{e}}(M)\times_{\rho}\mathbb{C}^{4}\nonumber\\
\mr\psi_i &\mapsto& f(\mr\psi_i):= \rho(\xi_i)\mr\psi_i = \psi_i
\eeq such that
   \beq\mr\nabla_X f(\mr\psi) = f({\nabla}_X\mr\psi) +\frac{1}{2}(X\lrcorner(\xi^{-1} d\xi))f(\mr\psi)\label{xi2}\noi\eeq
   holds for all sections $\psi\in P_{\mathrm{Spin}_{1,3}^{e}}(M)\times_{\rho}\mathbb{C}^{4}$ and all vector fields $X$.
Details on how to derive Eq.(\ref{xi2}) are comprehensively given in, e.g., \cite{choque,thomas,petry,isham0,isham1,avis0}.

 \section{Dark (ELKO) Spinor Fields}

This section is devoted to a brief review of the bilinear covariants through the programme introduced in
\cite{lou1,lou2,meu}. The spinor fields classification is provided by a brief review of \cite{osmano,meu,hof,nossoijg,ijmpdze}.

 Given a spinor field
$\psi$ $\in
\sec{P}_{\mathrm{Spin}_{1,3}^{e}}(M)\times_{\rho}\mathbb{C}^{4},$
the bilinear covariants are the following sections of
${\displaystyle\Lambda
}(TM)={\displaystyle\oplus_{r=0}^{4}}$ ${\displaystyle\Lambda^{r}%
}(TM)\hookrightarrow C\mathcal{\ell}(M,g)$ \cite{holl,moro}:
\begin{align}
\sigma &  =\psi^{\dagger}\gamma_{0}\psi,\quad\mathbf{J}=J_{\mu}\mathbf{e}%
^{\mu}=\psi^{\dagger}\gamma_{0}\gamma_{\mu}\psi\mathbf{e}^{\mu},\quad
\mathbf{S}=S_{\mu\nu}\mathbf{e}^{\mu\nu}=\frac{1}{2}\psi^{\dagger}\gamma
_{0}i\gamma_{\mu\nu}\psi\mathbf{e}^{\mu}\wedge\mathbf{e}^{\nu},\nonumber\\
\mathbf{K} &  =\psi^{\dagger}\gamma_{0}i\gamma_{0123}\gamma_{\mu}%
\psi\mathbf{e}^{\mu},\quad\omega=-\psi^{\dagger}\gamma_{0}\gamma_{0123}%
\psi,\label{fierz}%
\end{align}
with $\sigma,\omega\in\sec%
{\displaystyle\Lambda^{0}}
(TM)$, $\mathbf{J,K}\in\sec%
{\displaystyle\Lambda^{1}}
(TM)$ and $\mathbf{S}\in\sec%
{\displaystyle\Lambda^{2}}
(TM)$. In the formul\ae\,
appearing in Eq.(\ref{fierz}) the set $\{\gamma_{\mu}\}$ refers to
the Dirac matrices in
chiral representation (see Eq.(\ref{dirac matrices})). Also, $
\{1,\mathbf{e}^{\mu},\mathbf{e}^{\mu}\mathbf{e}^{\nu},\mathbf{e}^{\mu
}\mathbf{e}^{\nu}\mathbf{e}^{\rho},\mathbf{e}^{0}\mathbf{e}^{1}\mathbf{e}%
^{2}\mathbf{e}^{3}\},
$ where $\mu,\nu,\rho=0,1,2,3$, and $\mu<\nu<\rho$ is a
basis for $C\mathcal{\ell}(M,g)$, and
$
\{\mathbf{1}_{4},\gamma_{\mu},\gamma_{\mu}\gamma_{\nu},\gamma_{\mu}\gamma
_{\nu}\gamma_{\rho},\gamma_{0}\gamma_{1}\gamma_{2}\gamma_{3}\}
$
is a basis for M$(4,\CC)$. In addition, these bases satisfy
the respective Clifford algebra relations \cite{lou1} $
\gamma_{\mu}\gamma_{\nu}+\gamma_{\nu}\gamma_{\mu}   =2g_{\mu\nu}%
\mathbf{1}_{4}$ and $\mathbf{e}^{\mu}\mathbf{e}^{\nu}+\mathbf{e}^{\nu}\mathbf{e}^{\mu}
=2g^{\mu\nu}$,
where $\mathbf{1}_{4}$ is the identity
matrix. When there
is no opportunity for confusion we shall omit the $\mathbf{1}_{4}$
identity matrix in our formul\ae . For the orthonormal covector fields $\mathbf{e}^{\mu}$ and
$\mathbf{e}^{\nu}$, $\mu\neq\nu,$ their Clifford product
$\mathbf{e}^{\mu }\mathbf{e}^{\nu}$ is equal to the exterior
product of those vectors, i.e.,
$\mathbf{e}^{\mu}\mathbf{e}^{\nu}=\mathbf{e}^{\mu}\wedge\mathbf{e}^{\nu
}=\mathbf{e}^{\mu\nu}$.\ \ Also, for $\mu\neq\nu\neq\rho,$ $\mathbf{e}^{\mu}%
{}^{\nu}{}^{\rho}=$\
$\mathbf{e}^{\mu}\mathbf{e}^{\nu}\mathbf{e}^{\rho}$, etc. More
details on our notations, if needed, can be found in
\cite{moro,rod}.

 In Minkowski spacetime, the case of the electron is described by Dirac spinor
fields (classes 1, 2 and 3 below), $\mathbf{J}$ is a
future-oriented timelike current vector which gives the current of
probability, and $\mathbf{J}^{2} = J_{\mu}J^{\mu}>0.$
Furthermore, for the case of Dirac spinor fields, the bivector $\mathbf{S}$ is associated with the
distribution of intrinsic angular momentum, and the spacelike
vector $\mathbf{K}$ is associated with the direction of the
electron spin. For a detailed discussion concerning such entities,
their relationships and physical interpretation, and
generalizations, see, e.g., \cite{cra,lou1,lou2,holl}.

The bilinear covariants satisfy the Fierz identities\footnote{
Given the spacetime metric  $g$, it is possible to extend $g$ to
the exterior bundle $\Lambda(TM)$. Given $\psi=u^1\w\cdots\w u^k$ and
$\phi=v^1\w\cdots\w v^l$, for $u^i, v^j\in$ sec $TM$, one
defines $g(\psi,\phi)
 = \det(g(u^i,v^j))$ if $k=l$ and $g(\psi,\phi)=0$ if $k\neq l$. 
 Given $\psi,\phi,\xi\in\Lambda(TM)$, the  {\ left contraction} is
defined implicitly by
$g(\psi\lrcorner\phi,\xi)=g(\phi,\tilde\psi\w\xi)$.}
\cite{cra,lou1,lou2,holl}
\begin{equation}
\mathbf{J}^{2}=\omega^{2}+\sigma^{2},\quad\mathbf{K}^{2}=-\mathbf{J}^{2}%
,\quad\mathbf{J}\llcorner\mathbf{K}=0,\quad\mathbf{J}\wedge\mathbf{K}%
=-(\omega+\sigma\gamma_{0123})\mathbf{S}.  \nonumber
\end{equation}

A spinor field such that {not both} $\omega$ and $\sigma$ are
null is said to be regular. When $\omega=0=\sigma$, a spinor field
is said to be singular, and in this case the Fierz identities
are in general replaced by the more general conditions \cite{cra}
\begin{align}
Z^{2}  &  =4\sigma Z,\quad Z\gamma_{\mu}Z=4J_{\mu}Z,\quad
Zi\gamma_{\mu\nu
}Z=4S_{\mu\nu}Z,\quad Zi\gamma_{0123}\gamma_{\mu}Z    =4K_{\mu}Z,\quad
Z\gamma_{0123}Z=-4\omega Z, \nonumber
\end{align} where $
Z=\sigma+\mathbf{J}+i\mathbf{S}+i\mathbf{K}\gamma_{0123}+\omega\gamma_{0123}$.

Lounesto spinor field classification is given by the following
spinor field classes \cite{lou1,lou2}, where in the first three
classes it is implicit that $\mathbf{J}$\textbf{,
}$\mathbf{K}$\textbf{, }$\mathbf{S}$ $\neq0$:
\begin{itemize}
\item[1)] $\sigma\neq0,\;\;\; \omega\neq0$.
\item[2)] $\sigma\neq0,\;\;\; \omega= 0$.
\item[3)] $\sigma= 0, \;\;\;\omega\neq0$.
\item[4)] $\sigma= 0 = \omega, \;\;\;\mathbf{K}\neq0,\;\;\;
\mathbf{S}\neq0$.
\item[5)] $\sigma= 0 = \omega, \;\;\;\mathbf{K}= 0,
\;\;\;\mathbf{S}\neq0$.\label{elko1}
\item[6)] $\sigma= 0 = \omega, \;\;\; \mathbf{K}\neq0, \;\;\;
\mathbf{S} = 0$.
\end{itemize}
\noindent The current density $\mathbf{J}$ is always non-zero.
Type 1, 2 and 3 spinor fields are denominated {Dirac spinor
fields} for spin-1/2 particles and type 4, 5, and 6 are
respectively called {flag-dipole}, {flagpole} and
{Weyl spinor fields}. Majorana spinor fields are a
particular case of a type 5 spinor field. It is worthwhile to
point out a peculiar feature of types 4, 5 and 6 spinor fields:
although $\mathbf{J}$ is always non-zero,
$\mathbf{J}^{2}=-\mathbf{K}^{2}=0$. It shall be seen below that the
bilinear covariants related to an ELKO spinor field, satisfy
$\sigma=0=\omega,\;\;\mathbf{K}=0,\;\;\mathbf{S}\neq0$ and
$\mathbf{J}^{2}=0$.

Since Lounesto proved that there are {no} other classes based on
distinctions among bilinear covariants, ELKO spinor fields
must belong to one of the disjoint six classes. In \cite{meu} it is shown
that ELKO spinor fields are indeed in class 5 above.

Some properties of dark (ELKO) spinor fields\footnote{Hereon throughout the text the term dark spinor field and ELKO are alternatively used having the same meaning, since ELKO is a candidate to describe dark matter, as comprehensively proposed, derived, and investigated in, e.g., \cite{
allu,alu2,boe1,boe111,bo121,boe12,boe13,fabbri1,ahlu4,Ahluwalia:xa,hof,marcao,gau,osmano,meu,alu3,nossoijg}. We choose the acronym ELKO to denote field theoretical and more formal properties of such a spinor field, whereas the naming dark spinor fields shall be used hereon alternatively to ELKO, in order to present and investigate the potentially cosmological applications as well as its usefulness as an attempt to the dark matter problem.}, as introduced  in \cite{allu,alu2,alu3} can be now briefly reviewed. An ELKO  $\Psi$
corresponding to a plane wave with momentum $p=(p^{0},\mathbf{p)}$
can be written, without loss of generality, as $\Psi(p)=\lambda({\bf p})
e^{\pm i{p\cdot x}}$ where
\begin{equation}
\lambda({\bf p})=\binom{i\Theta\phi^{\ast}(\mathbf{p})}{\phi(\mathbf{p})},
\label{1lee}%
\end{equation}
\noindent and given the rotation generators
${\mathfrak{J}}$, the Wigner's spin-1/2 time reversal operator
$\Theta$ satisfies $\Theta
\mathfrak{J}\Theta^{-1}=-\mathfrak{J}^{\ast}$.
Hereon, as in \cite{allu}, the Weyl representation of
$\gamma^{\mu}$ is used
\begin{equation}
\gamma_{0}=\gamma^{0}=%
\begin{pmatrix}
\OO & \II\\
\II & \OO
\end{pmatrix}
,\quad-\gamma_{k}=\gamma^{k}=%
\begin{pmatrix}
\OO & -\sigma_{k}\\
\sigma_{k} & \OO
\end{pmatrix},\quad \gamma^{5}=-i\gamma^{0}\gamma^{1}\gamma^{2}\gamma^{3}=-i\gamma^{0123}=
\begin{pmatrix}
\II&\OO \\
\OO & -\II
\end{pmatrix}, \label{dirac matrices}%
\end{equation}
\noindent where
$
\II= {\scriptsize\begin{pmatrix}
1 & 0\\
0 & 1
\end{pmatrix}}
,\; \OO={\scriptsize\begin{pmatrix}
0 & 0\\
0 & 0
\end{pmatrix}}
,\;
\sigma_{1}=%
{\scriptsize\begin{pmatrix}
0 & 1\\
1 & 0
\end{pmatrix}}
,\;\sigma_{2}=%
{\scriptsize\begin{pmatrix}
0 & -i\\
i & 0
\end{pmatrix}}
,\;\sigma_{3}={\scriptsize
\begin{pmatrix}
1 & 0\\
0 & -1
\end{pmatrix}}$. The $\sigma_i$ are the Pauli matrices.

  ELKO spinor fields are eigenspinors of the
charge conjugation operator $C$
\begin{equation}
C\lambda({\bf{p}})=\pm \lambda({\bf
p}),\qquad{\rm  for}\quad C=%
\begin{pmatrix}
\OO & i\Theta \\
-i\Theta & \OO \nonumber
\end{pmatrix}
K .\label{conj}\end{equation} The operator $K$ $\mathbb{C}$-conjugates 2-component
spinor fields appearing on the right. The plus sign stands for
self-conjugate spinor fields, $\lambda^{S}({\bf p})$, while the minus
yields anti self-conjugate spinor fields,  $\lambda^{A}({\bf p})$.
Explicitly, the complete form of ELKO can be found by solving
the equation of helicity $(\sigma\cdot\widehat{\bf{p}})\phi^{\pm}=\pm \phi^{\pm}$ in the rest frame and
subsequently make a boost,  to recover the result for any ${\bf p}$
\cite{allu}. Here $\widehat{\bf{p}}:={\bf p}/\|{\bf p}\|$. The boosted four spinor fields are
\begin{equation}
\lambda^{S/A}_{\{\mp,\pm \}}({\bf
p})=\sqrt{\frac{E+m}{2m}}\Bigg(1\mp
\frac{p}{E+m}\Bigg)\lambda^{S/A}_{\{\mp,\pm \}}({\bf{0}}),\qquad{\rm  where}\qquad
\lambda^{S/A}_{\{\mp,\pm \}}(\bf{0})=%
\begin{pmatrix}
\pm i \Theta[\phi^{\pm}(\bf{0})]^{*} \\
\phi^{\pm}(\bf{0})\label{five}
\end{pmatrix}.
\end{equation}
One should notice that, since
$\Theta[\phi^{\pm}(\bf{0})]^{*}$ and $\phi^{\pm}(\bf{0})$ have
opposite helicities, ELKO cannot be an eigenspinor field of the helicity
operator. The ELKO dual is given by \cite{allu}
\begin{equation}
\overset{\;\;\;\;\;\;\;\neg{S/A}}{\lambda}{}_{\{\mp,\pm \}}({\bf p})=\pm i \Big[
\lambda^{S/A}_{\{\pm,\mp \}}({\bf p})\Big]^{\dag}\gamma^{0}
\label{dual}.\end{equation}

Now let one denotes the eigenspinors of the Dirac operator for particles and antiparticles respectively by $u_{\pm}(\p)$ and $v_\pm(\p)$.  The subindex $\pm$ regards the eigenvalues of the helicity operator $(\sigma\cdot\widehat{\bf{p}})$. The parity operator acts as
\[ P u_\pm(\p) = +\, u_\pm(\p),\qquad\qquad P v_\pm(\p) = -\, v_\pm(\p),\] \noi which implies that $ P^2= \,\mathbb{I}$ in this case.
The action of $C$ on these spinors is given by
\beq
C(u_{\pm 1/2}(\p)) = \mp v_{\mp}(\p),\qquad\qquad
 C(v_{\pm 1/2}(\p)) = \pm u_{\mp 1/2}(\p).
\eeq
which implies that $\{C,P\}=0$.

On the another hand
the parity operator $P$ acts on ELKO by
\beq
P\lambda^\mathrm{S}_{\mp,\pm} (\p)= \pm\, {i}\,
\lambda^\mathrm{A}_{\pm,\mp}(\p)\,,\qquad
P\lambda^\mathrm{A}_{\mp,\pm} (\p)= \mp \,{i}
\,\lambda^\mathrm{S}_{\pm,\mp}(\p),
\eeq\noi and it follows that $[C,P]=0$.

Denoting \cite{allu} for  Dirac spinor fields
\[u_+(\p) =  d_1,\quad
u_-(\p) = d_2,\quad
v_+(\p)= d_3\quad v_-(\p)= d_4,\] and for the ELKO
\[\lambda^\mathrm{S}_{\ua}(\p)= e_1, \quad \lambda^\mathrm{S}_{\da}(\p) = e_2,\quad
\lambda^\mathrm{A}_{\ua}(\p) = e_3, \quad
 \lambda^\mathrm{A}_{\da}(\p) = e_4,\] it is possible to write ELKO as \cite{Ahluwalia:xa}
\beq
e_i= \sum_{j=1}^4 \Omega_{ij} d_j,\quad i=1,2,3,4,\label{elko11}\;\;\text{where}\;\;
\Omega_{ij}=
 \begin{cases}
+\left({1}/{2 m}\right) \overline{d}_j\, e_i \mathbb{I}, & \quad{\rm for}
\,\, j =1,2\,,\cr
-\left({1}/{2 m}\right) \overline{d}_j \,e_i \mathbb{I}, & \quad{\rm for}
\,\,j =3,4\,.\cr
\end{cases}
\eeq
In matrix form,  $\Omega$ reads
\beq
\Omega =
\frac{1}{2}\left(
\begin{array}{cccc}
\mathbb{I} & - i \mathbb{I} & - \mathbb{I} & - i\mathbb{I} \\
i\mathbb{I} & \mathbb{I} & i\mathbb{I} & - \mathbb{I} \\
\mathbb{I} & i\mathbb{I} & - \mathbb{I} & i\mathbb{I} \\
- i\mathbb{I} & \mathbb{I} & - i\mathbb{I} & - \mathbb{I}
\end{array}
\right) = \frac{1}{2}\left(
\begin{array}{cc}\matB & -\matB^\ast \\
\matB^\ast & -\matB
\end{array}\right)
\otimes \mathbb{I},\label{omega}
\eeq
\noi where $\matB:=(\mathbb{I}+\sigma_2)$,
Such results show that ELKO can be expressed somehow as a linear combination
of the  Dirac particle and antiparticle
spinor fields. It reinforces the Lounesto theorems, showing that classes of spinor fields under Lounesto spinor fields classification
are not preserved by sum (for details see \cite{hopf,lou1,lou2}).
In order to obtain the ELKO evolution, a prescription where the momentum is written in terms of the covariant derivative as $p_\mu \mapsto i \nabla_\mu$ is regarded. 
As one shall see in the following Section, such a prescription is convenient when one considers the coordinate representation $\lambda^\mathrm{S/A}(x)=\lambda^\mathrm{S/A}(\p) \;\exp\left(\varepsilon^\mathrm{S/A}\;\,i p_\mu x^\mu\right)$.
However, that is not the only way to prescribe the ELKO evolution.
The momentum can also be replaced with the derivatives times the $\gamma^{5}$ matrix as performed, for instance,
in the investigation of ELKO auto interactions when one considers the ELKO field interacting with its own spin density via contorsional auto interactions \cite{Fabbri1008}.

\section{ELKO Dynamics in the Exotic Spin Structure}

In spacetimes with non-trivial topology it is well known that there is an additional degree of
freedom for fermionic particles \cite{Asselmeyer:1995jp}. Albeit in the classical level it might be naively suggested that
 exotic spinor fields describe different particles,  the breakthrough idea proposed is that, in the quantum framework, a new partition
function which is the sum over all possibilities must be taken into account. See \cite{isham0,Asselmeyer:1995jp} and references therein for more details.

In this Section it is thoroughly shown that dark spinor fields are a natural probe of the non-trivial topology and also provide, from their inherent dynamics, constraints either in the spacetime metric structure or in its topology, or in both.

Essentially,  exotic  spinor fields are parallel transported  like standard spinor fields, but an outstanding property distinguishes both kinds of spinor fields: the covariant derivative acting on these exotic spinor fields changes by an additional one-form field that is manifestation of the non-trivial topology, as it was shown in Section 2. The exotic structure endows the Dirac operator with an additional term $\xi^{-1}(x) d\xi(x), x\in M$,
where $d:\sec\Lambda^0(TM)\rightarrow\sec\Lambda^1(TM)$ denotes the exterior derivative operator.
The term $\frac{1}{2\pi i}\xi^{-1}(x) d\xi(x)$ is real and closed, but not exact, and defines an integer cohomology class in the  $\check{\rm C}$ech sense \cite{isham0,isham1,petry,avis0}. Using the relation between $\check{\rm C}$ech and de Rham cohomologies, the integral of $\frac{1}{2\pi i}\xi^{-1}(x) d\xi(x)$ around any closed curve is an integer.
In the context of the exotic Dirac equation, the electromagnetic vector potential $A$ term is affected by the transformation $A \mapsto A + \frac{1}{2\pi i} \xi^{-1} d\xi$, which exactly corresponds to
the addition of another electromagnetic potential, when Dirac spinor fields are taken into account. In such case the exotic term  may be then absorbed in an external electromagnetic potential, representing an element of $H^1(M, \mathbb{Z}_2)$ \cite{isham1,petry,choque,grimm}. Namely, in this case the interaction is encoded as a shift in the vector potential.

 The importance to analyze dark spinor fields in this context is that
this possibility is not present if ELKO spinor fields are employed, as they cannot carry
electromagnetic charge and the full topological treatment is
appropriate in this case \cite{choque}.

In addition to the ELKO spinor fields $\lambda(x)$ --- that was indeed defined
as sections in the bundle
$
{P}_{\mathrm{Spin}_{1,3}^{e}}(M)\times_{\rho}\mathbb{C}^{4},
$ in Section 2 --- one can get a second
type of ELKO $\mathring\lambda(x)$, which can be described
by sections in the inequivalent spin structure-induced
spinor bundle
\beq\label{pinp1}
\mr{P}_{\mathrm{Spin}_{1,3}^{e}}(M)\times_{\rho}\mathbb{C}^{4},
\eeq\noi with a variation of the covariant derivative, given by \cite{petry}
\bege
\mr{\nabla}_X{\mathring\lambda(x)} = \nabla_X{\mathring\lambda(x)}  -\frac{1}{2}\left[X\lrcorner\left(\xi^{-1}(x)d\xi(x)\right)\right]\mathring\lambda(x)\label{covvv},
\enge where $X$ denotes a vector field in $M$.

In general, the exotic term in Eq.(\ref{covvv}) is assumed --- in order to be an integer of a ${\check{\rm C}}$ech cohomology class ---
to be indeed $\frac{1}{2\pi i}\left(\xi^{-1}(x)d\xi(x)\right)$ \cite{petry,isham0,avis0,isham1}. We henceforth redefine $\xi(x)\mapsto\xi(x)/\sqrt{2\pi}$
in such a way that the exotic Dirac operator can be written as (see Eq.(\ref{xi2}))
 \beq
 i\gamma^\mu\mr\nabla_\mu = i\gamma^\mu\nabla_\mu + \xi^{-1}(x)d\xi(x).
 \eeq\noi 
The exotic Dirac equation is given by
\bege\label{exodirac}(i\gamma^\mu \nabla_\mu + (\xi^{-1}(x)\,d\xi(x)) - m\mathbb{I}) \psi(x) = 0,\qquad\text{where $\psi$ denotes a Dirac spinor field.}  \nonumber\enge \noi
The exotic Dirac spinor fields
are annihilated by $\left(i\gamma^\mu\nabla_\mu + (\xi^{-1}(x)\,d\xi(x)) \pm m \mathbb{I}\right)$
\beq
\begin{cases}
\mbox{For particles:}
\qquad\left(i\gamma^\mu\nabla_\mu + (\xi^{-1}(x)\,d\xi(x)) - m \mathbb{I}\right) u(x)=0\,,\qquad\cr
\mbox{For antiparticles:}\qquad
\left(i\gamma^\mu\nabla_\mu + (\xi^{-1}(x)\,d\xi(x)) + m \mathbb{I}\right) v(x)=0\,.\cr
\end{cases}\label{elko10}
\eeq
Hereon we denote $\xi^{-1}(x)\,d\xi(x)$ by $a(x)$ in order to shorten
all formul\ae\, notations.

Now it is straightforward to show that ELKO  can not be eigenspinors of the
exotic Dirac operator  $i\gamma^\mu\nabla_\mu + a(x)$. Indeed,
denoting
\beq
e :=
\left(
\begin{array}{c}
e_1\\
e_2\\
e_3\\
e_4
\end{array}
\right)\,,\qquad
d :=
\left(
\begin{array}{c}
d_1\\
d_2\\
d_3\\
d_4
\end{array}
\right)\,, \nonumber
\eeq
and
$
\Gamma := \mathbb{I}\otimes (i\gamma^\mu\nabla_\mu + a(x))$, Eq.(\ref{elko11}) becomes
$e=\Omega d$. Using
$\left[\Gamma,\Omega\right] =0$ yields
$\Gamma e = \Omega \Gamma d$. Eqs.(\ref{elko10}) imply
$\Gamma d = m\,\gamma^5\otimes\mathbb{I}\,d$ and then $
\Gamma e =
\Omega \left( m\,\gamma^5\otimes\mathbb{I}\right) \Omega^{-1} e$.
An explicit evaluation of
$\mu := \Omega \left( m\,\gamma^5\otimes\mathbb{I}\right) \Omega^{-1}$
reveals
\beq
\hspace*{-200pt}
\mu= m\, \left(
\begin{array}{cc}
\sigma_2 & \mathbb{O} \\
\mathbb{O} & - \sigma_2
\end{array}
\right)\otimes\mathbb{I} \nonumber\,.
\eeq
Thus, making the direct product explicit again,
finally one reaches the result
\beq
\hspace*{-2.2cm}\begin{pmatrix}
i\gamma^\mu\nabla_\mu + a(x)& \mathbb{O} & \mathbb{O} & \mathbb{O} \\
 \mathbb{O} & i\gamma^\mu\nabla_\mu + a(x)& \mathbb{O} & \mathbb{O}  \\
  \mathbb{O} & \mathbb{O}& i\gamma^\mu\nabla_\mu + a(x)& \mathbb{O}   \\
  \mathbb{O} & \mathbb{O} & \mathbb{O} & i\gamma^\mu\nabla_\mu + a(x)  \\
\end{pmatrix}
\begin{pmatrix}
\mathring\lambda^\mathrm{S}_{\ua} \\
\mathring\lambda^\mathrm{S}_{\da} \\
\mathring\lambda^\mathrm{A}_{\ua} \\
\mathring\lambda^\mathrm{A}_{\da}
\end{pmatrix}
-
i m  \mathbb{I}
\begin{pmatrix}
- \,\mathring\lambda^\mathrm{S}_{\da} \\
 \,\mathring\lambda^\mathrm{S}_{\ua} \\
 \,\mathring\lambda^\mathrm{A}_{\da} \\
- \,\mathring\lambda^\mathrm{A}_{\ua}
\end{pmatrix} = 0 \label{elko12}
\eeq
which establishes that
$\left(i\gamma^\mu\nabla_\mu + a(x)\pm m \mathbb{I}\right)$ do not annihilate
the ELKO (dark) spinor fields.
The antisymmetric symbol
defined as $\varepsilon^{\{-,+\}}_{\{+,-\}}:=-1$, the above equations
reduces to
\beq
\left((i\gamma^\mu\nabla_\mu + a(x))\delta_\alpha^\beta +  m\mathbb{I}
\varepsilon_\alpha^\beta\right)\mathring\lambda_\beta^\mathrm{S}(x)=0,\,\quad
\left((i\gamma^\mu\nabla_\mu + a(x))\delta_\alpha^\beta -  m\mathbb{I}
\varepsilon_\alpha^\beta\right)\mathring\lambda_\beta^\mathrm{A}(x)=0,\label{aaa}
\eeq
which are the inherent counterparts of Eqs.(\ref{elko10}). The term
of $\delta_\alpha^\beta$ is
$i\gamma^\mu\nabla_\mu + a(x)$, and the existence of
$\varepsilon_\alpha^\beta$ in the mass term forbids ELKO spinor fields to be eigenspinors of the
$i\gamma^\mu\nabla_\mu + a(x)$ operator.  Namely, the mass terms carry opposite signs and consequently
ELKO
cannot be annihilated by
$\left(i\gamma^\mu\nabla_\mu + a(x) \pm m \mathbb{I}\right)$, because the term
$\varepsilon_\alpha^\beta$ in Eq.(\ref{aaa}), which implies that
$\epsilon^\mathrm{S} = -1$ and $\epsilon^\mathrm{A} = +1$.

Furthermore,  as comprehensively discussed in, e.g., \cite{petry,hess}, we can express $\xi(x) = \exp(i\theta(x))\in$ U(1), $x\in M$. The exotic spin structure term in this way  reads
\beq\label{tu1}
\xi^{-1}(x) d\xi(x) = \exp(-i\theta(x))(i\gamma^\mu\nabla_\mu\theta(x))\exp(i\theta(x)) = i\gamma^\mu\partial_\mu\theta(x).
\eeq
From Eq.(\ref{tu1}), Eqs.(\ref{aaa}) are written as
\beq
\left((i\gamma^\mu \nabla_\mu + i\gamma^\mu\pa_\mu\theta) \delta_\alpha^\beta \pm m\mathbb{I}
\varepsilon_\alpha^\beta\right)\mathring\lambda_\beta^\mathrm{S/A}(x)=0\,.\label{elko3}
\eeq
The exotic Dirac operator
$i\gamma^\mu {\nabla}_\mu + i\gamma^\mu\pa_\mu\theta -  m \mathbb{I}$,
annihilates each of the four exotic Dirac spinor fields $u_\pm(x)$ and  $v_\pm(x)$, but as the wave operator in (\ref{elko3}) couples the $\{\pm,\mp\}$
degrees of freedom such exotic Dirac operator does not annihilate ELKO.

Much has been extensively discussed about the subtle differences between Majorana and ELKO spinor fields, see e. g., \cite{meu}. Both in the Lounesto spinor field classification are type-(5) spinor fields, satisfying (\ref{elko1}). 

We now shall discuss whether the exotic Dirac operator can be considered as a square root of the
Klein\textendash Gordon operator
\textendash~ in the sense that
$(i\gamma^\mu\nabla_\mu+i\gamma^\mu\pa_\mu\theta-m\mathbb{I})(i\gamma^\mu\nabla_\mu+i\gamma^\mu\pa_\mu\theta+m\mathbb{I}
)=(g^{\mu\nu}\nabla_\mu \nabla_\nu + m^2)\mathbb{I}$. This feature must remain true for the ELKO and its exotic partner:
\beq\label{propa}((i\gamma^\mu\nabla_\mu +i\gamma^\mu\pa_\mu\theta)\delta_\alpha^\beta \pm m\mathbb{I}
\varepsilon_\alpha^\beta)((i\gamma^\mu\nabla_\mu+i\gamma^\mu\pa_\mu\theta)
\delta_\alpha^\beta\mp  m\mathbb{I}
\varepsilon_\alpha^\beta)=(g^{\mu\nu}\nabla_\mu \nabla_\nu + m^2)\mathbb{I}\,
\delta_\alpha^\beta,\eeq \noi since the introduction of an exotic spin structure
does not modify the Klein\textendash Gordon propagator fulfillment by dark spinor fields.

The corresponding Klein-Gordon equation is given by
\bege\label{exo1}
(\Box + m^2 +g^{\mu\nu}\nabla_\mu\nabla_\nu\theta + \pa^\mu\theta\nabla_\mu + \pa^\mu\theta\pa_\mu\theta)\mathring\lambda(x)^{S/A}_{\{\pm,\mp\}} = 0,\enge
where $\Box$ denotes the square of the spin-Dirac operator, that can be related to the Laplace-Beltrami operator by the Lichnerowicz formula \cite{lich,lichn,wal}.
 In order that the Klein-Gordon propagator for the exotic ELKO remains the same as the standard Klein-Gordon propagator for the ELKO spinor field, from Eq.(\ref{exo1}) it follows  that

\bege(\Box\theta + \pa^\mu\theta\nabla_\mu + \pa^\mu\theta\pa_\mu\theta)\mathring\lambda^{S/A}_{\{\pm,\mp\}}(x)=0.\label{kg1}\enge
Explicitly, for consistency with the standard formalism it can be written that
\begin{equation}
\mathring\lambda(x)=\binom{\sigma_{2}\phi^{\ast}(x)}{\phi(x)},\qquad
\text{where}\;\;\;
\phi(x)=\binom{\alpha(x)}{\beta(x)},\quad\alpha(x%
),\beta(x)\in\mathbb{C},  \label{011}%
\end{equation} implying that
\bege
\begin{pmatrix}
\Box\theta +i \pa^\mu\theta\nabla_\mu - i\pa^\mu\theta\pa_\mu\theta&0\\
0&\Box\theta +i \pa^\mu\theta\nabla_\mu - i\pa^\mu\theta\pa_\mu\theta\end{pmatrix}\begin{pmatrix}\beta-i\beta^*\\\alpha + i\alpha^*\end{pmatrix} = \begin{pmatrix}0\\0\end{pmatrix}.\label{c1}
\enge
Still, the carrier of the representation space can be written as
\bege
\mr\lambda^{S/A}_{\{\pm,\mp\}}(x) = \binom{(\beta_1 \mp \beta_2)\exp(\mp i\frac{\pi}{4})}{(\alpha_1 \pm \alpha_2)\exp(\pm i\frac{\pi}{4})},\qquad \beta = \beta_1 + i\beta_2,\quad \alpha = \alpha_1 + i\alpha_2.\label{300}\enge
\noindent Note that the condition in Eq.(\ref{c1}) is independent of the function $\theta(x)$ in the case where Im($\alpha$) = $-$Re($\alpha$) and Im($\beta$) = Re($\beta$), by Eq.(\ref{300}). As this condition is too stringent, since we want to analyze the function $\theta(x)$ for an arbitrary ELKO and not for such so particular case, we demand the most general condition given by Eq. (\ref{kg1}) when  arbitrary exotic dark spinor fields are taken into account, since the general case must be formulated without restricting the theory on any particular case as in Eq.(\ref{300}).

Our analysis hereon sheds new light on the character of the function $\theta$ --- that is \emph{a priori} arbitrary --- that defines the exotic topology. Furthermore, it delves into the way how the exotic topology can be constrained to the spacetime metric structure, via the dynamics of exotic ELKO spinor fields.

Since Eq.(\ref{kg1}) holds for every exotic dark spinor field $\mr\lambda^{S/A}_{\{\pm,\mp\}}(x)$, in particular let us analyze the solutions of Eq.(\ref{kg1}) applied to, for instance, $\mathring\lambda^{S}_{\{-,+\}}(x)$. We omit hereon the argument ``($x$)" for simplicity. Using the expression\footnote{Assume a metric compatible covariant derivative operator, we emphasize here that the connection is not required to be symmetric, and it is decomposed into the Christoffel symbol and the contortion tensor.}
\beq\nabla_\mu\mathring\lambda^{S/A}_{\{ \mp,\pm \}} = \pa_\mu \mathring\lambda^{S/A}_{\{ \mp,\pm \}} - \frac{1}{4}\Gamma_{\mu\rho\sigma}\gamma^\rho\gamma^\sigma\mathring\lambda^{S/A}_{\{\mp,\pm \}},\label{choice}\eeq\noi for such case,
after some calculation\footnote{Here we consider the torsionless connection. For the torsion case it must be written ${\nabla}_\mu \lambda = \partial_\mu \lambda - \frac{1}{4} \Gamma_{\mu\rho\sigma} [\gamma^\mu,\gamma^\sigma] \lambda +
      \frac{1}{4} K_{\mu\rho\sigma} \gamma^\rho \gamma^\sigma \lambda$, where $K_{\mu\rho\sigma}$ are the contorsion tensor coefficients. 
Such case is used in the analysis of dark spinor fields in Cosmology in \cite{boe1, boe12}, 
and it is important to remark that in the presence of torsion an
additional dynamical term appears \cite{Pfabbri}. Such more general formalism for while is unnecessary here, since
      our main aim now is to verify that exotic dark spinor fields dynamics  indeed can constraint the metric spacetime structure. By now, we just call some attention to the fact that the torsion fields may act in order to cancel the connection effects in the constraint equation.}  --- denoting $x^0 = t$ --- it follows that
\beq
&&   (\Box\theta)\left. \mathring\lambda^{S/A}_{\{\mp,\pm\}} + (\pa_0\theta)\Big[\pa_0\mathring\lambda^{S/A}_{\{\mp,\pm\}}-\frac{1}{4}\Big((\Gamma_{000}-\Gamma_{011}-\Gamma_{022}-\Gamma_{033})\mathring\lambda^{S/A}_{\{\mp,\pm\}}+i\Gamma_{001}\mathring\lambda^{A/S}_{\{\pm,\mp\}}  \right.\nonumber\\&& \left. + \Gamma_{002} \mathring\lambda^{S/A}_{\{\pm,\mp\}}\mp \Gamma_{003} \mathring\lambda^{S/A}_{\{\mp,\pm\}}\pm i\Gamma_{012}\mathring\lambda^{A/S}_{\{\mp,\pm\}}+ i\Gamma_{013}\mathring\lambda^{A/S}_{\{\pm,\mp\}} \mp \Gamma_{023}\mathring\lambda^{S/A}_{\{\pm,\mp\}}\Big)\Big]\right.\nonumber\\&& \left.-g^{00}(\pa_0\theta)^2\mathring\lambda^{S/A}_{\{\mp,\pm\}}\right.=0\label{4r}
\eeq
\noi The equation above couples \emph{again} all the four exotic spinor fields $\mathring\lambda^{S/A}_{\{\pm,\mp\}}$, in the case of spacetimes
which the associated connection are non zero.

As proposed in, e. g., \cite{boe111,bo121,boe12}, it is possible for cosmological applications, to assume that the dark spinor fields
depend only on the time variable $t$ via a matter field $\kappa(t)$  compatible with homogeneity and isotropy \cite{boe12} and acts as the only dynamical cosmological variable, in such a way that
$\mathring\lambda^{S/A}_{\{\pm,\mp\}}(x)$ can be explicitly written as
\beq\label{zet}
\mathring\lambda^{A/S}_{\{-,+\}}(x)\,=\,\kappa(t)\,\chi^{A/S}_{\{-,+\}},\qquad\quad\quad\mathring\lambda^{A/S}_{\{+,-\}}(x)\,=\,\kappa(t)\,\zeta^{A/S}_{\{+,-\}},
\eeq
where $\zeta^{S/A}$ and $\chi^{S/A}$ are linearly independent constant spinor fields given by \cite{boe12}
\beq\label{zeta}
\chi^{S}_{\{-,+\}} = \begin{pmatrix}0\\ i\,\\1\\0\end{pmatrix},\qquad
\chi^{A}_{\{-,+\}} = \begin{pmatrix}0\\-i\\1\\0\end{pmatrix},\quad \zeta^{S}_{\{+,-\}} = \begin{pmatrix}1\\0\,\\0\\-i\end{pmatrix},\qquad
\zeta^{A}_{\{+,-\}} = -\begin{pmatrix}1\\0\\0\\i\end{pmatrix}.
\eeq\noi
The matter field $\kappa(t)$ was introduced and satisfies a first order
ordinary differential equation in time derivative, involving the time component of the total energy-momentum tensor $\Sigma_{tt}$, the Planck mass, and the Hubble constant. In the limits proposed in \cite{boe12} we can write
\beq\label{oop}
\frac{\dot\kappa}{\kappa} = - \frac{1}{3}\sqrt{{\frac{1}{3M_{\rm Pl}^2}\Sigma_{tt}}} + \mathcal{O}(\kappa^4).\eeq\noi
where $M_{\rm Pl}^ {-2} = 8\pi G$ is the coupling constant. The last term in the right hand side of the equation above is in Ref.\cite{boe12} kept apart, and such an approximation gives robust
 cosmological results in full compliance with the references therein. The most general case shall be 
 considered still in this Section.
 
 Therefore we can write $\kappa(t) = \exp(at)$, where $a$ is the constant given in the equation above.
Using now Eqs.(\ref{zet}) and (\ref{zeta}), and considering each one of the four exotic dark spinor $\lambda^{S/A}_{\{-,+\}}$ components in Eq.(\ref{4r}), we have the system
\beq
\begin{cases}\;\; \qquad\,\qquad\qquad(-i\Gamma_{001}+\Gamma_{002}-  i\Gamma_{013} -\Gamma_{023})\pa_0\theta = 0
\label{cop1}\\\
 i\Box\theta+\pa_0\theta\left(i -\frac{1}{4}\left(i\Gamma_{000}-i\Gamma_{011}-i\Gamma_{022}-i\Gamma_{033}+\Gamma_{012}-i\Gamma_{003}\right)\right)- i(\pa_0\theta)^2 = 0\label{cop2}\\\
\qquad \Box\theta+\pa_0\theta\left(1 -\frac{1}{4}\left(\Gamma_{000}-\Gamma_{011}-\Gamma_{022}-\Gamma_{033}+i\Gamma_{012}-\Gamma_{003}\right)\right)- (\pa_0\theta)^2 =0\label{cop3}\\
\qquad\qquad\qquad\qquad\qquad (\Gamma_{001}-i\Gamma_{002}+\Gamma_{013} +i\Gamma_{023})\pa_0\theta= 0.\label{cop4}\end{cases}
\eeq\noi
The first and fourth equations above together imply that
\beq\label{st11}(\pa_0\theta)\Gamma_{012} = 0,\eeq\noi what means that if $\theta$ is time dependent, it necessarily means that $\Gamma_{012} = 0$. Otherwise, in the case where
$\theta$ does not depends on time, it implies that $\pa_0\theta=0$, and then we obtain the Laplace equation for $\theta$
\beq\label{nabla}\nabla^2\theta=0.\eeq
It is worthwhile to note, by passing, that Eq.(\ref{4r}) in spacetimes where the connection symbols above are zero --- in
the Minkowski space with Cartesian coordinates, for instance --- is reduced to
\beq
&&   \left(\Box\theta+(\pa_0\theta)a -(\pa_0\theta)^2\right)\mathring\lambda^{S/A}_{\{\mp,\pm\}}=0\label{4r1},
\eeq
\noi and in this way the dark spinor field dynamics imposes constraints only on the topological sector determined by $\theta$, and there is no coupling among the four exotic spinor fields.

On the another hand, the second and the third equations in the system above together imply that
\beq\label{st22}\Box\theta + (\pa_0\theta)\left(1-\frac{1}{2}(\Gamma_{000}-\Gamma_{011}-\Gamma_{022}-\Gamma_{033}-\Gamma_{003})\right) - (\pa_0\theta)^2 = 0,\label{thrid}\eeq\noi what means that if $\theta=\theta(t)$, so necessarily $4-(\Gamma_{000}-\Gamma_{011}-\Gamma_{022}-\Gamma_{033}-\Gamma_{003})$ = $\pa_0\theta$. Otherwise, again Eq.(\ref{nabla}) holds.

Now, using Eqs.(\ref{zet}) and (\ref{zeta}) and considering each one of the four exotic dark spinor $\lambda^{S/A}_{\{+,-\}}$ components in Eq.(\ref{4r}), we have the same results as for the $\lambda^{S/A}_{\{-,+\}}$.
It shows that the exotic topology induces constraints in the spacetime geometry, coming from the dynamics
of dark spinor fields. Indeed it is the case in such an approach when the function $\theta(x)$ that generates the exotic structure
--- realized by Eq.(\ref{tu1}) --- is most general, time dependent.

As previously observed, Eq.(\ref{choice}) was solved for $\mr\lambda^{S/A}_{\{\mp,\pm \}}(x)$, in order to illustrate the exotic dark spinor fields dynamics. It evinces the constraints either on the spacetime metric structure --- given an arbitrary 1-form field in spacetime, manifestation of the exotic topology encrypted in the term $\theta(x)$ in Eq.(\ref{tu1}) --- or on the exotic parameter $\theta(x)$.

To the most general case, it is not necessary indeed to consider any particular case about $\kappa(t)$,
and the system (\ref{cop4}) is written as
\beq
\begin{cases}\;\; \qquad\qquad(-i\Gamma_{001}+\Gamma_{002}-  i\Gamma_{013} -\Gamma_{023})\pa_0\theta = 0
\label{cop11}\\
 \Box\theta-\pa_0\theta\left(\frac{1}{4}\left(\Gamma_{000}-\Gamma_{011}-\Gamma_{022}-\Gamma_{033}+i\Gamma_{012}-\Gamma_{003}\right)\right)- (\pa_0\theta)^2 = -(\pa_0\theta)\frac{\dot\kappa(t)}{\kappa(t)}\label{cop21}\\ \Box\theta+\pa_0\theta\left(-\frac{1}{4}\left(\Gamma_{000}-\Gamma_{011}-\Gamma_{022}-\Gamma_{033}+i\Gamma_{012}-\Gamma_{003}\right)\right)- (\pa_0\theta)^2 =-(\pa_0\theta)\frac{\dot\kappa(t)}{\kappa(t)}\label{cop31}\\
\qquad\qquad\qquad (\Gamma_{001}-i\Gamma_{002}+\Gamma_{013} +i\Gamma_{023})\pa_0\theta= 0.\label{cop41}\end{cases}
\eeq\noi The analysis that evinces the constraints among topological and geometrical terms is similar, except for the term
$-(\pa_0\theta)\frac{\dot\kappa(t)}{\kappa(t)}$ on the right hand side of the second and third equations above.
For instance, in the case previously analyzed, all terms in Eq.(\ref{oop}) are given by
 \begin{align}
      \frac{\dot{\kappa}}{\kappa} = -\frac{\sqrt{1/M_{\rm pl}^2}}{4\sqrt{3}}
      \left(\frac{8 + 3 \kappa^4/M_{\rm pl}^4}{12 -\kappa^4/M_{\rm pl}^4}\right)
      \sqrt{4 - \kappa^4/M_{\rm pl}^4},
      \label{oop1}
\end{align}\noi 
and cogently the exotic dark spinor fields dynamics constraints the spacetime topology, or the spacetime metric structure, 
or both, whatever the form of $\kappa(t)$, and also even for the most general dark spinor fields $\mathring\lambda^{S/A}_{\{\mp,\pm\}}$, 
predicted by Eq.(\ref{kg1}).

\section{Concluding Remarks and Outlook}

Given an a priori arbitrary manifold M with non-trivial topology, and
using the fact that the inequivalent spin structures give rise to the
exotic term endowing the Dirac operator --- in our analysis, the exotic
term in the Dirac operator (evinced when an arbitrary inequivalent spin
structure is taken into account) --- we have shown that  the exotic dark spinor fields dynamics  indeed can constraint the metric spacetime structure. Such constraints can be mitigated
for some particular choices of
the exotic term $\theta$ in (\ref{tu1}) --- but in the most general case both the spacetime metric structure and the non-trivial topology are constrained by the exotic dark spinor field dynamics.

Much has been discussed the about equations constraining the dynamics and the spinor structures, and some questions were addressed about the validity of Klein-Gordon propagator globally, but not locally, for dark spinor fields \cite{bo121}. The formalism here introduced is promising
to derive and provide open questions on the dark spinor fields models structures and their subsequent application in cosmology --- in particular the dark matter problem. 

Eq.(\ref{zet}) is successful to decouple topological terms evinced by the exotic $\theta$ function and the geometrical terms given by the connection symbols, in some particular cases analyzed from Eq.(\ref{cop4}) on. As in such situations 
the connection symbols are constrained, it also induces constraints among Christoffel symbols and contorsion tensor components, in the case where torsion is taken into account in the covariant derivative.
In addition, Eq.(\ref{4r}) is the most general coupling between topological and geometrical terms when no particular exotic dark spinor field is considered. 

Besides analyzing the exotic dark spinor fields elicited from a non-trivial topology endowed manifold, such additional term in the Dirac operator may be useful to solve some open questions addressed in the current
literature \cite{allu,alu2,alex,alex1,alu3,horv1,bo121,boe111,boe12,boe13,schritt}.

In certain sense, that idea that is in the background of the theoretical tools through which one set the quoted constraints can be identified with the problem of constraints in gauge theories known as the Velo-Zwanziger problem \cite{VeloZwan1}.
In the context of such theories, to avoid algebraic inconsistencies originated from a kind of exotic interactions, one sets the constraints independently of the equations of motion. Furthermore, in \cite{VeloZwan2} it is used a similar prescription to introduce a more suitable Dirac operator, that can be also related to the one introduced in \cite{Fierz}. In some cases, the Lagrangian device by itself does not provide satisfactory wave equations \cite{VeloZwan2}, a problem that is given an adequate interpretation, and we expect to have overcome.

\section{Acknowledgment}
R. da Rocha is grateful to Conselho Nacional de Desenvolvimento Cient\'{\i}fico e Tecnol\'ogico (CNPq) grants 472903/2008-0 and
304862/2009-6 for financial support. A. E. B. would like to thank the financial support FAPESP 2008/50671-0 and CNPq grant 300233/2010-8.

\appendix

\section{Clifford Bundles}
One thus introduces the Clifford bundle of differential forms
$\mathcal{C\ell(}M,g)$, which is a vector bundle associated with $P_{\mathrm{Spin}%
_{1,3}^{e}}(M\mathbf{)}$ \cite{ro4,wal}. 
Sections of the Clifford bundles are sums of
non-homogeneous differential forms,  called Clifford
fields. Remember that
$\mathcal{C\ell(}M,g)=P_{\mathrm{SO}_{1,3}^{e}}(M)\times
_{\mathrm{Ad}^{\prime}}\cl_{1,3}$, where $\cl_{1,3}%
\simeq$ M(2,${\mathbb{H}})$ is the spacetime algebra and $\mathbb{H}$ denotes the quaternions. Details of
the bundle structure are as follows:\\

(1) Let $\mathbf{\pi}_{c}:\mathcal{C}\ell(M,g)\rightarrow M$ be
the canonical projection of $\mathcal{C}\ell(M,g)$ and let  $\cup_{i\in I}U_i$ be an open simple covering of $M$, together with a set of transition functions $a_{ij}: U_i\cap U_j \rightarrow$ SO$_{1,3}^e$ such that $a_{ij}\circ a_{jk} = a_{ik}$ in $U_i\cap U_j\cap U_k$ and $a_{jj} = $ id.  There are
trivialization mappings
$\mathbf{\psi}_{i}:\mathbf{\pi}_{c}^{-1}(U_{i})\rightarrow U_{i}%
\times\cl_{1,3}$ of the form $\mathbf{\psi}_{i}(p)=(\mathbf{\pi}%
_{c}(p),\psi_{i,x}(p))=(x,\psi_{i,x}(p))$. If $x\in U_{i}\cap
U_{j}$ and $p\in\mathbf{\pi}_{c}^{-1}(x)$, then
\begin{equation}
\psi_{i,x}(p)=h_{ij}(x)\psi_{j,x}(p),\nonumber
\end{equation}
for $h_{ij}(x)\in\mathrm{Aut}(\cl_{1,3})$, where $h_{ij}:U_{i}\cap
U_{j}\rightarrow\mathrm{Aut}(\cl_{1,3})$ are the transition
mappings of $\mathcal{C}\ell(M,g)$. Since every
automorphism of $\cl_{1,3}$ is {inner, }then
$
h_{ij}(x)\psi_{j,x}(p)=a_{ij}(x)\psi_{i,x}(p)a_{ij}(x)^{-1} \nonumber
$
for some $a_{ij}(x)\in\cl_{1,3}^{\star}$, the group of invertible
elements of $\cl_{1,3}$. In particular, a spin structure ($P_{\mathrm{Spin}_{1,3}^{e}}(M),s$) on $M$ is precisely comprised by the system of transition functions $h_{ij}:U_i\cap U_j\rightarrow$ Spin$^e_{1,3}$ such that
\beq\label{jgij}
\varsigma\circ {h_{ij}}=a_{ij},\qquad\quad {h_{ij}}\circ {h_{jk}} = {h_{ik}},\qquad {h_{ii}}=\, {\rm id},\eeq where $\varsigma$ is defined in Eq.(\ref{sigma}).

(2) Since
$\cl_{1,3}^{\star}$ acts naturally on $\cl_{1,3}$ as an algebra automorphism
through its adjoint representation, the group $\mathrm{SO}_{1,3}^{e}$ has a
natural extension in the Clifford algebra $\cl_{1,3}$.  A set of {lifts} of the
transition functions in $\mathcal{C}\ell(M,{g})$ is a set
of elements $\{a_{ij}\}\subset$ $\cl_{1,3}^{\star}$ such that, if\footnote{It is well known
that $\mathrm{Spin}_{1,3}^{e}=\{\phi\in\cl_{1,3}^{0}:\phi\tilde{\phi}%
=1\}\simeq\mathrm{SL}(2,\mathbb{C)}$ is the universal covering
group of the
restricted Lorentz group $\mathrm{SO}_{1,3}^{e}$. Notice that $\cl%
_{1,3}^{0}\simeq\cl_{3,0}\simeq$ M(2,$\mathbb{C})$, the even
subalgebra of $\cl_{1,3}$ is the Pauli algebra.}
\begin{eqnarray}
&&\mathrm{Ad} :\phi\mapsto\mathrm{Ad}_{\phi} \nonumber \\
&&\mathrm{Ad}_{\phi}(\Xi) =\phi \Xi\phi^{-1}, \quad \forall
\Xi\in\cl_{1,3}, \nonumber
\end{eqnarray}
then $\mathrm{Ad}_{a_{ij}}=h_{ij}$ in all intersections.

(3) The application $\mathrm{Ad}|_{\mathrm{Spin}_{1,3}^{e}}$ defines a
group homomorphism
\beq\label{sigma}
\varsigma:\mathrm{Spin}_{1,3}^{e}\rightarrow\mathrm{SO}_{1,3}^{e},\qquad\text{which is onto and ker $\varsigma = \mathbb{Z}_{2}$}.
\eeq\noi Then
Ad$_{\pm 1}=$ identity,
and $\mathrm{Ad}:\mathrm{Spin}_{1,3}^{e}\rightarrow\mathrm{Aut}%
(\cl_{1,3})$ descends to a representation of
$\mathrm{SO}_{1,3}^{e}$.
Let us call $\mathrm{Ad}^{\prime}$ this representation, i.e., $\mathrm{Ad}%
^{\prime}:\mathrm{SO}_{1,3}^{e}\rightarrow\mathrm{Aut}(\cl_{1,3})$.
Then  $\mathrm{Ad}_{\varsigma(\phi)}^{\prime}\Xi=\mathrm{Ad}_{\phi}%
\Xi=\phi\Xi\phi^{-1}$.

(4) It is clear that the structure group of the Clifford bundle
$\mathcal{C}\ell(M,g)$ is reducible from $\mathrm{Aut}%
(\cl_{1,3})$ to $\mathrm{SO}_{1,3}^{e}$. The transition maps of
the principal bundle $P_{\mathrm{SO}_{1,3}^{e}
}(M)$ can thus be --- through $\mathrm{Ad}^{\prime}$ --- taken as
transition maps for the Clifford bundle. It follows that \cite{lawmi,41mosna1}
\begin{equation}
\mathcal{C}\ell(M,g)=P_{\mathrm{SO}_{1,3}^{e}}(M)\times
_{\mathrm{Ad}^{\prime}}\cl_{1,3}, \nonumber
\end{equation}
i.e., the Clifford bundle is a vector bundle associated with the
principal bundle $P_{\mathrm{SO}_{1,3}^{e}}(M)$ of orthonormal coframes.

\section{Principal Bundles and Associated Vector Bundles}

In this Section it is reviewed the main definitions and
concepts of the theory of principal bundles and their associated vector
bundles, which is needed to introduce the
Clifford and spin-Clifford bundles used in this paper. Propositions
are in general presented without proofs, which can be found, e.g., in
\cite{wal,choquet,koni}.

A fiber bundle on a manifold $M$ with Lie group $G$ is denoted by
$(E,M,\mathbf{\pi},G,F)$. $E$ is a topological space called the total space of
the bundle, $\mathbf{\pi}:E\rightarrow M$ is a continuous surjective map,
called the canonical projection, and $F$ is the typical fiber. The following
conditions must be satisfied:\\
a) $\mathbf{\pi}^{-1}(x)$, the fiber over $x$, is homeomorphic to $F$.\\
b) Let $\{U_{i},$ $i\in\mathfrak{I}\}$, where $\mathfrak{I}$ is an index set,
be a covering of $M$, such that:
\begin{enumerate}
\item[b1)] Locally a fiber bundle $E$ is trivial, namely it is diffeomorphic to a
product bundle $\mathbf{\pi}^{-1}(U_{i})\simeq U_{i}\times F$ for all
$i\in\mathfrak{I}$.
\item[b2)] The diffeomorphisms $\Phi_{i}:\mathbf{\pi}^{-1}(U_{i})\rightarrow
U_{i}\times F$ have the form
\begin{equation}
\Phi_{i}(p) = (\mathbf{\pi}(p),\phi_{i,x}(p)),\qquad\quad  \phi_{i}\vert _{\pi^{-1}(x)}   \equiv\phi_{i,x}:\mathbf{\pi
}^{-1}(x)\rightarrow F\text{ is onto}%
\end{equation}
The collection $\{(U_{i},\Phi_{i})\}$, $i\in\mathfrak{I}$, are said to be a
family of local trivializations for $E$.
\item[b3)] The group $G$ acts on the typical fiber. Considering $x\in U_{i}\cap U_{j}$, then, $\phi_{j,x}\circ\phi_{i,x}^{-1}:F\rightarrow F$
must coincide with the action of an element of $G$, for all $x\in U_{i}\cap
U_{j}$ and $i,j\in\mathfrak{I}$.
\item[b4)] One calls transition functions of the bundle the continuous induced
mappings
\begin{equation}
a_{ij}:U_{i}\cap U_{j}\rightarrow G\text{, where }a_{ij}(x)=\phi_{i,x}%
\circ\phi_{j,x}^{-1}.
\end{equation}
\end{enumerate}
For consistence of the theory the transition functions must satisfy the
cocycle condition $
a_{ij}(x)a_{jk}(x)=a_{ik}(x)$.

The 5-tuple $(P,M,\mathbf{\pi},G,F\equiv G)\equiv
(P,M,\mathbf{\pi},G)$ is called a principal fiber bundle {(PFB)} if all
the conditions about fiber bundles are fulfilled and, moreover, if there
is a right action of $G$ on elements $p\in P$, such that:\\
a) the mapping (defining the right action) $P\times G\ni$ $(p,g)\mapsto pg\in
P$ is continuous.\\
b) given $g,g^{\prime}\in G$ and $\forall p\in P$, $(pg)g^{\prime
}=p(gg^{\prime}).$\\
c) $\forall x\in M,\mathbf{\pi}^{-1}(x)$ is invariant under the action of $G$: each element of $p\in\mathbf{\pi}^{-1}(x)$ is mapped into $pg\in
\mathbf{\pi}^{-1}(x)$, i.e., it is mapped into an element of the same fiber.\\
d) $G$ acts free and transitively on each fiber $\mathbf{\pi}^{-1}(x)$, which
means that all elements within $\mathbf{\pi}^{-1}(x)$ are obtained by the
action of all the elements of $G$ on any given element of the fiber
$\mathbf{\pi}^{-1}(x)$. This condition is, of course, necessary for the
identification of the typical fiber with $G$.

A bundle $(E,M,\mathbf{\pi}$, $G$ = GL($n,\mathbb{K}),V$), where
$\mathbb{K}=\mathbb{R}$ or $\mathbb{C}$, and ${V}$ is an
$n$-dimensional vector space over $\mathbb{K}$ is called a vector bundle.

A vector bundle $(E,M,\mathbf{\pi},G,F)$ denoted
$E = P\times_{\rho} F$ is said to be associated to a \emph{PFB} bundle
$(P,M,\mathbf{\pi},G)$ by the linear representation $\rho: G\rightarrow$ GL($V$) --- which is called the {carrier space} of the representation --- if
its transition functions are the images under $\rho$ of the corresponding
transition functions of the \emph{PFB} $(P,M,\mathbf{\pi},G)$. This precisely means the
following: consider the following local trivializations of $P$ and $E$
respectively
\begin{align}
\Phi_{i}: &\mathbf{\pi}^{-1}(U_{i})\rightarrow U_{i}\times G,\quad
\label{fb4.6n}\qquad\qquad \Xi_{i}:\mathbf{\pi}_{1}^{-1}(U_{i})\rightarrow U_{i}\times
F,\\
\Xi_{i}(q)  &  =(\mathbf{\pi}_{1}(q),\chi_{i}(q))=(x,\chi_{i}%
(q)), \qquad\quad  \chi_{i}\vert_{\mathbf{\pi}_{1}^{-1}(x)}\equiv\chi
_{i,x}:\mathbf{\pi}_{1}^{-1}(x)\rightarrow F, \label{fb4.8}%
\end{align}
where $\mathbf{\pi}_{1}:P\times_{\rho}F\rightarrow M$ is the projection of the
bundle associated to $(P,M,\mathbf{\pi},G)$. Then, for all $x\in U_{i}\cap
U_{j}$, $i,j\in\mathfrak{I}$, it follows that
\begin{equation}
\chi_{j,x}\circ\chi_{i,x}^{-1}=\rho(\phi_{j,x}\circ\phi_{i,x}^{-1}).
\end{equation}
In addition, the fibers $\mathbf{\pi}^{-1}(x)$ are vector spaces isomorphic to
the representation space $V$.

Let $(E,M,\mathbf{\pi},G,F)$ be a fiber bundle and $U\subset M$ an open set. A
local section of the fiber bundle $(E,M,\mathbf{\pi},G,F)$ on $U$ is a
mapping
\begin{equation}
s:U\rightarrow E\quad\text{such that}\quad\pi\circ s=Id_{U},
\end{equation}
If $U=M$  $s$ is said to be a {global section}.

There is a relation between sections and local trivializations for principal
bundles. Indeed, each local section $s$ (on $U_{i}\subset M$) for a principal
bundle $(P,M,\mathbf{\pi},G)$ determines a local trivialization $\Phi
_{i}:\mathbf{\pi}^{-1}(U)\rightarrow U\times G,$ of $P$ by setting
$
\Phi_{i}^{-1}(x,g)=s(x)g=pg=R_{g}p.
$
Conversely, $\Phi_{i}$ determines $s$ since
\begin{equation}
s(x)=\Phi_{i}^{-1}(x,e).
\end{equation}

A principal bundle is trivial if and only if it has a global cross section.
A vector bundle is trivial if and only if its associated principal bundle is trivial.
Any fiber bundle $(E,M,\mathbf{\pi},G,F)$ such that $M$ is a
paracompact manifold and the fiber $F$ is a vector space admits a cross section.
Then, any vector bundle associated to a trivial principal bundle has non-zero
global sections. Note however that a vector bundle may admit a non-zero global
section even if it is not trivial. Indeed, as shown in the main text, any
Clifford bundle possesses a global identity section, and some spin-Clifford
bundles admits also identity sections once a trivialization is given.

\end{document}